\documentclass{aa}
\makeatletter
\renewcommand*\aa@pageof{, page \thepage{} of \pageref*{LastPage}}
\makeatother
\usepackage{graphicx}
\usepackage{txfonts}
\usepackage{enumitem}
\usepackage{xcolor}
\usepackage{hyperref}
\usepackage{siunitx}
\usepackage{amsmath,amssymb}
\usepackage{textcomp}
\usepackage{gensymb}
\newcommand{\orcid}[1]{\href{https://orcid.org/#1}{\includegraphics[width=8pt]{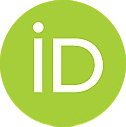}}}
\usepackage{scalerel}
\newcommand\scht[1]{\stretchrel*{$\textsc{#1}$}{\textsc{x}}}
\newcommand{\project}[1]{\textsl{#1}}
\newcommand{\thecannon}{\project{The~Cannon}}
\newcommand{\cannon}{\project{Cannon}}
\newcommand{\apogee}{\project{\textsc{apogee}}}
\newcommand{\argos}{\project{\textsc{argos}}}
\newcommand{\atoa}{\project{\textsc{a\protect\scht{2}a}}}

\newcommand{\twomass}{\project{\textsc{\protect\scht{2}mass}}}
\newcommand{\Gaia}{\project{Gaia}}
\newcommand{\Gaiaeso}{\project{Gaia-\textsc{eso}}}
\newcommand{\galah}{\project{\textsc{galah}}}

\newcommand{\feh}{\mbox{$\rm [Fe/H]$}}
\newcommand{\alphafe}{\mbox{$\rm [\alpha/Fe]$}}
\newcommand{\mgfe}{\mbox{$\rm [Mg/Fe]$}}
\newcommand{\logg}{\mbox{$\log (\mathrm{g})$}}
\newcommand{\teff}{\mbox{$\mathrm{T}_{\rm eff}$}}

\defcitealias{Debattista_2017}{DB17}
\defcitealias{DMatteo_2019}{DM19}
\defcitealias{Fragkoudi_2018}{F18}

\begin{document} 
   \title{A2A: 21,000 bulge stars from the ARGOS survey with stellar parameters on the APOGEE scale}

  \author{S.M. Wylie \inst{1} \orcid{0000-0001-9116-6767} \thanks{swylie@mpe.mpg.de} \and O.E. Gerhard \inst{1} \orcid{0000-0003-3333-0033} \and M.K. Ness \inst{2,3} \orcid{0000-0001-5082-6693} \and J.P. Clarke \inst{1} \orcid{0000-0002-2243-178X} \and K.C. Freeman \inst{4,6} \and J. Bland-Hawthorn \inst{5,6}}

  \institute{Max-Planck-Institut fur Extraterrestrische Physik, Gießenbachstraße, D-85748 Garching, Germany
  \and
  Department of Astronomy, Columbia University, Pupin
  Physics Laboratories, New York, NY 10027, USA
  \and 
  Center for Computational Astrophysics, Flatiron Institute, 162 Fifth Avenue, New York, NY 10010, USA
  \and
  Research School of Astronomy \& Astrophysics, Australian National University, ACT 2611, Australia
  \and
  Sydney Institute for Astronomy, School of Physics, A28, The University of Sydney, NSW 2006, Australia
  \and
  Centre of Excellence for All-Sky Astrophysics in Three Dimensions (ASTRO 3D), Australia
             }
             
   \date{Received-; accepted -}

  \abstract
   {}
   {Spectroscopic surveys have by now collectively observed tens of thousands of stars in the bulge of our Galaxy. However, each of these surveys had unique observing and data processing strategies which led to distinct stellar parameter and abundance scales. Because of this, stellar samples from different surveys cannot be directly combined.}
   {Here we use the data-driven method, $\thecannon$, to bring 21,000 stars from the $\argos$ bulge survey, including 10,000 red clump stars, onto the parameter and abundance scales of the cross-Galactic survey, $\apogee$, obtaining rms precisions of $0.10$ dex, $0.07$ dex, $74$ K, and $0.18$ dex for $\feh$, $\mgfe$, $\teff$, and $\logg$, respectively. The re-calibrated $\argos$ survey - which we refer to as the $\atoa$ survey - is combined with the APOGEE survey to investigate the abundance structure of the Galactic bulge.}
   {We find X-shaped $\feh$ and [Mg/Fe] distributions in the bulge that are more pinched than the bulge density, a signature of its disk origin. The mean abundance along the major axis of the bar varies such that the stars are more $\feh$-poor and $\mgfe$-rich near the Galactic center than in the long bar/outer bulge region. The vertical $\feh$ and $\mgfe$ gradients vary between the inner bulge and long bar with the inner bulge showing a flattening near the plane that is absent in the long bar. The $\feh$-$\mgfe$ distribution shows two main maxima, an ``$\feh$-poor $\mgfe$- rich'' maximum and an ``$\feh$-rich $\mgfe$-poor'' maximum, that vary in strength with position in the bulge. In particular, the outer long bar close to the Galactic plane is dominated by super-solar $\feh$, $\mgfe$-normal stars. Stars composing the $\feh$-rich maximum show little kinematic dependence on $\feh$, but for lower $\feh$ the rotation and dispersion of the bulge increase slowly. Stars with $\feh<-1$ dex have a very different kinematic structure than stars with higher $\feh$.}
   {Comparing with recent models for the Galactic boxy-peanut bulge, the abundance gradients and distribution, and the relation between $\feh$ and kinematics suggest that the stars comprising each maximum have separate disk origins with the ``$\feh$-poor $\mgfe$-rich'' stars originating from a thicker disk than the ``$\feh$-rich $\mgfe$-poor'' stars.}


   \maketitle


\section{Introduction}
The Milky Way bulge is notoriously difficult and expensive to observe due to the high extinction along our sightline to the Galactic Centre. Nevertheless, over the past two decades, the number of spectroscopically observed bulge stars has increased from a few hundred to tens of thousands thanks to multiple spectroscopic stellar surveys such as $\argos$ \citep{Freeman_2012}, $\Gaiaeso$ \citep{Gilmore_2012}, GIBS \citep{Zoccali_2014}, $\apogee$ \citep{Majewski2016}, and $\Gaia$ \citep{Cropper_2018}.

The extensive coverage of these spectroscopic surveys has led to many novel discoveries and vastly improved our understanding of bulge formation and evolution. We know from its wide, multi-peaked metallicity distribution function (MDF) that the bulge is composed of a mixture of stellar populations. This is further supported by the different populations, defined by their metallicities, exhibiting different kinematics \citep{Hill_2011, Ness_2013_IV, Rojas_Arriagada_2014, Rojas_Arriagada_2017, Zoccali_2017}. Through careful chemodynamical dissection, the bulge has been found to contain stars that are part of the bar, inner thin and thick disks, and a pressure supported component \citep{Queiroz_2020b}. Furthermore, there is evidence that the bulge also contains a remnant of a past accretion event, the inner Galaxy structure \citep{Horta_2021}. Multiple age studies of the bulge have reported that while the bulge is mainly composed of old stars (${\sim}10$ Gyr), it contains a non-negligible fraction of younger stars \citep{Bensby_2013, Bensby_2017, Schultheis_2017, Bovy_2019, Hasselquist_2020}. 

While analysis of these surveys has greatly improved our understanding of the bulge, direct comparisons of studies that use different survey data, as well as combinations of the measurements of the stars from different survey pipelines is problematic. This is because different surveys use different selection criteria, wavelength coverage, and spectral resolution. Furthermore, they employ different data analysis methods, assume different underlying stellar models, and make different approximations to derive stellar parameters and individual element abundances from their spectra (see \citet{Jofre_2019} for a review).

Despite these inconsistencies, analyses that employ stars from different surveys are often compared, leading to uncertainty as to whether the results reflect intrinsic properties of the Galaxy or if they are simply due to different observing and data processing strategies. For example, \citet{Zoccali_2017} and \citet{Rojas_Arriagada_2017} found bi-modal bulge MDFs using data from the $\Gaiaeso$ and GIBS surveys, \citet{Rojas_Arriagada_2020} found a three component bulge MDF using data from the $\apogee$ survey, and \citet{Ness_2013_III} found a five component bulge MDF using data from the $\argos$ survey. Because the stars in these surveys have not been observed and analyzed in the exact same manner, it is unclear whether these differences in the bulge MDF arise because of different parameter and abundance scales or because of different selection functions.

In this paper, we use the data driven method, $\thecannon$ \citep{Ness_2015}, to put 21,000 stars from the Galactic bulge survey $\argos$ onto the parameter and abundance scales of the cross-Galactic survey $\apogee$. Of these 21,000 stars, there are roughly 10,000 red clump (RC) stars with accurate distances. By rectifying the scale differences between the two surveys, we can combine them and gain a deeper coverage of the Galactic bulge. We call the re-calibrated $\argos$ catalog the $\atoa$ catalog as we are putting $\argos$ stars onto the $\apogee$ scale. Then, using the combined $\atoa$ and $\apogee$ surveys, we investigate the chemodynamical structure of the bulge. Specifically, we examine how the Iron abundance and Magnesium enhancement vary over the bulge as well as their kinematic dependencies.

The paper is structured as follows: In Section \ref{Data} we describe the $\argos$ and $\apogee$ surveys as well as highlight the inconsistencies between them which make directly combining them questionable. In Section \ref{The Cannon} we summarize the technical background of $\thecannon$. In Section \ref{a2acat} we explain how we apply $\thecannon$ to the $\argos$ catalog to create the $\atoa$ catalog. In Section \ref{a2acat} we also describe the three validation tests we perform to verify that the label transfer was successful. In Section \ref{selfunc} we discuss the selection functions of the $\atoa$ and $\apogee$ surveys. In Section \ref{msb} we use the $\atoa$ and $\apogee$ catalogs to examine the abundance structure of the Galactic bulge. In Section \ref{Discussion} we discuss the results of the paper in more detail and finally in Section \ref{Conclusion} we end the paper with our conclusions.

\section{Data} \label{Data}
In this section, we provide some background on the data used in this paper before discussing their main properties. 

\subsection{ARGOS}\label{ARGOS}
\begin{figure}
\centering
\includegraphics{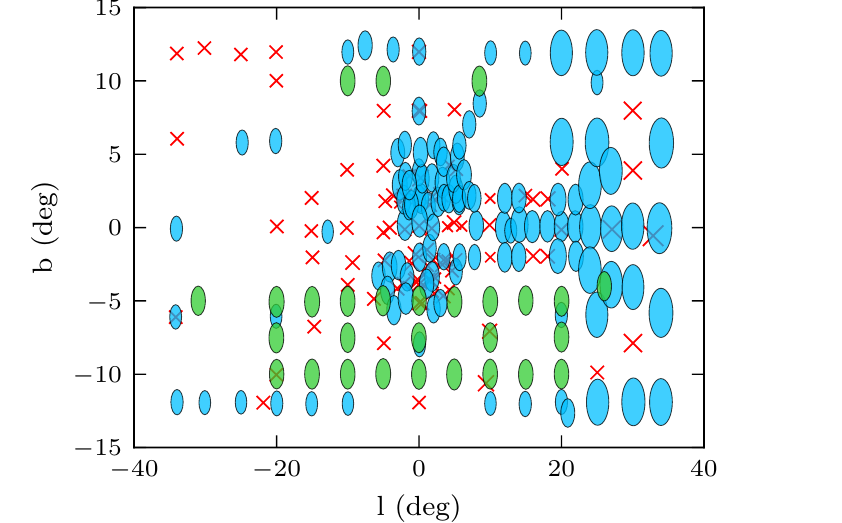} 
\caption{Locations of the $\apogee$ (blue ellipses and red crosses) and $\argos$ (green ellipses) bulge fields. The red crosses indicate $\apogee$ fields that either have no or poor survey selection function estimates. The marker size indicates the field size.} 
\label{fig:fieldloc}
\end{figure}

The Abundance and Radial Velocity Galactic Origin Survey ($\argos$; \citet{Freeman_2012, Ness_2013_III, Ness_2013_IV}) is a medium resolution spectroscopic survey designed to observe RC stars in the Galactic bulge. Using the AAOmega fibre spectrometer on the Anglo-Australian Telescope, ARGOS observed nearly $28,000$ stars located in $28$ fields directed towards the bulge. The field locations are shown as green ellipses in Figure \ref{fig:fieldloc}. The observations were performed across a wavelength region of 840$-$885 nm at a resolution of $\mathrm{R}  = \lambda/\delta\lambda\simeq 11,000$, where $\delta\lambda$ is the spectral resolution element. 

The $\argos$ team determined the Iron abundance ($\feh$), surface gravity ($\logg$), and alpha enhancement ($\alphafe$) of each star in their catalog using $\chi^{2}$ minimisation to find the best fit between the observed spectra and a library of synthetic spectra. The Local Thermodynamic Equilibrium stellar synthesis program MOOG \citep{Moog2009} was used to generate the library of spectra. The effective temperature ($\teff$) was determined from the stellar colours $(\mathrm{J}-\mathrm{K}_{\mathrm{s}})_{0}$ using the calibration by \citet{Bessell_1998}. For more information about the $\argos$ parameter and abundance determination process see \citet{Freeman_2012}. 

The $\argos$ catalog used in this paper contains $25,712$ stars/spectra from the original $28,000$ observed. The missing stars/spectra were removed because they had low signal-to-noise (SNR) and poor quality spectra \citep{Freeman_2012}. The number of pixels of each $\argos$ flux array is 1697. To process the data, we re-normalized the remaining $\argos$ spectra by dividing each by a Gaussian-smoothed version of itself, with the Calcium-triplet lines removed, using a smoothing kernel of 10 nm. We also transformed each spectrum to a common rest frame and masked out the diffuse interstellar band at 8621 nm. We also masked out the region around 8429 nm as we found that this region had a strong residual between the mean spectra of positive and negative velocity stars, indicating that it is systematically affected by the velocity shift.

\subsection{APOGEE}\label{APOGEE}
The Apache Point Observatory Galactic Evolution Experiments ($\apogee$; \citet{Majewski2016}) is a program in the Sloan Digital Sky Survey (SDSS) that was designed to obtain high resolution spectra of red giant stars located in all major components of the Galaxy. The survey operates two telescopes ; one in each hemisphere with identical spectrographs which observe in the near-infrared between $1.5\mu \mathrm{m}$ to $1.7\mu \mathrm{m}$ at a resolution of $\mathrm{R} \simeq 22,500$. In this work, we use the latest data release, DR16 \citep{ahumada2019sixteenth}, which is the first data release to contain stars observed in the southern hemisphere. The locations of the $\apogee$ fields used in this work are shown in Figure \ref{fig:fieldloc} as blue ellipses and red crosses.

Stellar parameters and abundances of the $\apogee$ stars used in this work were obtained from the $\apogee$ Stellar Parameters and Chemical Abundance Pipeline (ASPCAP; \citet{Garc_a_P_rez_2016, Holtzman_2018, jnsson2020apogee}). This pipeline uses the radiative transfer code Turbospectrum \citep{Plez_1992, Plez_2012} to build a grid of synthetic spectra. The parameters and abundances were determined using the code FERRE \citep{Allende_Prieto_2006} which iteratively calculates the best-fit between the synthetic and observed spectra. The fundamental atmospheric parameters such as $\logg$, $\teff$, and overall metallicity were determined by fitting the entire $\apogee$ spectrum of a star. Individual elemental abundances were determined by fitting spectral windows within which the spectral features of a given element are dominant.

We obtain spectrophotometric distances for the $\apogee$ stars from the AstroNN catalog \citep{Leung_2019, Mackereth_2019}, which derives them from a deep neural network trained on stars common to both $\apogee$ and $\Gaia$ \citep{gaia_colab_2018}.

In this work, we specifically focus on $\apogee$ stars located in fields directed towards the bulge with $|\mathrm{l}_{\mathrm{f}}|<35\degree$ and $|\mathrm{b}_{\mathrm{f}}|<13\degree$ where $\mathrm{l}_{\mathrm{f}}$ and $\mathrm{b}_{\mathrm{f}}$ are the Galactic longitude and latitude locations of the fields. We remove 6 fields that were designed to observe the core of Sagittarius. We require that the stars are part of the $\apogee$ main sample (MSp) by setting the $\apogee$ flag, EXTRATARG, to zero. We refer to this sample as the $\apogee$ bulge MSp. To ensure that the stars we use have trustworthy parameters and abundances, we also require the stars to have valid ASPCAP parameters and abundances, $\mathrm{SNR}\geq60$, $\teff\geq3200$ K, and no $\mathrm{Star}\_\mathrm{Bad}$ flag set (23rd bit of ASPCAPFLAG = 0). After applying these cuts, there are $172$ remaining bulge fields containing $37,313$ stars. For reference, we refer to this sample as the HQ $\apogee$ bulge MSp.

In the analysis sections of this work, unless explicitly stated otherwise, we further restrict our $\apogee$ sample to only stars for which we can obtain good selection function estimates. This sample contains $23,512$ stars and we refer to it as the HQSSF $\apogee$ bulge MSp. See Section \ref{apogee_sf_sec} for further details on this sample.

\subsection{Survey Inconsistencies}\label{Survey Inconsistencies}
\begin{figure}
\centering
\includegraphics{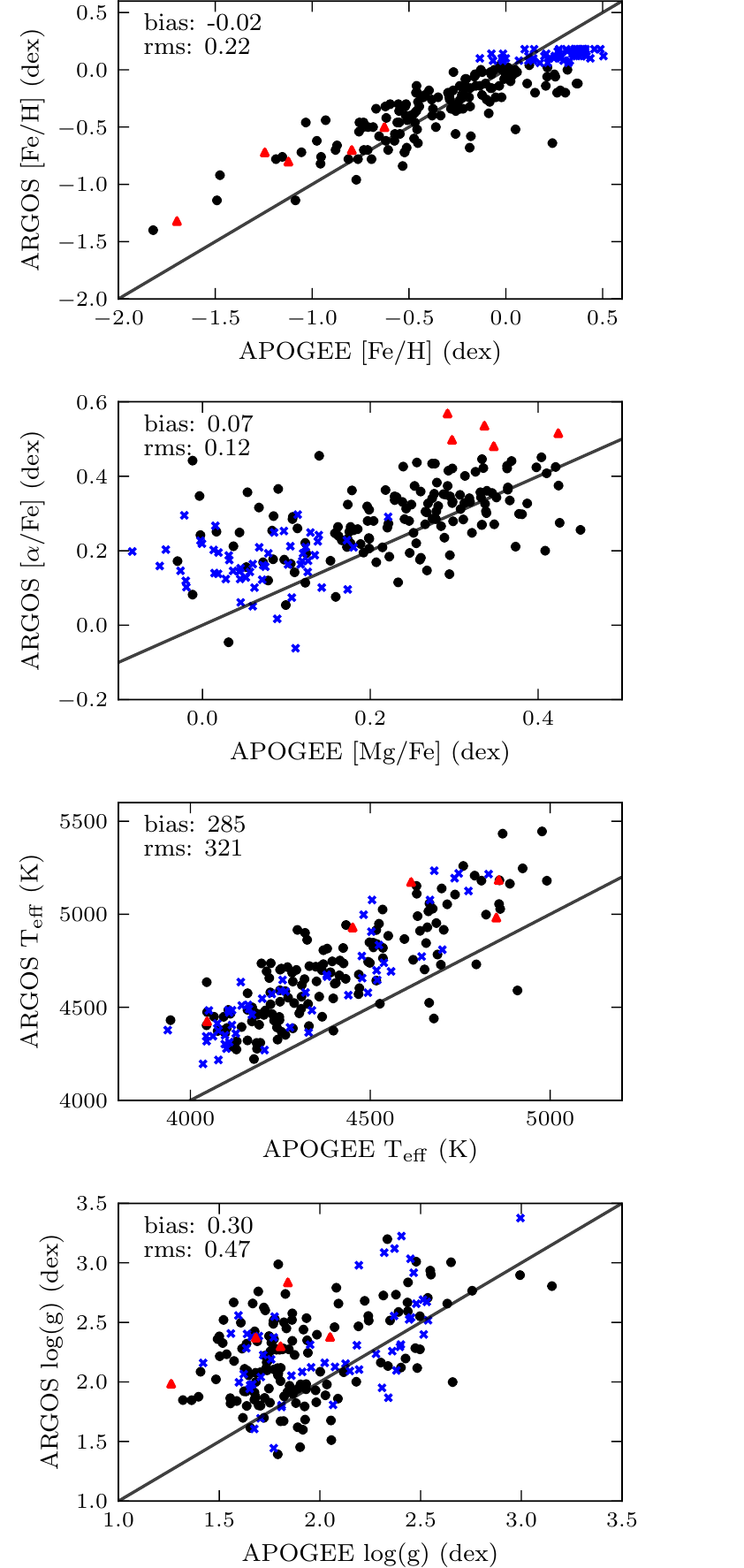}
\caption{$\apogee$ derived parameters (x-axis) versus $\argos$ derived parameters (y-axis) for the 204 reference set stars observed by both surveys. The bias (mean of the differences) and rms of the distributions are given in the upper left hand corner of each plot. The reference set stars that are within one $\argos$ observational error, $0.13$ dex ($0.1$ dex), from the maximum $\feh$ ($\alphafe$) value reached by the reference set are plotted as blue crosses (red triangles).}
\label{fig:orig_compare}
\end{figure}

After the removal of potential binaries though visual inspection of individual spectra, we find $204$ stars that are observed by both the $\apogee$ and $\argos$ surveys. Using these stars we can determine whether the surveys are consistent by checking that they derive the same parameters and abundances for the same stars. 

In Figure \ref{fig:orig_compare}, we compare the $\feh$, $\alpha$-enhancements, $\teff$, and $\logg$ of the common stars. The $\argos$ $\alpha$-enhancement is the average of the individual $\alpha$-elements over Iron ($\alphafe$). For $\apogee$ stars, ASPCAP provides individual $\alpha$-elements with respect to Iron ($\mgfe$, $[\mathrm{O}/\mathrm{Fe}]$, $[\mathrm{Ca}/\mathrm{Fe}]$, ...) as well as an average of the $\alpha$-elements to metallicity ($[\alpha/\mathrm{M}]$). For our comparisons, we choose the ASPCAP Magnesium enhancement ($\mgfe$) because Magnesium is produced only by SN-II with no contribution from SN-Ia. The bias and rms of each distribution are given in the upper left hand corner of each plot. The bias is calculated by subtracting the $\apogee$ values from the $\argos$ values and taking the mean of the differences. 

The $\feh$ comparison in the first plot of Figure \ref{fig:orig_compare} shows that the surveys roughly agree between ${\sim}-0.75$ dex and ${\sim}0$ dex (limits in $\apogee$ $\feh$) with a scatter of ${\sim}0.16$ dex. However, beyond these limits, the deviation between the surveys increases, reaching up to ${\sim}0.4$ dex. The $\alpha$-enhancement Magnesium-enhancement comparison in the second plot shows that the $\argos$ $\alphafe$ estimates are on average ${\sim}0.07$ dex larger than the $\apogee$ $\mgfe$ estimates. If we instead compare the $\argos$ $\alphafe$ estimates to the $\apogee$ $\alphafe$ estimates then the bias and rms of the distribution are larger at $0.12$ dex and $0.15$ dex respectively. The $\teff$ comparison in the third plot shows that the $\argos$ $\teff$ estimates are on average ${\sim}300$ K hotter than the $\apogee$ $\teff$ estimates. Finally, while there is a lot of scatter, the $\logg$ comparison in the last plot shows that the $\argos$ $\logg$ values are generally higher than the $\apogee$ $\logg$ values.

The parameter comparisons show that for most of the common stars, the $\apogee$ and $\argos$ parameters differ significantly. This could be due to a number of factors such as observing in different wavelength regions (e.g. optical in $\argos$ versus infrared in $\apogee$), their use of different data analysis methods (e.g. photometric temperatures in $\argos$ versus spectroscopic temperatures in $\apogee$), or their use of different stellar models. In the following sections, we use the data driven method, $\thecannon$, and this set of common stars to bring the $\apogee$ and $\argos$ surveys on to the same parameter and abundance scales, thereby correcting the deviations we see in Figure \ref{fig:orig_compare}.

\section{The Cannon Method}\label{The Cannon}
$\thecannon$ is a data-driven method that can cross-calibrate spectroscopic surveys. It has the advantage that it is very fast, requires no direct spectral model, and has measurement accuracy comparable to physics based methods even at lower SNR. $\thecannon$ has been previously used to put different surveys on the same parameter and abundance scales using common stars \citep{Casey_2017, Ho_2017, Birky_2020, Galgano_2020, wheeler2020abundances}.

$\thecannon$ uses a set of reference objects with known labels (i.e. $\teff$, $\logg$, $\feh$, [X/Fe] ...), that well describe the spectral variability, to build a model to predict the spectrum from the labels. This model is then used to re-label the remaining stars in the survey. The word "label" is a machine learning term that we use here to refer to stellar parameters and abundances together with one term. The set of common stars used to build the model is called the reference set.

$\thecannon$ is built on two main assumptions:
\begin{enumerate}
\item Stars with the same set of labels have the same spectra.
\item Spectra vary smoothly with changing labels.
\end{enumerate}

Consider two surveys, A and B, where we want to put the stars in survey A onto the label scales of survey B using $\thecannon$. Assume we also have the required set of common stars between the two surveys to form the reference set. To cross calibrate the surveys, $\thecannon$ performs two main steps: the training step and the application step. During the training step, the spectra from survey A and the labels from survey B of the reference set stars are used to train a generative model. Then, given a set of labels, this model predicts the probability density function for the flux at each wavelength/pixel. During the application step, the spectra of a new set of survey A stars (not the reference set) are relabeled by the trained model. We call this set of stars the application set. If the region of the spectra fit to carries the label information and the reference set well represents the application set, then the new labels of survey A's stars should be on survey B's label scales. The success of the re-calibration can be quantified with a cross-validation procedure, the pick-on-out test described in Section \ref{validation_tests}, which returns the systematic uncertainty with which labels can be inferred from the data, as well as a comparison of the generated model spectra to the observational spectra for individual stars, using a $\chi^2$ metric.

In the next two subsections, we will describe the main steps of $\thecannon$ in more detail. 

\subsection{The Training Step}\label{The Training Step}
During the training step, a generative model is trained such that it takes the labels as input and returns the flux at each wavelength/pixel of the spectrum. The functional form of the generative model, $f_{n\lambda}$, can be written as a matrix equation:
\begin{equation}\label{modelequ}
f_{n\lambda} = \theta^{T}_{\lambda} \cdot l_{n} + \sigma,
\end{equation}
where $\theta_{\lambda}$ is a coefficient matrix, $l_{n}$ is a label matrix, and $\sigma$ is the noise. The subscript $n$ indicates the reference set star while subscript $\lambda$ indicates the wavelength/pixel. 

The coefficient matrix, $\theta_{\lambda}$, contains the coefficients that control how much each label affects the flux at each pixel. The coefficients are calculated during the training step. 

Here the label matrix, $l_{n}$, is quadratic in the labels and for each star (each column in the matrix) has the form:
\begin{equation} 
l_{n} \equiv [1,\, l_{(1...n)},\, l_{(1...n)}  \cdot l_{(1...n)}].
\end{equation}
If, for example, the generative model is trained on the labels $\teff$ and $\logg$, then each column in the label matrix would be:
\begin{equation} 
l_{n} \equiv  [1,\, \teff,\, \logg,\, \teff \cdot \teff,\, \logg \cdot \logg,\, \teff \cdot \logg ]
\end{equation}

The noise, $\sigma$, is the rms combination of the uncertainty in the flux at each wavelength due to observational errors, $\sigma_{n\lambda}$, and the intrinsic scatter at each wavelength in the model, $s_{\lambda}$.

Equation \ref{modelequ} corresponds to the single-pixel log-likelihood function:
\begin{equation} \label{loglike}
\ln p(f_{n\lambda} | \theta_{\lambda}^{T}, l_{n},s_{\lambda}) = -\frac{1}{2}\frac{[f_{n\lambda} - \theta_{\lambda}^{T} \cdot l_{n}]^{2}}{s_{\lambda}^{2} + \sigma_{n\lambda}^{2}}\\
 - \frac{1}{2}\ln (s_{\lambda}^{2} + \sigma_{n\lambda}^{2}) .
\end{equation}
During the training step, the coefficient matrix  $\theta_{\lambda}$ and the model scatter $s_{\lambda}$ are determined by optimising the single-pixel log-likelihood in Equation \ref{loglike} for every pixel separately:
\begin{equation}\label{loglike_2}
\theta_{\lambda}, s_{\lambda} \leftarrow \operatorname*{argmax}_{\theta_{\lambda}, s_{\lambda}} \sum_{n = 1}^{\mathrm{N}_{\mathrm{stars}}} \ln p(f_{n\lambda} | \theta_{\lambda}^{T}, l_{n},s_{\lambda}).
\end{equation}
During this step $\thecannon$ uses the reference set stars to provide the label matrix, $l_{n}$. The label matrix is held fixed while the coefficient matrix and model scatter are treated as free parameters.

\subsection{The Application Step} \label{The Test Step}
In the training step, we have the label matrix, $l_{n}$, and we solve for the coefficient matrix $\theta_{\lambda}$ and the scatter $s_{\lambda}$. In the application step, we do the opposite: we have the coefficient matrix $\theta_{\lambda}$ and the scatter $s_{\lambda}$ and we solve for a new label matrix, $l_{m}$. The subscript $m$ is used in this step because the label matrix now corresponds to stars in the application set, not the reference set. 

The label matrix, $l_{m}$, is solved for by optimising the same log-likelihood function as Equation \ref{loglike}. However, here this optimisation is performed using a non-linear least squares fit over the whole spectrum, instead of per pixel:
\begin{equation}
l_{m} \leftarrow \operatorname*{argmax}_{l_{m}} \sum_{\lambda = 1}^{\mathrm{N}_{\mathrm{pix}}} \ln p(f_{m\lambda} | \theta_{\lambda}^{T}, l_{m},s_{\lambda}).
\end{equation}

\section{A2A Catalog}\label{a2acat}
In this paper, we use $\thecannon$ to put the stars from the $\argos$ survey onto the $\apogee$ survey's label scales for the following labels: $\feh$, $\mgfe$, $\teff$, $\logg$, and K-band extinction ($\mathrm{A}_{\mathrm{k}}$). After applying $\thecannon$ to the $\argos$ survey, we obtain a new catalog containing the same stars observed by the $\argos$ survey but with new label values. Other labels, such as line-of-sight velocity or apparent magnitude, remain unchanged. We call this new catalog the $\atoa$ catalog. In this Section, we describe the reference and application sets used to build the $\atoa$ catalog and perform three validation tests to confirm that the $\atoa$ catalog is on the $\apogee$ scale. Lastly, we compare the $\atoa$ catalog to the $\argos$ catalog and explain how we extract the $\atoa$ RC and corresponding distances.

\subsection{Reference and Application Sets}\label{refapp}
\begin{figure}
\centering
\includegraphics{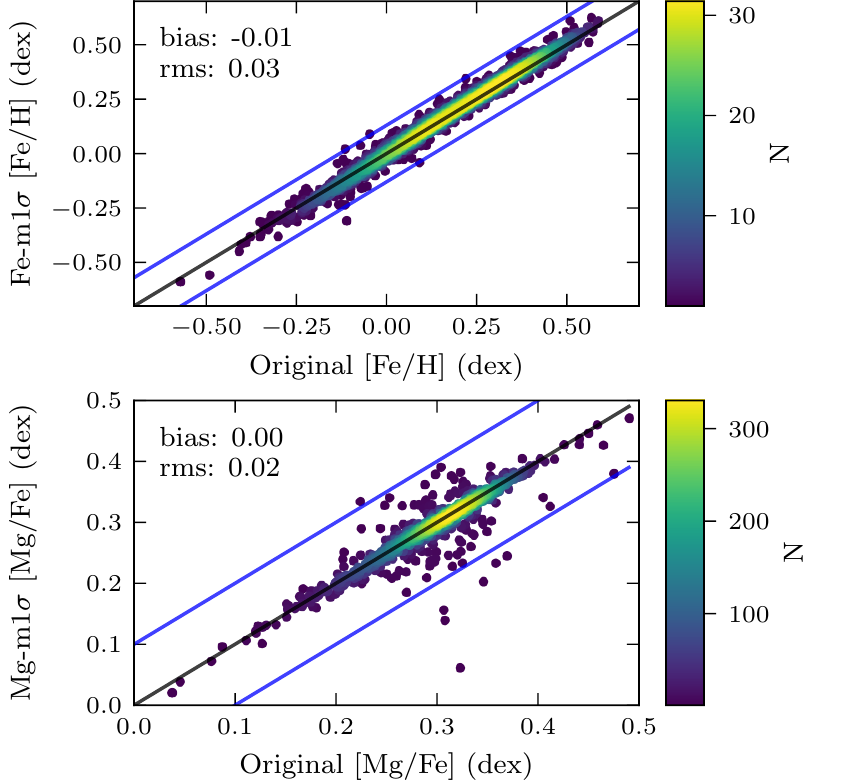} 
\caption{Comparison of the $\feh$ (top) and $\mgfe$ (bottom) labels generated by the original $\cannon$ model (x-axis) and the m1$\sigma$ models (y-axis). In the top plot, only stars with $\argos$ $0.05<\feh \, (\mathrm{dex}) < 0.18$ are compared (the region in $\feh$ spanned by the blue crosses in the top plot of Figure \ref{fig:orig_compare}). In the bottom plot, only stars with $\argos$ $0.47<\alphafe \, (\mathrm{dex}) < 0.57$ are compared (the region in $\alphafe$ spanned by the red triangles in the plot second from the top in Figure \ref{fig:orig_compare}). The points are coloured by the point density.The black lines are one-to-one lines and blue lines indicate $\pm 1\sigma_{\mathrm{ARG}}$. The bias and rms of each distribution are given in the top left corner of each plot.}
\label{fig:extrapolation}
\end{figure}

A reference set of stars, which is used to train $\thecannon$'s model, is comprised of the stars that are observed by both surveys. The labels for this reference set come from the survey with the desired label scale (in our case $\apogee$), while the spectra are taken from the other (in our case $\argos$). The model that is learned at training time should only be applied to stars that are well represented by the reference set. That is, applied to stars that span the label region of the training data, within which the model can interpolate but need not extrapolate. This can also be thought of as a selection in spectra. In our case, the 204 stars that are common to both the $\apogee$ and the $\argos$ surveys (discussed in Section \ref{Survey Inconsistencies}) form the reference set for our $\cannon$ model. The average SNR of the reference set is $46$ for $\argos$ and $107$ for $\apogee$.

These reference set stars are found in the following intervals in the $\argos$ parameter space:
\begin{subequations}\label{first:main}
\begin{gather}
4195 \leq \mathrm{T}_{\mathrm{eff}} \,\, (\mathrm{K}) \leq 5444 \label{first:a}\\
1.393 \leq \mathrm{log}({\mathrm{g}}) \,\, (\mathrm{dex}) \leq 3.376 \label{first:b}\\
-1.4 \leq [\mathrm{Fe}/\mathrm{H}]  \,\, (\mathrm{dex})\leq 0.18 \label{first:c}\\
-0.062 \leq [\alpha/\mathrm{Fe}]  \,\, (\mathrm{dex}) \leq 0.569 \label{first:d}
\end{gather}
\end{subequations}
We ignore the limits in $\mathrm{A}_{\mathrm{k}}$ as this label was only included to stabilize the fits to the other labels. As such, we do not use the learned $\mathrm{A}_{\mathrm{k}}$ label for science.  

There are 20,435 $\argos$ stars (${\sim}79\%$ of the $\argos$ catalog) within the 4D parameter space defined by intervals \ref{first:a} to \ref{first:d}. These stars are considered to be well represented by the reference set and normally would form our application set. However, there are many stars with parameter values close to but just outside of the reference set limits. For example, if we could extend all the limits by 1$\sigma_{\mathrm{ARG}}$, equal to the $\argos$ observational error of each label, then we would include 2704 more stars in the application set (${\sim}10.5\%$ more of the $\argos$ catalog). Because these stars are still close to the reference set stars, the labels returned by $\thecannon$ for these stars may be correct to the first order. To test whether we can extend any of the limits we use the following procedure:
\begin{enumerate}
  \item Remove reference set stars 1$\sigma_{\mathrm{ARG}}$ from each limit. This decreases the number of reference set stars.
  \item Train a new $\cannon$ model on the reduced reference set. For clarity, we refer to this model as the minus-one-sigma (m1$\sigma$) model.
  \item Reprocess $\argos$ spectra using the m1$\sigma$ model to obtain new $\cannon$ parameters for each star. The application set remains the same as the one processed by the original $\cannon$ model.
  \item Compare the new $\cannon$ labels from the m1$\sigma$ model to the labels given by the original $\cannon$ model.
\end{enumerate}
For reference, the $\argos$ observational errors, $\sigma_{\mathrm{ARG}}$, for $\feh$, $\alphafe$, $\teff$, and $\logg$ are: $0.13$ dex, $0.1$ dex, $100$ K, and $0.3$ dex respectively.

We applied this test to each label limit separately and found that the extrapolation works best for the high $\feh$ and high $\alphafe$ limits. In Figure \ref{fig:extrapolation}, we compare the output labels produced by the original $\cannon$ model against the output labels produced by the m1$\sigma$ models trained on the reduced reference sets. The original $\cannon$ model was trained on all 204 reference set stars, shown as the black, blue, and red markers in Figure \ref{fig:orig_compare}. The Fe-m1$\sigma$ model (top plot) was trained on 149 reference set stars, shown as just the black and red markers in Figure \ref{fig:orig_compare}. The Mg-m1$\sigma$ model (second plot) was trained on 199 reference set stars, shown as just the  black and blue markers in Figure \ref{fig:orig_compare}. For each comparison (or plot) we only compare stars that have $\argos$ parameter values in the 1$\sigma_{\mathrm{ARG}}$ region we removed (regions occupied by the blue and red markers in Figure \ref{fig:orig_compare}). For both labels, less than $1\%$ of stars have original $\cannon$ model and m1$\sigma$ model labels that differ by more than 1$\sigma_{\mathrm{ARG}}$.

We also compared the labels produced by the original $\cannon$ model to the labels produced by a $\cannon$ model that was trained on a reference set that was simultaneously reduced by 1$\sigma_{\mathrm{ARG}}$ in $\feh$ and $\mgfe$. The reference set of this model consists of only 144 stars, shown as the black markers in Figure \ref{fig:orig_compare}. We find that less than $1\%$ of stars from this model have labels that differ from the original $\cannon$ model by more than 1$\sigma_{\mathrm{ARG}}$ in $\feh$ and $2\%$ in $\mgfe$.

\begin{figure}
\centering
\includegraphics{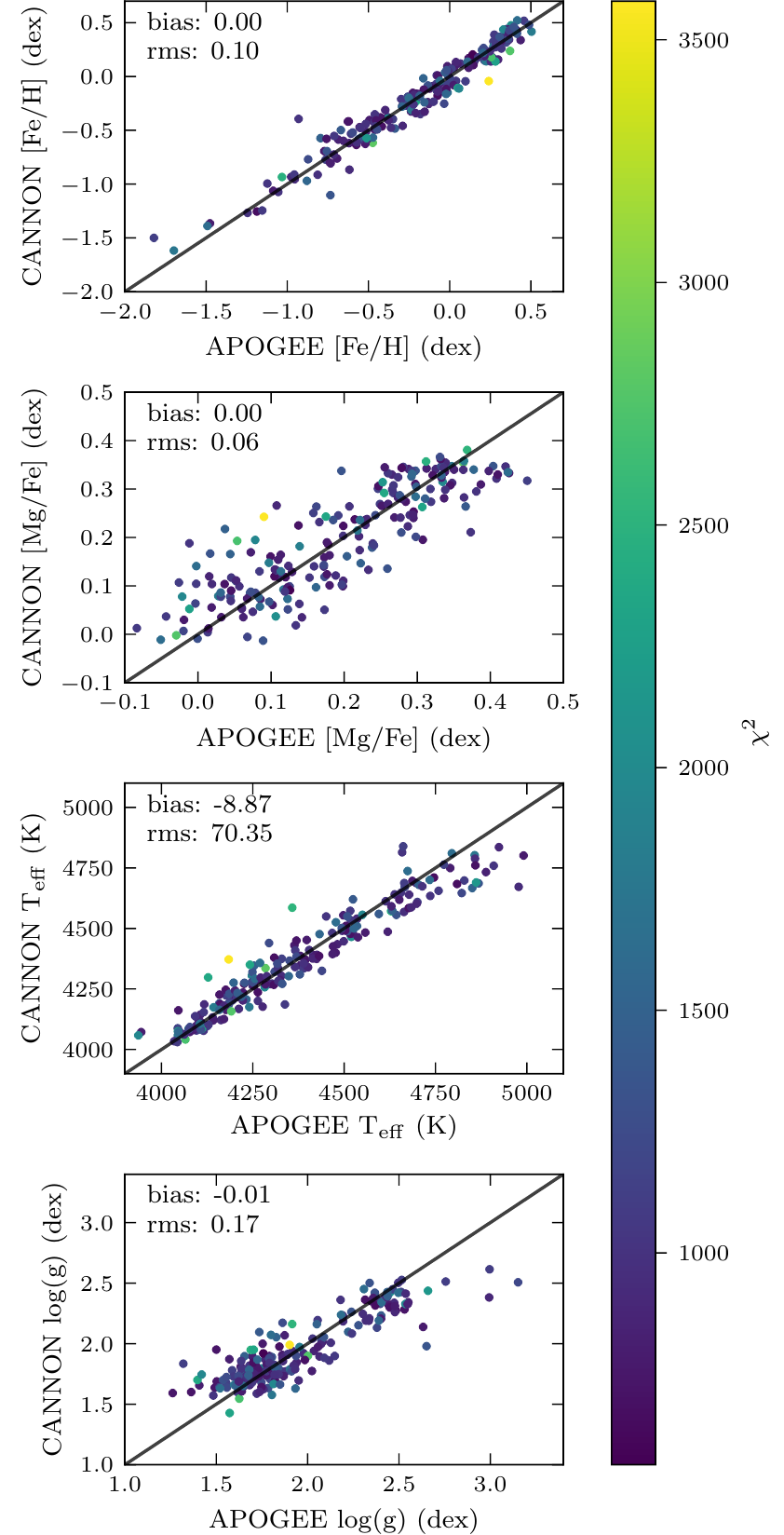} 
\caption{Pick-one-out test. For each plot, each point represents a different reference set star. For a given point in a plot, the x-axis value is the $\apogee$ derived label while the y-axis value is the label prediction from a $\cannon$ model trained on all other (203) reference set stars. Therefore, for each point in each plot the applied $\cannon$ model is different than that of every other point. The points are coloured by their model $\chi^{2}$ values. The bias and rms of each distribution are given in the top left corner of each plot.}
\label{fig:pickoneout}
\end{figure}

\begin{figure}
\centering
\includegraphics{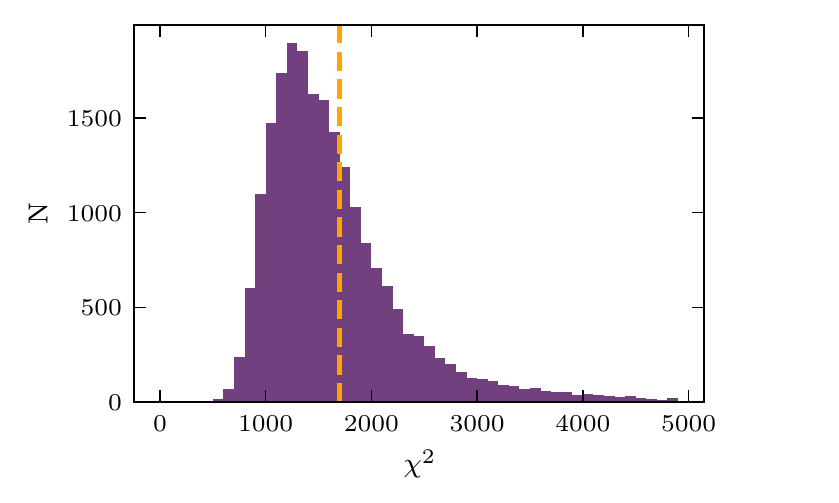} 
\caption{The model $\chi^{2}$ distribution of $\atoa$ stars. The dashed orange line gives the number of pixels in each $\atoa$/$\argos$ spectrum.}
\label{fig:chi2}
\end{figure}

Because we find that we can accurately predict the $\feh$ and $\mgfe$ labels of stars in the 1$\sigma_{\mathrm{ARG}}$ regions of the parameter space removed from the reference set, we make the assumption that we can apply the $\cannon$ model trained on all 204 reference set stars to stars with $\argos$ parameters 1$\sigma_{\mathrm{ARG}}$ beyond the high $\feh$ and $\alphafe$ limits and still get approximately correct labels. Thus, we extend the limits of \ref{first:c} and \ref{first:d} to be:
\begin{subequations}\label{third:main}
\begin{gather}
-1.4 \leq [\mathrm{Fe}/\mathrm{H}] \,\, (\mathrm{dex}) \leq 0.31 \label{third:a}\\
-0.062 \leq [\alpha/\mathrm{Fe}] \,\, (\mathrm{dex}) \leq 0.669 \label{third:b}
\end{gather}
\end{subequations}
Limits \ref{first:a},\ref{first:b}, \ref{third:a}, and \ref{third:b} enable us to process $~85\%$ of the $\argos$ catalog, or $21,577$ stars. This increases the number of stars in the $\atoa$ catalog with $\feh$ above $0.5$ dex by roughly $45\%$, the number of stars with $\feh$ between 0 dex and $0.5$ dex by roughly $23\%$, and the stars with $\feh$ below $-1$ dex by roughly $10\%$. 

In Section \ref{seca2aargos}, we define our $\atoa$ catalog used for the bulge analysis. This final catalog has an additional colour cut applied (Equation \ref{second}) which removes an additional $252$ stars leaving $21,325$ stars in the final catalog. If the same colour cut is applied to the original $\argos$ catalog then the $\argos$ catalog would contain $23,487$ stars. The final colour cut $\atoa$ catalog is then $91\%$ complete compared to the colour cut $\argos$ catalog. The parameter and abundance errors of the colour cut $\atoa$ catalog are calculated in Section \ref{validation_tests}. For $\feh$, $\mgfe$, $\teff$, and $\logg$, the rms of the errors, $\sigma_{\mathrm{A2A}}$, are $0.10$ dex, $0.07$ dex, $74$ K, and $0.18$ dex respectively.

\subsection{Validation Tests}\label{validation_tests}
\begin{figure*}
\centering
\includegraphics{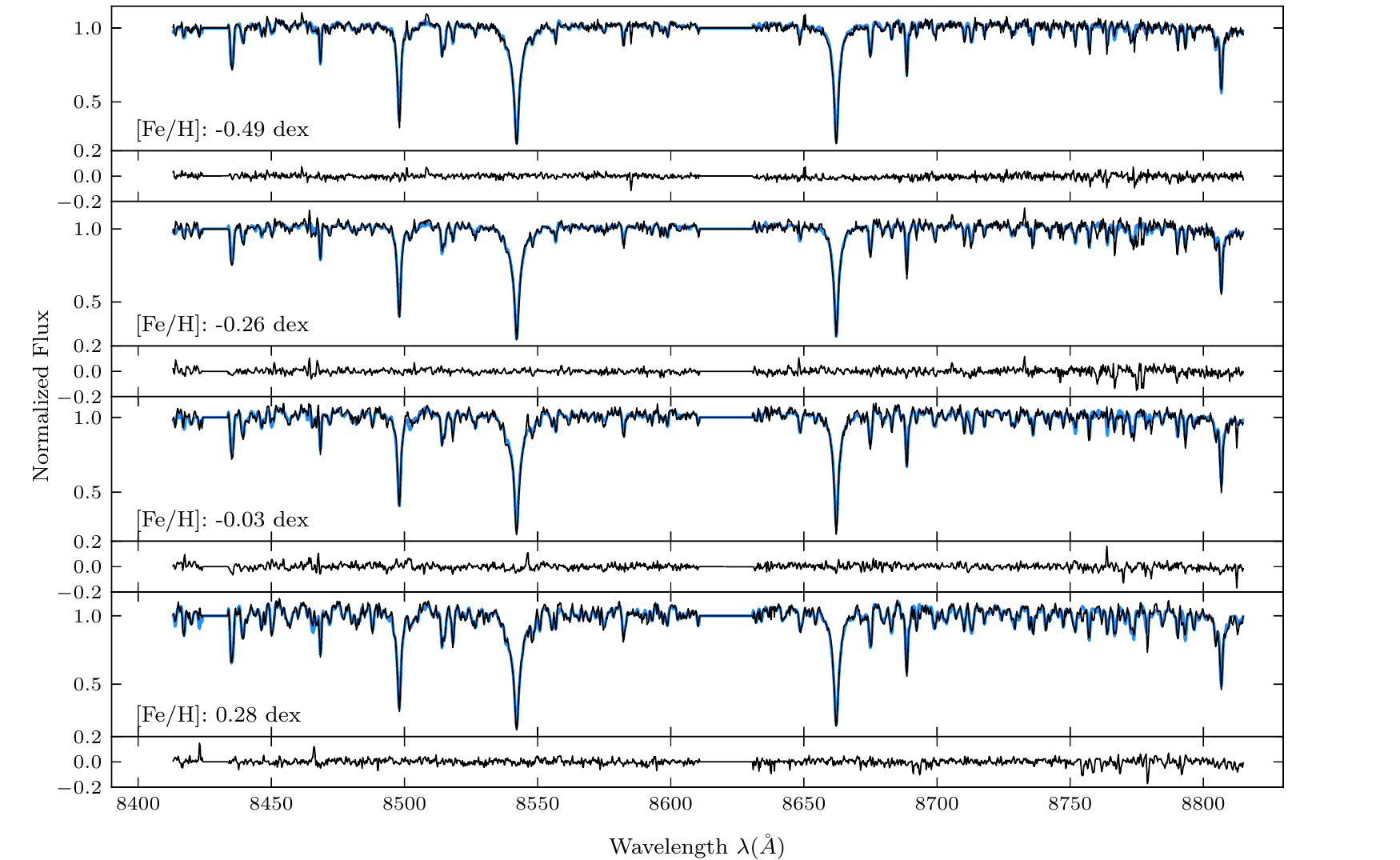} 
\caption{The normalized $\argos$ spectra (black) versus the normalized model spectra (blue) generated by $\thecannon$ for $\atoa$ stars with $\feh$ values between $-0.5 \lesssim \feh \lesssim 0.25$. The plotted line thicknesses of the model spectra indicate the scatter of each fit by $\thecannon$. The residuals between the normalized $\argos$ spectra and normalized model spectra are also shown in the panels below the spectra.}
\label{fig:model_data}
\end{figure*}

\begin{figure}
\centering
\includegraphics{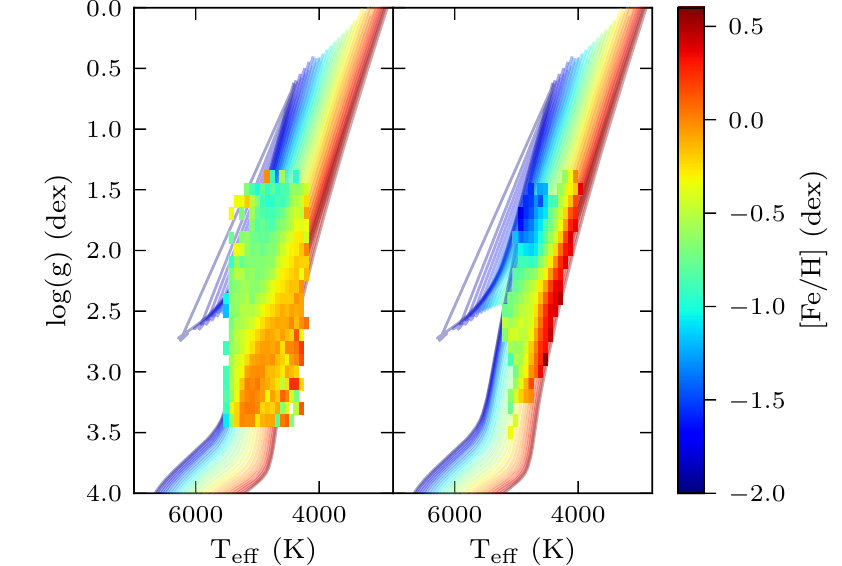} 
\caption{$\teff$-$\logg$ distribution of $\argos$ (left) and $\atoa$ (right) stars coloured by mean $\feh$. 10Gyr PARSEC isochrones with $-2<\feh \, (\mathrm{dex})<0.6$ are plotted beneath. Note, the isochrones are plotted at $35\%$ transparency in order to visually differentiate them from the 2D histograms.}
\label{fig:hrd}
\end{figure}

Given a reference set and an application set, $\thecannon$ will always return new labels for the stars in the application set. However, if one is not careful, the returned labels can have large errors. In this section, we perform three validation tests to verify that the labels returned by $\thecannon$ are reasonable. 

The first validation test we perform is a common machine learning test called the pick-one-out test. In this test, we create $204$ models, each of which is trained on $203$ stars from the reference set. The single star that is left out from the reference set changes between each model. Every model is then applied to the spectra of the respective left out star to obtain a new set of labels for it. How similar the new set of labels are to the original $\apogee$ labels indicates how well $\thecannon$ can learn the $\apogee$ labels given the reference set. In Figure \ref{fig:pickoneout} we compare the new $\cannon$ labels of these stars to their $\apogee$ labels. For all four labels, the bias and rms, given in the upper left hand corner of each plot, are much lower than those from the $\argos$-$\apogee$ comparisons in Figure \ref{fig:orig_compare}. The strong agreement indicates that $\thecannon$ can successfully learn the $\apogee$ labels from the $\argos$ spectra using the reference set composed of the 204 common stars. The error on each parameter for each star in the $\atoa$ catalog is calculated by adding in quadrature the rms value from the pick-one-out test and the small error that is output by the optimiser of the $\thecannon$ (see Section \ref{The Cannon}).

As a second validation test we compare the model and observational spectra. The shape of the spectrum of a star can be affected by many different stellar parameters and abundances. Ideally, when training a model to describe a stellar spectrum with labels one would like to include all stellar labels that affect the spectrum's shape. However, this would require a huge number of reference set stars, which we do not have. Instead, we are making the approximation that the $\argos$ spectra ($8400 \, \si{\angstrom}$ - $8800 \, \si{\angstrom}$) can be well described by the five labels: $\feh$, $\mgfe$, $\teff$, $\logg$, and $\mathrm{A}_{\mathrm{k}}$. To test this, we compare the model spectra generated by $\thecannon$ against the true observational spectra. This can be done because $\thecannon$ trained model returns the flux at each pixel when given the labels (Equation \ref{modelequ}). In Figure \ref{fig:model_data} we plot the $\argos$ spectra of a few example stars with a range of $\feh$ values and cumulative $\chi^{2}$ values (sum of the pixel $\chi^{2}$ values) around the $\argos$ pixel number (1697, see Figure \ref{fig:chi2}) versus their model spectra generated by $\thecannon$. For the model spectra, line thicknesses show the scatter of the fit by $\thecannon$ at each wavelength. Figure \ref{fig:model_data} shows that the model spectra closely reproduce the true observational spectra. The overall good fit between the model spectra and the true spectra indicates that the spectra of the $\argos$ stars can be well described by the variation of the five labels.

The third test we do is comparing the $\teff$-$\logg$-$\feh$ distribution of $\atoa$ stars to theoretical distributions. In the right hand plot of Figure \ref{fig:hrd}, we show the $\teff$-$\logg$ distribution of $\atoa$ stars coloured by mean $\feh$ on top of 10 Gyr PARSEC isochrones with metallicities ranging from $-2 \, \mathrm{dex}$ to $0.6 \, \mathrm{dex}$ \citep{Tang_2014,Chen_2014,Chen_2015,Bressan_2012}. The $\atoa$ stars tightly follow the PARSEC isochrones. Furthermore, even though no isochrone information was input into $\thecannon$, there are no $\atoa$ stars in nonphysical regions of the diagram. The close fit of the $\atoa$ stars to the PARSEC isochrones supports that the label transfer was successful.

The success of these three tests shows that it is possible to train a $\cannon$ model on a moderate number ($204$) of common stars and still obtain a set of labels with good precisions (see Section \ref{refapp}).

\subsection{A2A versus ARGOS}
\begin{figure}
\centering
\includegraphics{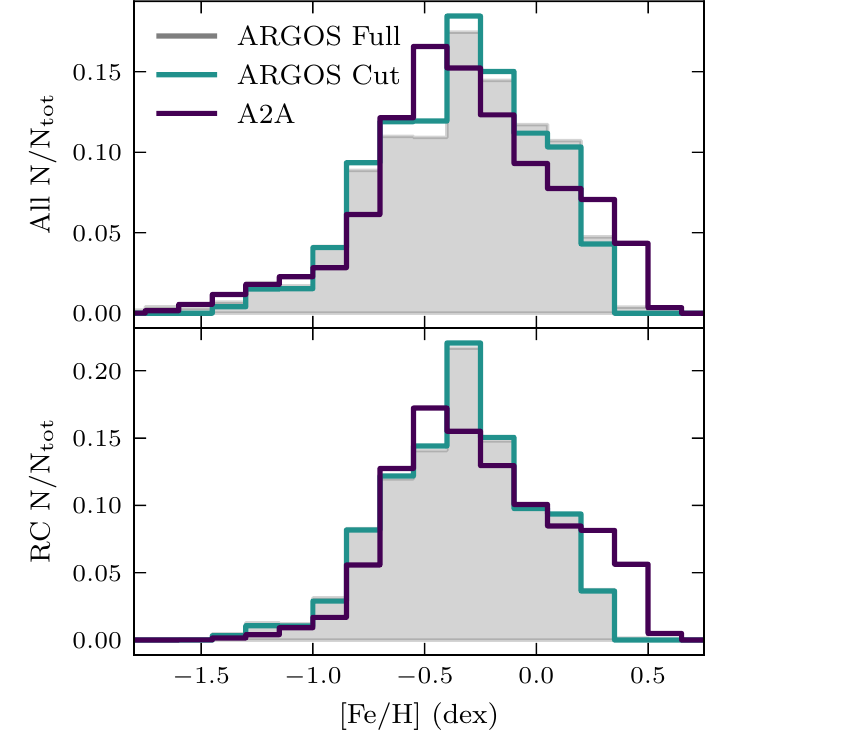}
\caption{The normalised $\argos$ and $\atoa$ MDFs of all stars (top) and RC stars (bottom). The grey histogram includes stars from the full $\argos$ catalog while the teal histogram includes only $\argos$ stars that could be processed by $\thecannon$ (same stars as in the $\atoa$ catalog but with their old $\argos$ labels).}
\label{fig:argatoamdf}
\end{figure}

In this section we compare the $\atoa$ catalog to the $\argos$ catalog. In Figure \ref{fig:hrd} we show the $\teff$-$\logg$-$\feh$ distributions of $\argos$ stars (left) and $\atoa$ stars (right) on top of 10 Gyr PARSEC isochrones. The $\argos$ stars very roughly follow the PARSEC isochrones. Many $\argos$ stars also fall in nonphysical regions of the parameter space. As discussed in the previous section, $\atoa$ stars have a much tighter alignment with the PARSEC isochrones with no stars falling in nonphysical regions.

In Figure \ref{fig:argatoamdf} we show the $\argos$ and $\atoa$ MDFs of all stars (top) and only RC stars (bottom). The most prominent difference between the MDFs of the surveys is that $\atoa$ obtains more very $\feh$-rich stars than $\argos$ for all stars as well as when we restrict to only the RC. $\argos$ has more solar to sub-solar stars until ${\sim}-0.5$ dex where $\atoa$ has more stars. Between ${\sim}-1$ and ${\sim}-0.7$ dex $\argos$ has more stars for all stars and the RC. Below ${\sim}-1$ dex, the difference between $\argos$ and $\atoa$ is small.

\subsection{Red Clump Extraction and A2A Distances}\label{aprcext}
We statistically extract RC stars from the $\atoa$ catalog using the following probabilistic method: First, we determine the spectroscopic magnitudes, $\mathrm{M}_{\mathrm{K}_{\mathrm{s}}}$, of each $\atoa$ star by fitting their $\logg$, $\teff$, and $\feh$ parameters to theoretical isochrones. Then, using the spectroscopic magnitudes, we calculate a weight for each star which gives the probability that it is part of the RC. The functional form of this weight is a Gaussian:
\begin{equation} \label{rcweight}
\omega_{\mathrm{rc}}(\mathrm{M}_{\mathrm{K}_{\mathrm{s}}}) =
\frac{1}{\sigma_{\mathrm{M}_{\mathrm{K}_{\mathrm{s}}}}\sqrt{2\pi}}\mathrm{exp}\left(-\frac{1}{2}\frac{(\mathrm{M}_{\mathrm{K}_{\mathrm{s}}}-\mathrm{M}_{\mathrm{rc}})^2}{\sigma_{\mathrm{M}_{\mathrm{K}_{\mathrm{s}}}}^{2}}\right),
\end{equation}
where $\mathrm{M}_{\mathrm{rc}} = -1.61 \pm 0.22$ mag is the intrinsic magnitude of the RC \citep{Alves_2000}. We find that for 10 Gyr old PARSEC isochrones \citep{Tang_2014,Chen_2014,Chen_2015,Bressan_2012} the spectroscopic magnitude varies with $\logg$ as $\mathrm{d}\mathrm{M}_{\mathrm{K}_{\mathrm{s}}}/\mathrm{d}(\log g) = 2.33$. The average $\atoa$ $\logg$ error is $0.18$ dex, giving an average magnitude error of $0.42$ mag. We add this in quadrature with the intrinsic width of the RC magnitude to obtain a total magnitude error of $0.47$ mag for our RC sample. This is $\sigma_{\mathrm{M}_{\mathrm{K}_{\mathrm{s}}}}$ in Equation \ref{rcweight}. This magnitude dependent weighing method extracts the RC from $\atoa$ by giving higher weights to stars that are likely to be part of the RC and lower weights to stars that are unlikely to be part of the RC.

To obtain distances for the $\atoa$ stars, we first treat each star as a RC star and assume their absolute magnitudes are that of the RC ($-1.61$ mag). We then compare the RC absolute magnitude to the de-reddened apparent magnitude of each star which we obtain using the \citet{Schlegel_1998} extinction maps re-calibrated by \citet{Schlafly_2011}. This method gives us the distance of each star assuming that it is a RC star. To account for the fact that not every $\atoa$ star is a RC star, we weigh the stars by how likely they are to be RC stars using the weight in Equation \ref{rcweight}. By weighing the stars in this manner, we treat all stars as RC stars but effectively remove the stars that are unlikely to be part of the RC by strongly de-weighting them. 

The $\atoa$ catalog contains $10,357$ RC stars. We obtain this number by summing the RC weights (Equation \ref{rcweight}).

\section{Selection Functions}\label{selfunc}
The probability that any given star in the Galaxy is observed by a large survey program is called the survey selection function (SSF); see \cite{Sharma_2011} for a detailed discussion. In order to obtain unbiased parameter and abundance distributions of the Galactic bulge using the $\atoa$ and $\apogee$ surveys we must correct for their SSFs. Otherwise, it would not be clear if the distributions we obtain are the true distributions of the Galactic bulge, or whether they are biased by the selection choices of the surveys. In the next two sections, we discuss the $\atoa$ and $\apogee$ SSFs.

\subsection{A2A Selection Function}
The stars composing the $\atoa$ catalog were selected from the $\argos$ catalog which in turn was selected from a high quality (HQ) subset of the Two Micron All Sky Survey ($\twomass$; \citep{Skrutskie_2006}). In the following subsections we discuss the selection of the $\argos$ survey from the HQ $\twomass$ subset and the selection of the $\atoa$ survey from the $\argos$ survey.  

\subsubsection{Selection of ARGOS from HQ 2MASS}\label{Arg_sf}
\begin{figure}
\centering
\includegraphics{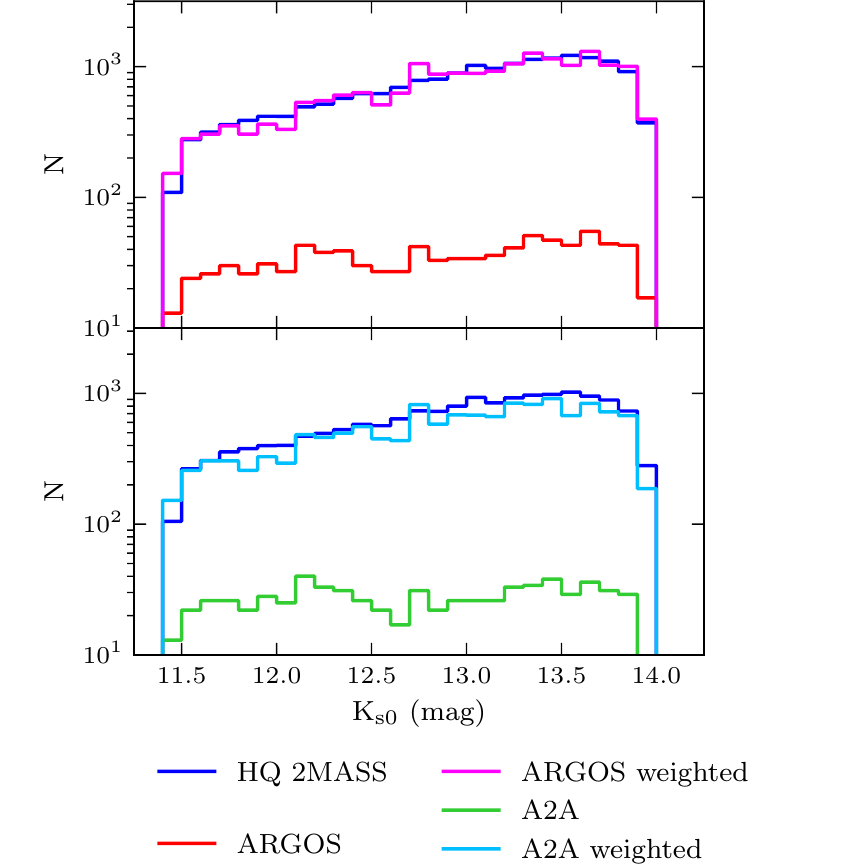} 
\caption{The original and weighted $\mathrm{K}_{\mathrm{s}0}$-band LFs of $\argos$ (top) and $\atoa$ (bottom) stars in the field $(l, b) = (-10\degree, -10\degree)$. The HQ $\twomass$ LF is also plotted. The colour limits $0.45 \leq \mathrm{(J-\mathrm{K}_{\mathrm{s}})_{0}}  \,\, (\mathrm{mag}) \leq 0.86$ are applied to the LFs in the bottom plot. The $\atoa$ LF is slightly below the HQ $\twomass$ LF due to the parametric limits applied during the creation of the $\atoa$ catalog.}
\label{fig:ibandsel}
\end{figure}

The $\argos$ stars were selected from a HQ sub-sample of the $\twomass$ survey, requiring the stars to have high photometric quality flags (see \citet{Freeman_2012}), magnitudes between $11.5\leq \mathrm{K}_{\mathrm{s}} \, (\mathrm{mag})\leq14.0$, and colours $(\mathrm{J}-\mathrm{K}_{\mathrm{s}})_{0} \ge 0.38$ mag. For each $\twomass$ star that met these requirements, its $\mathrm{I}_{0}$-band magnitude was estimated using the equation:
\begin{equation}\label{Iband}
\mathrm{I}_{0} = \mathrm{K}_{\mathrm{s}} + 2.095(\mathrm{J}-\mathrm{K}_{\mathrm{s}}) + 0.421\mathrm{E}(\mathrm{B}-\mathrm{V}).
\end{equation}
Then, for each field, the $\argos$ team randomly selected approximately $1000$ stars roughly evenly distributed among the $\mathrm{I}_{0}$-band bins: 13-14 mag, 14-15 mag, and 15-16 mag. This was done in order to sample a roughly equal number of stars from the front, middle, and back regions of the bulge.

We use the following procedure to correct for the $\mathrm{I}_{0}$-band selection (similar to \citet{Portail_dyn_2017}, their Section 5.1.1). First we take all $\twomass$ stars in a given field and apply the colour, magnitude, and quality cuts described above. Then, we estimate the $\mathrm{I}_{0}$-band magnitude of each remaining $\twomass$ star as well as of each $\argos$ star using Equation \ref{Iband}. We can then correct for the $\mathrm{I}_{0}$-band selection by weighing each $\argos$ star by the ratio of the number of HQ $\twomass$ stars to the number of $\argos$ stars in each $\mathrm{I}_{0}$-band bin and field:
\begin{equation}\label{ibandw}
\omega_{\mathrm{f}, \mathrm{I}_{0}} = \mathrm{N}^{\mathrm{HQ \,\, 2MASS}}_{\mathrm{f}, \mathrm{I}_{0}}/\mathrm{N}^{\mathrm{ARGOS}}_{\mathrm{f}, \mathrm{I}_{0}}.
\end{equation}
After the application of the weights in Equation \ref{ibandw} to the $\argos$ luminosity function (LF), we statistically recover the HQ $\twomass$ LF within the respective colour and magnitude limits. The upper plot of Figure \ref{fig:ibandsel} shows this for the field $(l, b) = (-10\degree, -10\degree)$.

\subsubsection{Selection of A2A from ARGOS}\label{seca2aargos}
\begin{figure}
\centering
\includegraphics[width=\columnwidth]{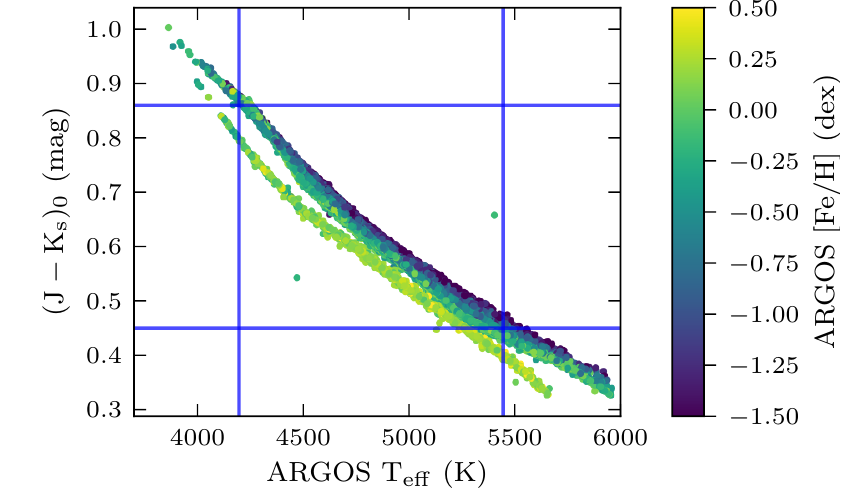} 
\caption{$\argos$ $\teff$ versus de-reddened colour distribution of the full $\argos$ catalog. The point colour indicates the $\argos$ $\feh$. The blue vertical lines show the reference set $\teff$ limits (\ref{first:a}). The blue horizontal lines show the colour limits (\ref{second}) used to approximate the $\teff$ limits.} 
\label{fig:teff_vs_color}
\end{figure}

As the $\atoa$ stars were selected from the $\argos$ catalog, we also similarly correct the $\atoa$ catalog for the $\mathrm{I}_{0}$-band selection using the weights from Equation \ref{ibandw}. However, the weighted $\atoa$ LFs are systematically below the HQ $\twomass$ LFs because of the $\atoa$ selection from the $\argos$ catalog which removes $4,135$ stars. These stars were removed because their spectra could not be processed by $\thecannon$ (did not satisfy limits \ref{first:a}, \ref{first:b}, \ref{third:a}, and \ref{third:b}). Because of this, we know the label values of these stars on the $\argos$ scale but only have approximate knowledge of where they are on the $\apogee$ scale. To replicate this selection, we examine if the limits that removed these stars can be described using parameters that do not change during the label transfer.

The $\teff$ limits (see \ref{first:a}) are simple to approximate as there is a near linear relationship between $\argos$ $\teff$ and colour, shown in Figure \ref{fig:teff_vs_color}. However, this substitution is not perfect and the colour limits must be chosen carefully as the $\teff$-colour distribution has some spread due to variations in the other labels. For example, stars with lower $\argos$ $\feh$ are hotter for constant colour (see the point colour in Figure \ref{fig:teff_vs_color}). If chosen incorrectly, the colour limits can remove many stars that satisfy limits \ref{first:a},\ref{first:b}, \ref{third:a}, and \ref{third:b}. We find that the $\teff$ limits are well approximated by the colour limits:
\begin{equation}\label{second}
0.45 \leq \mathrm{(J-\mathrm{K}_{\mathrm{s}})_{0}}  \,\, (\mathrm{mag}) \leq 0.86.
\end{equation}
We show in the lower plot of Figure \ref{fig:ibandsel} the weighted $\atoa$ and HQ 2MASS LFs in the field $(l, b) = (-10\degree, -10\degree)$; both of which have the colour cut in Equation \ref{second} applied. While the two LFs are close, there is still a slight deviation due to the other parametric limits.

Unfortunately, the other parametric limits are not as easily replaced using alternative parameters that remain constant during the label transfer. We take the final $\atoa$ catalog to include all stars processed by $\thecannon$ that:
\begin{enumerate}
  \item Have model $\chi^{2}$ values below 5000 (see Figure \ref{fig:chi2}).
  \item Satisfy the limits \ref{first:a},\ref{first:b}, \ref{third:a}, and \ref{third:b}.
  \item Are within the colour limits in Equation \ref{second}.
\end{enumerate}
Within these conditions, the $\atoa$ catalog contains $21,325$ stars. If we apply the colour cut (Equation \ref{second}) to the $\argos$ catalog then the $\argos$ catalog would contain $23,487$ stars. Thus, the colour cut $\atoa$ catalog is $91\%$ complete compared to the colour cut $\argos$ catalog. 
 
In the subsequent analysis of the bulge's chemodynamical structure, we often select and plot $\atoa$ RC stars to obtain good distance estimates (see Section \ref{aprcext} for a discussion on RC extraction). We make the assumption that the reference set limits affect RC and red giant branch stars equally such that the $\atoa$ RC catalog is also ${\sim} 91\%$ complete. We test this assumption in Appendix \ref{aprccomp}.

\subsection{APOGEE Selection Function}\label{apogee_sf_sec}
\begin{figure}
\centering
\includegraphics{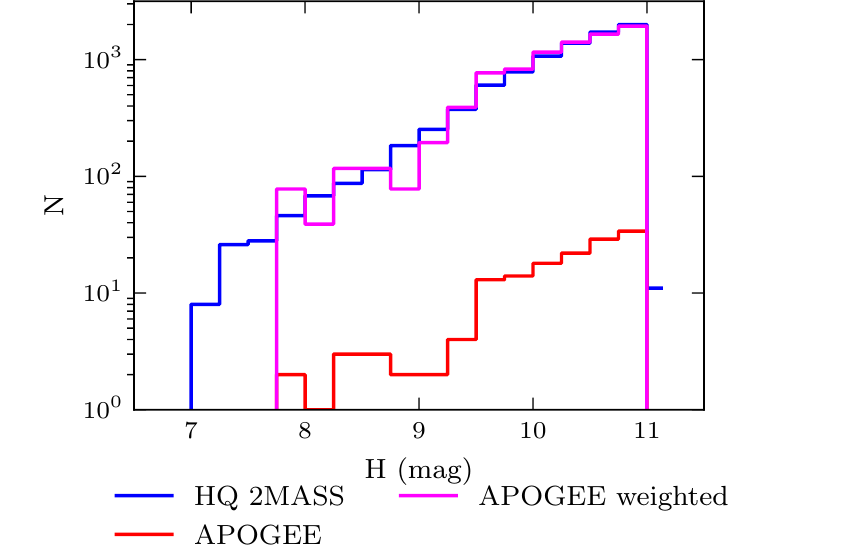} 
\caption{The original and weighted $\apogee$ H-band LFs of the HQSSF bulge MSp for a cohort in the field $(l, b) = (-2\degree, 0\degree)$. The HQ $\twomass$ LF is also plotted.}
\label{fig:aposel}
\end{figure}

The $\apogee$ sample we use for most of this work's analysis is the HQSSF $\apogee$ bulge MSp. It is a sub-sample of the full $\apogee$ bulge MSp in that we also require the stars to have high quality ASPCAP parameters and abundances (see Section \ref{APOGEE}) and good SSF estimates. In this sample, only stars that are part of complete cohorts, i.e. group of stars observed together during the same visits, have SSF estimates. Estimating the SSF for this sample proceeds in two steps:

\begin{enumerate}
\item To account for the selection of the $\apogee$ {bulge} MSp from the HQ $\twomass$ subset, we use the publicly available python package \emph{APOGEE} \citep{Bovy_2014, Bovy_2016, Mackereth_2020}. For each {complete} cohort, the program returns the ratio of the number of $\apogee$ MSp stars to the number of HQ $\twomass$ stars within the respective colour and magnitude limits of the cohort\footnote{Cohort magnitude limits are set depending on the planned number of visits. Most cohorts used in this work have the magnitude limits $7<\mathrm{H}_{0}\, \mathrm{(mag)}\!<\! 11$, $7\!<\mathrm{H}_{0}\, \mathrm{(mag)}\!<\! 12.2$, or $7<\mathrm{H}_{0}\, \mathrm{(mag)}\!<\! 12.8$, although some have fainter limits. The colour limits of the cohorts are  $(\mathrm{J}-\mathrm{K}_{\mathrm{s}})_{0}\geq 0.5$ mag in bulge and $\apogee$-1 disk fields, and $0.5\!\leq\!(\mathrm{J}-\mathrm{K}_{\mathrm{s}})_{0}\, \mathrm{(mag)}\!\leq\! 0.8$ and $(\mathrm{J}-\mathrm{K}_{\mathrm{s}})_{0}>0.8$ mag in $\apogee$-2 disk fields; see \citet{Zasowski_2017}.}. Then, we weight each star in each cohort, $\mathrm{c}_{\mathrm{i}}$, by the inverse of this ratio:
\begin{equation}\label{apow}
\omega_{\mathrm{c}_{\mathrm{i}}} = \mathrm{N}^{\mathrm{HQ \,\, 2MASS}}_{\mathrm{c}_{\mathrm{i}}} / \mathrm{N}^{\mathrm{APOGEE\,\,MSp}}_{\mathrm{c}_{\mathrm{i}}}.
\end{equation}

\item Restricting our sample to $\apogee$ stars with HQ ASPCAP parameters and abundances (Section \ref{APOGEE}) removes ${\sim}14\%$ of the $\apogee$ bulge MSp. To correct for this selection we bin all $\apogee$ bulge MSp stars (including the stars with poor ASPCAP estimates) and all HQ ASPCAP MSp stars in magnitude, colour, and cohort. Then, we weight each HQ ASPCAP MSp star by the ratio of the number of MSp stars to the number of HQ ASPCAP MSp stars in the colour and magnitude bin in which it falls:
\begin{equation}\label{jkbandw}
\omega_{\mathrm{c}_{\mathrm{i}},\mathrm{H},\mathrm{(\mathrm{J}-\mathrm{K}_{\mathrm{s}})_{0}}}  = \mathrm{N}^{\mathrm{APOGEE\,\,MSp}}_{\mathrm{c}_{\mathrm{i}},\mathrm{H},\mathrm{(\mathrm{J}-\mathrm{K}_{\mathrm{s}})_{0}}} / \mathrm{N}^{\mathrm{HQ \,\, ASPCAP}}_{\mathrm{c}_{\mathrm{i}},\mathrm{H},\mathrm{(\mathrm{J}-\mathrm{K}_{\mathrm{s}})_{0}}}.
\end{equation}
\end{enumerate}

Figure \ref{fig:aposel} shows the result of the application of the weights in Equations \ref{apow} and \ref{jkbandw} to the $\mathrm{H}$-band LF of a cohort in the field $(l, b) = (-2\degree, 0\degree)$. We see that after the application of the weights, the LFs of the HQSSF $\apogee$ bulge MSp and HQ $\twomass$ subset approximately match.

In Figure \ref{fig:fieldloc}, the red crosses indicate $\apogee$ field locations for which we cannot use the \emph{APOGEE} python package to obtain good SSF estimates of the observed stars. This occurs either because the cohorts composing the fields are not currently complete or because they do not contain any MSp stars. Removing these fields, as well as a few cohorts for which the weighted LF poorly reproduces the LF of its HQ $\twomass$ parent sample, leaves $23,512$ stars in the HQSSF $\apogee$ bulge MSp.

In the subsequent analysis we restrict the HQSSF $\apogee$ bulge MSp further by requiring stars to have AstroNN distance errors less than $20\%$. This roughly removes $5\%$ of the HQSSF $\apogee$ bulge MSp leaving $22,340$ $\apogee$ stars.

\subsection{Selection of HQ 2MASS Catalogs}\label{hq2mass_sf}
\begin{figure}
\centering
\includegraphics{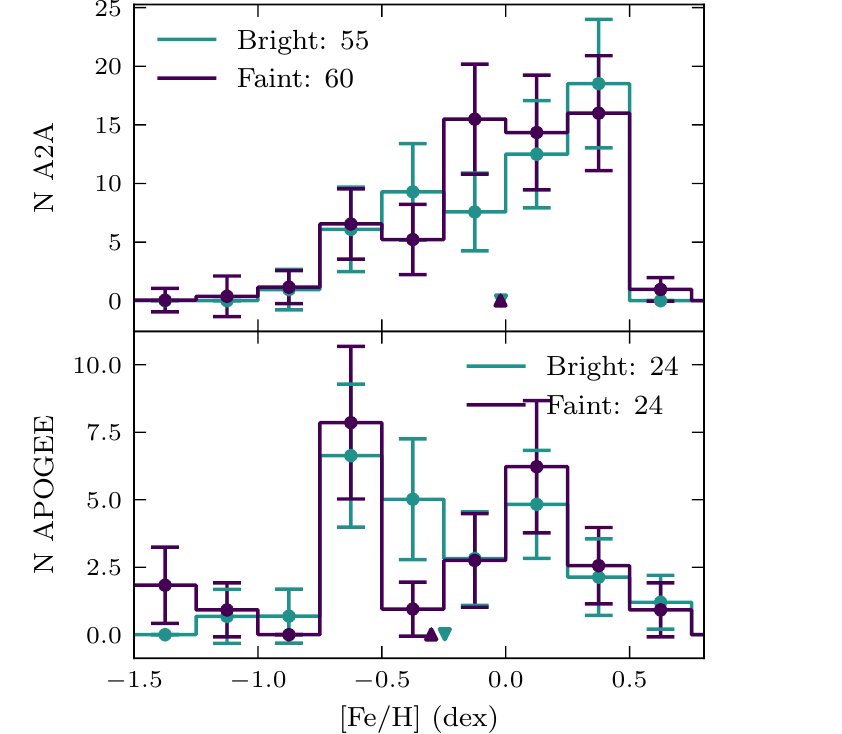} 
\caption{The MDFs of bright and faint stars in the same distance bins. Top: $\atoa$ RC stars from the field $(l, b) = (0\degree, -5\degree)$ and the distance bin $6$ to $8$ kpc with the weights from Equation \ref{ibandw} applied. Bottom: $\apogee$ stars from the field $(l, b) = (0\degree, -2\degree)$ and the distance bin $6$ to $8$ kpc with the weights from Equations \ref{apow} and \ref{jkbandw} applied. The number of stars in each MDF is given in the legend of each plot. The means of each histogram are given by the triangular markers.}
\label{fig:brightfaint}
\end{figure}

So far we have described the $\atoa$ and $\apogee$ SSFs as well as the corresponding weights that are needed to statistically correct each survey to the magnitude and colour distributions of their respective HQ $\twomass$ parent samples. This is similar to the procedure done by \citet{Rojas_Arriagada_2020}, who used simple stellar populations to determine the fraction of giants with fixed distance modulus and metallicity that fall with in the $\apogee$ magnitude and colour ranges. Then using these fractions, they re-weighted the observed stars to the weights they had in the survey input sample. However, the input HQ $\twomass$ sample of each survey itself has a SSF relative to the real Galaxy (in practise the current deepest photometric survey, VVV \citep{Minniti_2010, Surot_2019}) due to photometric criteria, crowding, and extinction. The $\twomass$ SSF is strongest at low latitudes and is illustrated in \citet[][their Section 5.1.1]{Portail_dyn_2017}. This SSF would be additionally required when comparing (or weighting by) the relative number densities of stars in different fields, especially those with different latitudes. 

In this paper, we confine our analysis to small spatial bins, making use of RC distances for $\atoa$ and AstroNN distances for $\apogee$. When we do this, the observed stars in a given bin are representative of the stellar population at that distance making further corrections of the HQ $\twomass$ survey magnitude distribution unnecessary. However, in practise, the bins we use have sizes ${\sim} 2$ kpc, thus if there is a line-of-sight abundance gradient in a field, the fainter stars in a given bin could have a slightly different abundance distribution than the brighter stars as they trace somewhat larger distance. Figure \ref{fig:brightfaint} shows, for example bins, that no such effect is seen within the errors in either survey.

An additional effect could arise due to fields at different latitudes/heights contributing stars to the same distance bins. Specifically, lower latitude fields are generally more $\feh$-rich and have higher crowding than higher latitude fields due to the $\feh$ and density gradients in the bulge. In such cases, by not correcting for the HQ $\twomass$ SSF we may introduce a slight bias against the lower latitude, higher $\feh$ stars in each spatial bin. However, the effects of field mixing would be small in $\atoa$ as its fields are well separated and generally located at latitudes with low crowding ($|\mathrm{b}|\geq 5\degree$). Whereas for $\apogee$, the effects of field mixing would also be small because at low latitudes ($|\mathrm{b}|< 4\degree$), where the incompleteness of the HQ $\twomass$ catalog is largest, the $\feh$ gradient is nearly flat \citep{Rich_2012, Ness_2016}, and at high latitudes ($|\mathrm{b}|> 4\degree$), where the $\feh$ gradient is negative, the incompleteness of the HQ $\twomass$ catalog is small. When we vary the width of our distance bins in $|\mathrm{Z}|$, we do not find significant changes in the bulge $\feh$ gradient. Therefore we neglect field mixing effects in this paper.

\subsection{Application of the SSF-corrections}\label{appSFcorr}\begin{figure*}
\centering
\includegraphics{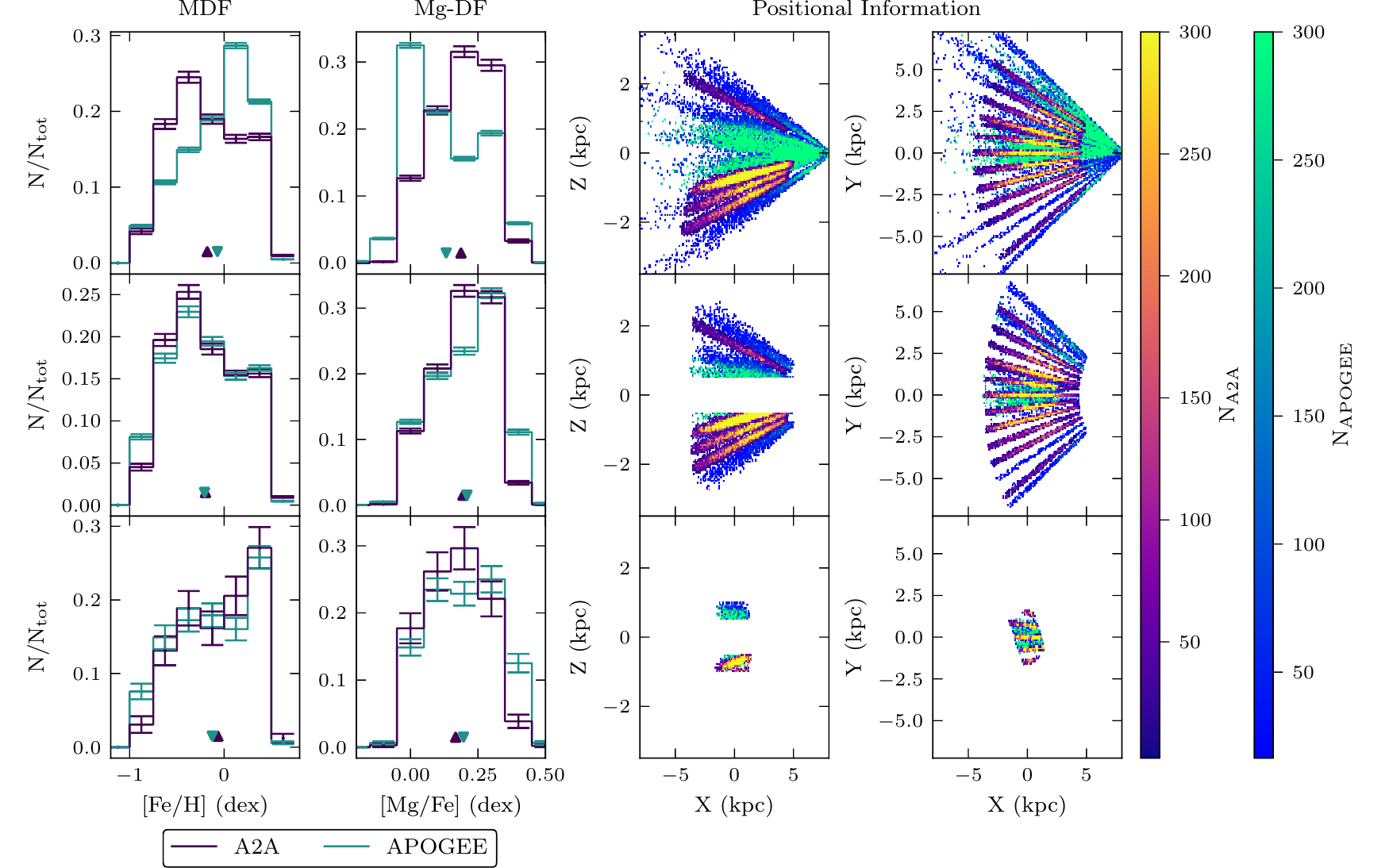}
\caption{SSF-corrected MDFs (first column), Mg-DFs (second column), and their respective positional information (third and fourth columns) of $\atoa$ and $\apogee$ stars with $\feh>-1$ dex. The top row includes all stars in each catalog, while the stars in the second and third rows are restricted to successively smaller areas. The $\atoa$ stars are restricted to RC stars. The mean $\feh$ and $\mgfe$ values of each MDF and Mg-DF are shown by the triangular markers in each plot.}
\label{fig:MDFpos}
\end{figure*}

Here we illustrate the effect of the different spatial selections of the two surveys in the inner Galaxy, and then compare their SSF-corrected MDFs and $\mgfe$ distribution functions (Mg-DFs) in regions of spatial overlap.

The first two plots in the top row of Figure \ref{fig:MDFpos} show the $\apogee$ and $\atoa$ MDFs and Mg-DFs of all stars (RC for $\atoa$) in each catalog. $\apogee$ observes many stars near the Galactic plane and in the nearby disk that $\atoa$ misses, as illustrated in the right two plots of this row. These stars tend to be more $\feh$-rich and $\mgfe$-poor than stars at larger heights, causing much stronger $\feh$-rich and $\mgfe$-poor peaks in the $\apogee$ histograms than in $\atoa$. In the second row of Figure \ref{fig:MDFpos}, the samples are restricted to smaller regions of overlap between the surveys, demanding  $|\mathrm{Z}|\geq 0.5$ kpc and distances from the Sun between 4 and 12 kpc, and thereby removing many of the in-plane $\feh$-rich/$\mgfe$-poor stars in the $\apogee$ catalog. This causes the $\feh$-rich and $\mgfe$-poor peaks in the $\apogee$ MDF and Mg-DF to decrease, leading to better agreement with $\atoa$. Some differences in the MDF and Mg-DF shapes are still expected, due to differences both in detailed coverage and in number density along the line-of-sight, as we do not correct each survey past the HQ $\twomass$ catalogs they were selected from. 

However, if we restrict the sample to a smaller distance bin as shown in the third row of Figure \ref{fig:MDFpos}, such effects are significantly weakened. Now the MDFs and Mg-DFs agree within the errors except for the most $\feh$-poor bin in the MDF and the most $\mgfe$-rich bin ($>0.35$ dex) in the Mg-DF where $\apogee$ observes a larger fraction of stars.

\subsection{The High [Mg/Fe] and Low [Fe/H] Stars}\label{inconsis}
\begin{figure*}
\centering
\includegraphics{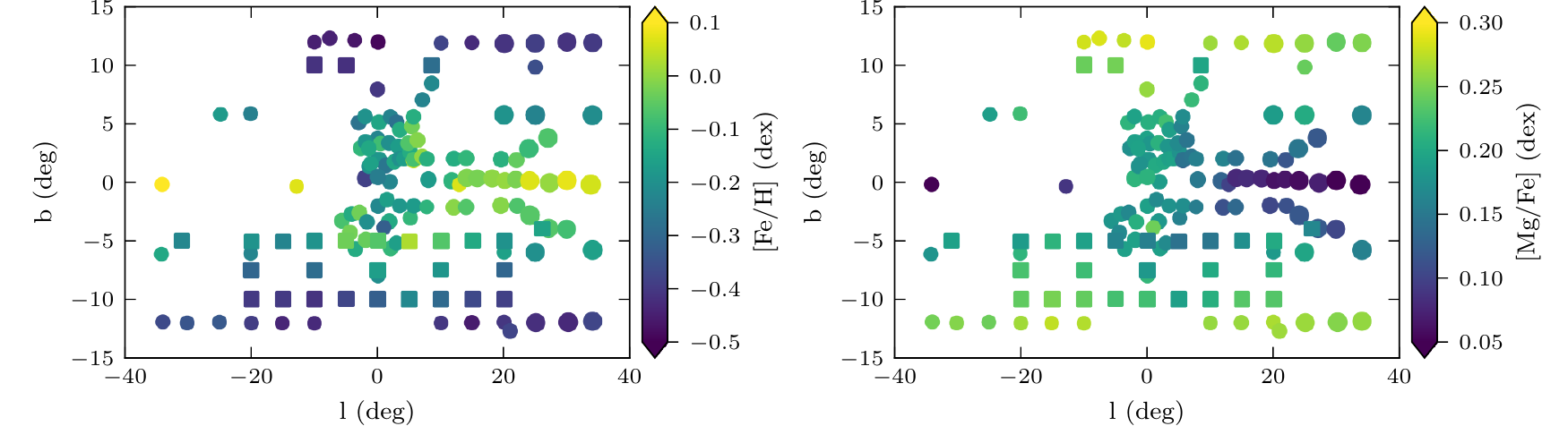} 
\caption{SSF-corrected mean $\feh$ (left) and $\mgfe$ (right) in $\apogee$ (circles) and $\atoa$ (squares) fields for stars with $[\mathrm{Fe}/\mathrm{H}]>-1$ dex and distances from the Sun between $4$ and $12$ kpc.}
\label{fig:l_b_femg}
\end{figure*}

In the following sections, we will see that the discrepancy seen in Figure \ref{fig:MDFpos} at the high $\mgfe$ end is widespread in the bulge occurring in both the inner and outer bulge and at various heights from the plane, even after the SSF corrections are applied. We believe that this discrepancy can be at least partially explained by the limited $\teff$ range spanned by the reference set (see Figure \ref{fig:teffcompare}) coupled with systematic trends between the ASPCAP $\teff$ and abundances of the $\apogee$ stars \citep{jonsson_2018, Jofre_2019}. Figure \ref{fig:tmgtrend} in Appendix \ref{abundtrends} shows the trends between ASPCAP $\teff$ and $\mgfe$ in the $\apogee$ bulge sample for a range of $\feh$ bins and roughly fixed stellar distance, height from the plane, and SNR. From this figure, we can see that regardless of $\feh$, the average $\mgfe$ of the $\apogee$ stars generally increases with increasing $\teff$ until ${\sim} 4000$ K, after which it decreases with increasing $\teff$. The $\teff$ range of the reference set, shown by the blue shaded region in Figure \ref{fig:tmgtrend}, does not reach below ${\sim} 4000$ K. Because of this, $\thecannon$ cannot learn the trends between $\teff$ and the abundances in ASPCAP that exist below ${\sim}4000$ K. Furthermore, this $\teff$ cut means that the $\atoa$ catalog would not contain many of these $\mgfe$-rich stars with $\teff$ values just below ${\sim} 4000$ K. Together, this could explain why the $\apogee$ and $\atoa$ Mg-DFs disagree at the high $\mgfe$ end. As we will see in Section \ref{mgfedist}, $\mgfe$-rich stars are typically also $\feh$-poor. This could then explain why $\atoa$ also observes fewer $\feh$-poor stars as compared to $\apogee$.

We cannot currently be sure whether the trends we observe between ASPCAP $\teff$ and $\mgfe$ are physical or systematic and therefore whether the lack of these trends in $\atoa$ is problematic or not.

\section{Abundance Structure of the Bulge}\label{msb}
We now present how the abundances and kinematics vary over the Galactic bulge using the combined $\apogee$ and $\atoa$ catalogs.

For all figures in this section, we restrict the $\atoa$ stars to RC stars and require the $\apogee$ stars to have AstroNN distance errors less than $20\%$. Unless mentioned otherwise, we use the HQSSF $\apogee$ bulge MSp and, correct each survey to the HQ $\twomass$ catalog they were selected from, and limit stars to $\feh> -1$ dex. Furthermore, when combining stars from different spatial bins, we weight the stars in each distance bin to correct for the SSF effects on the abundance distributions but then in each bin we re-weight both surveys such that the sum of their weights is equal to the number of stars (RC for $\atoa$) contributed by each survey.

\subsection{Mean Abundance Maps}\label{mstructure}
\begin{figure*}
\centering
\includegraphics{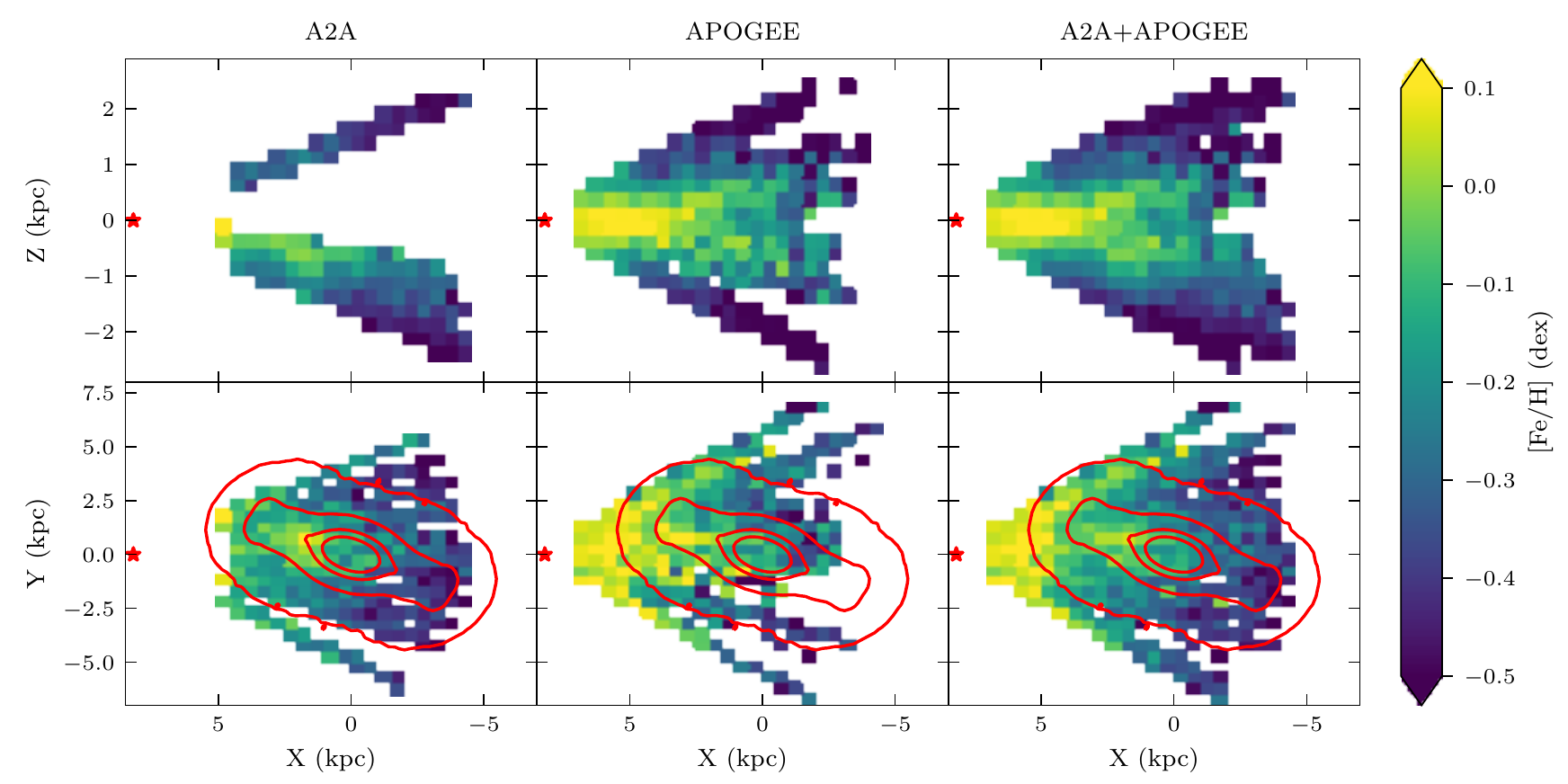} 
\caption{SSF-corrected mean $\feh$ distributions in the X-Z plane (top row) and X-Y plane (bottom row) of $\atoa$ (left column), $\apogee$ (middle column), and combined (right column) stars with $[\mathrm{Fe}/\mathrm{H}]>-1$ dex. The red lines trace the density distribution of the Milky Way's bar obtained from a \citet{Portail_dyn_2017} bulge/bar model. The red star in each plot marks the position of the Sun.}
\label{fig:Z_Xbar_nosym}
\end{figure*}

\begin{figure*}
\centering
\includegraphics{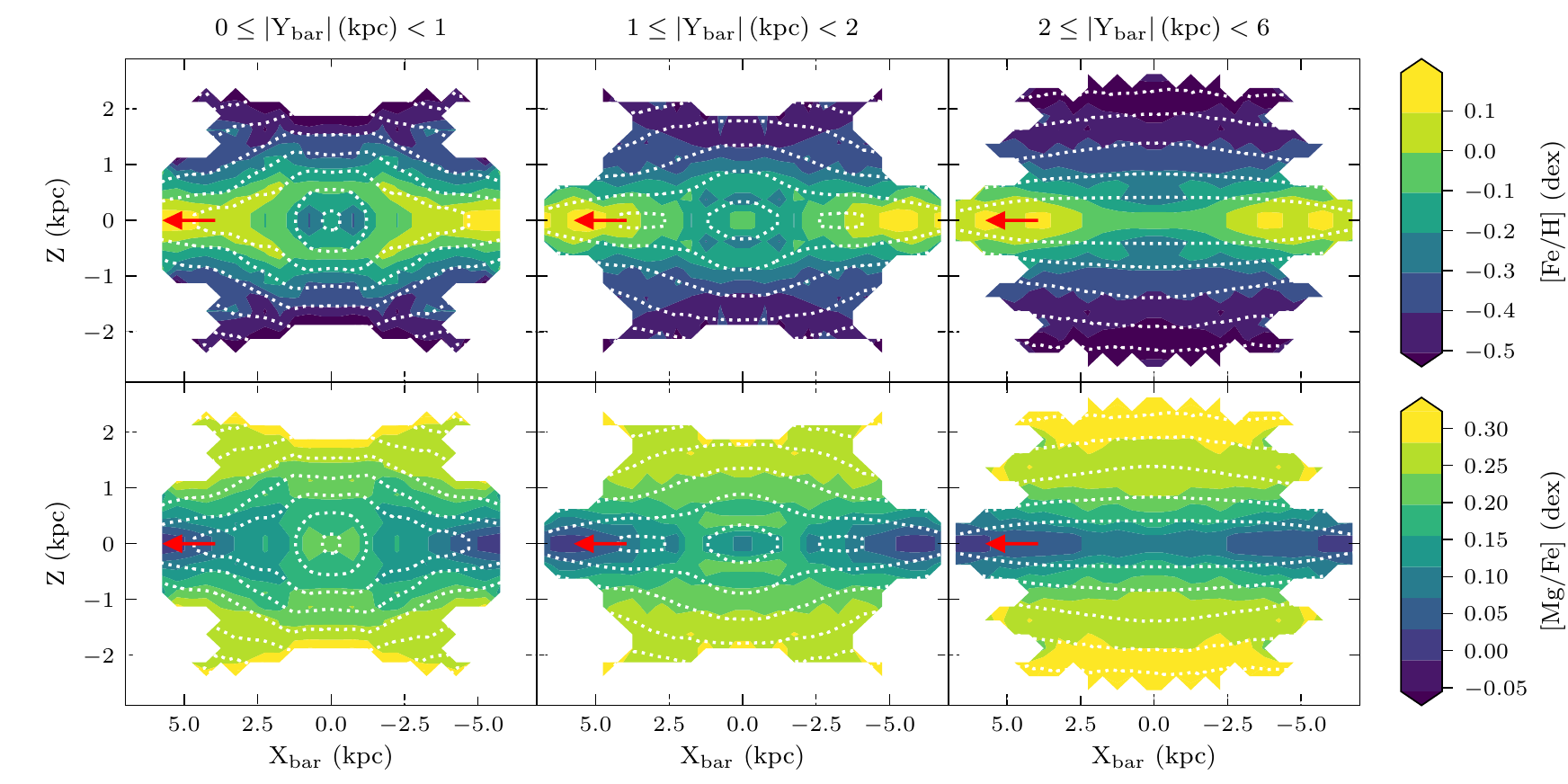} 
\caption{SSF-corrected symmetrised mean $\feh$ (top row) and $\mgfe$ (bottom row) distribution in the $\mathrm{X}_{\mathrm{bar}}$-Z plane for combined $\atoa$ and $\apogee$ stars with $[\mathrm{Fe}/\mathrm{H}]> -1$ dex in slices of $|\mathrm{Y}_{\mathrm{bar}}|$. The dotted white lines trace the density distribution obtained from a \citet{Portail_dyn_2017} bulge/bar model. The red arrow points in the direction of the Sun.}
\label{fig:Z_Xbar}
\end{figure*}

\begin{figure}
\centering
\includegraphics{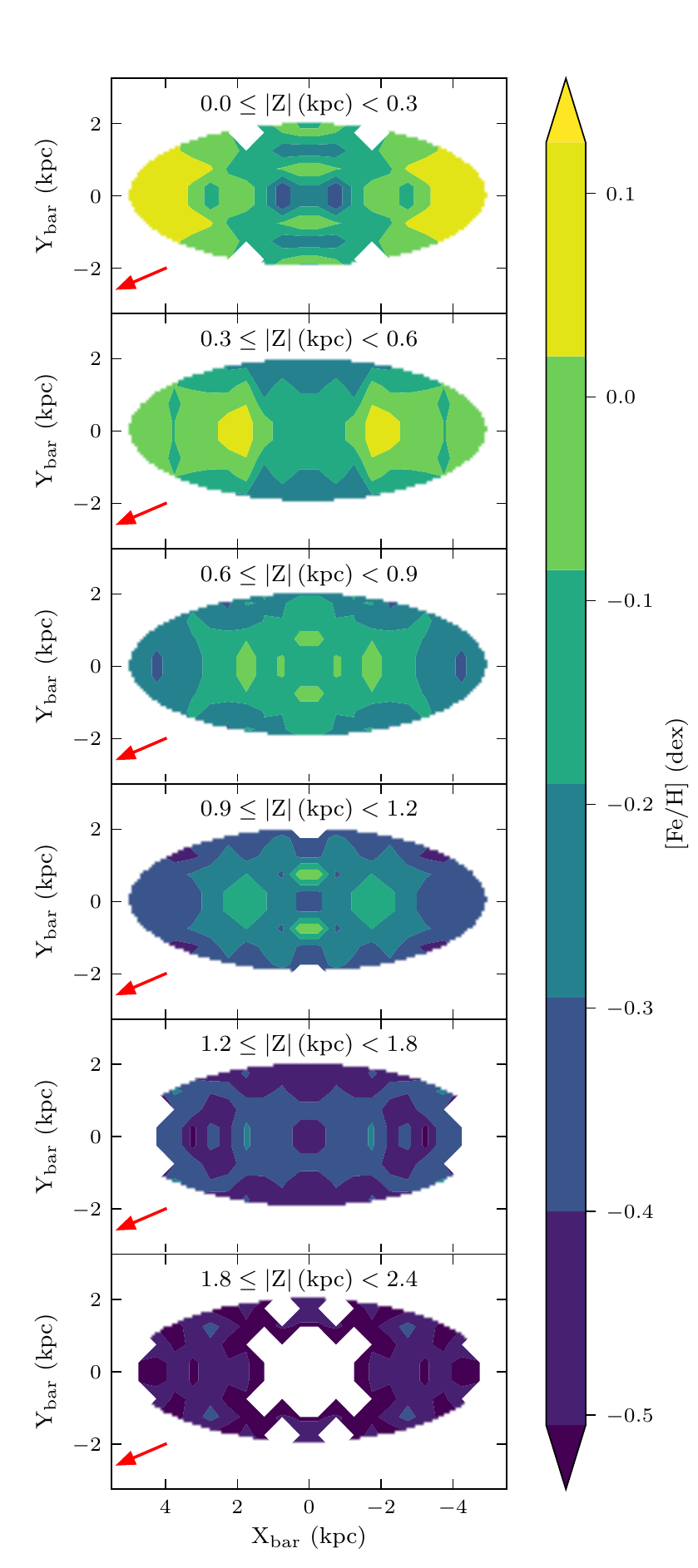} 
\caption{SSF-corrected symmetrised mean $\feh$ distribution in vertical slices along the Galactic bar for combined $\atoa$ and $\apogee$ stars with $\feh>-1$ dex. The red arrow points in the direction of the Sun.}
\label{fig:Ybar_Xbar}
\end{figure}

We first examine the overall variation of $\feh$ and $\mgfe$ with position in the Galactic bulge. Figure \ref{fig:l_b_femg} shows the mean $\feh$ and $\mgfe$ values in each field of stars with distances from the Sun between $4$ and $12$ kpc. The $\apogee$ and $\atoa$ surveys generally agree on the overall $\feh$ and $\mgfe$ trends with Galactic longitude and latitude. As expected, the high latitude fields are more $\feh$-poor and $\mgfe$-rich than the low latitude fields. Additionally, at low latitudes, the stars are more $\feh$-poor and $\mgfe$-rich near the Galactic center (GC) than they are in the long bar/disk. Because of this, the vertical abundance gradients at large absolute longitudes are steeper near the plane than at small absolute longitudes. Similar abundance trends with Galactic longitude and latitude were seen by \citet{Ness_2016} using $\apogee$ DR12 data.

Figure \ref{fig:Z_Xbar_nosym} shows illustrative mean X-Z and X-Y $\feh$ maps built using $\atoa$ and $\apogee$ stars separately and combined.
Here we use a Galactocentric left-handed coordinate system with positive X directed towards the Sun, Y along positive longitude ($l$), and Z along positive latitude ($b$). The assumed value of the solar distance is $\mathrm{R}_{0}=8.2$ kpc \citep{Bland_Hawthorn_2016}. In order to show all our data, we do not restrict the third dimension in each plot. From the X-Z plots in the top row, we see that the stars from both surveys become more $\feh$-rich towards the plane. Additionally, both the individual and combined maps show that the more $\feh$-rich stars dominate at larger $|\mathrm{Z}|$ at small $|\mathrm{X}|$ than they do at larger $|\mathrm{X}|$. Lastly, the stars at the GC are more $\feh$-poor than their immediate surroundings. 

In the bottom row of Figure \ref{fig:Z_Xbar_nosym}, on top of the X-Y $\feh$ maps, we plot the bulge's density distribution obtained from one of the \citet{Portail_dyn_2017} bulge/bar models. These models were fit to the RC density of VVV, UKIDSS and 2MASS and to the stellar kinematics of BRAVA, OGLE, and $\argos$. The model we use has a pattern speed of $\Omega_{\mathrm{b}} = 37.5 \, \mathrm{km} \, \mathrm{s}^{-1} \, \mathrm{kpc}^{-1}$ as that was found to give the best visual match to the VIRAC proper motion data \citep{clarke2019milky}. In both the separate and combined X-Y maps, the near side of the bulge appears to be more $\feh$-rich than the far side. This is an effect of the field viewing angles which cause the nearer stars to be preferentially sampled closer to the plane than the farther stars.

Because we do not restrict the surveys to small bins in the projection direction in Figure \ref{fig:Z_Xbar_nosym}, the relative weighting by number density is incorrect, especially at low heights in the face-on view (see Section 5). In the following plots of this section, we restrict the abundance maps to smaller bins in vertical height and distance in order to minimize
this effect.

The bar causes an asymmetry in the spatial maps. To remove this asymmetry, we reorient the following plots to the bar reference system taking the bar angle to be $25\degree$ \citep{Bovy_2019}. The coordinate system is: the bar long axis ($\mathrm{X}_{\mathrm{bar}}$), the bar short axis ($\mathrm{Y}_{\mathrm{bar}}$), and the height from the Galactic plane ($\mathrm{Z}$). For these figures we also symmetrize the distribution of stars in order to fill in gaps in our spatial coverage as well as increase the statistics. The symmetrisation is done by reflecting each star into each projected quadrant.  

The top row of Figure \ref{fig:Z_Xbar} shows the symmetrised mean $\feh$ maps in the $|\mathrm{X}_{\mathrm{bar}}|$-$|\mathrm{Z}|$ plane for different slices in $|\mathrm{Y}_{\mathrm{bar}}|$ using stars from both $\apogee$ and $\atoa$. On top of the map, we plot the bulge density distribution (white dotted lines) obtained from the \citet{Portail_dyn_2017} model. In all $|\mathrm{Y}_{\mathrm{bar}}|$ slices, the mean $\feh$ generally increases towards the plane. However, for $|\mathrm{Y}_{\mathrm{bar}}|<1$ kpc and $|\mathrm{X}_{\mathrm{bar}}| < 1$ kpc, the mean $\feh$ increases rapidly towards the plane, then remains roughly constant between $0.3\lesssim|\mathrm{Z}| \, (\mathrm{kpc})\lesssim 0.7$, and finally decreases within the inner few 100 pc. This is not the case well outside the boxy-peanut (b/p) bulge lobes ($|\mathrm{X}_{\mathrm{bar}}| > 3$ kpc) where, within $1$ kpc from the Galactic plane, the mean $\feh$ values increase rapidly towards the plane with no large regions of constant mean $\feh$ or inversions of the $\feh$ gradient. Furthermore, the $\feh$ structure in the $|\mathrm{Y}_{\mathrm{bar}}|<1$ kpc slice (top left panel of Figure \ref{fig:Z_Xbar}) is puffed up and X-shaped with the more $\feh$-rich stars dominating at large $|\mathrm{Z}|$ inside the b/p bulge lobes ($|\mathrm{X}_{\mathrm{bar}}|{\sim}2$ kpc). The X-shape is seen in mean $\feh$ values between 0 dex and $-0.4$ dex. For $|\mathrm{Y}_{\mathrm{bar}}| > 1$ kpc, the mean $\feh$ structure becomes increasingly flat with increasing $|\mathrm{Y}_{\mathrm{bar}}|$ and the difference between large and small $|\mathrm{X}_{\mathrm{bar}}|$ decreases. In the $1\leq|\mathrm{Y}_{\mathrm{bar}}| \, (\mathrm{kpc})<2$ slice (middle panel of Figure \ref{fig:Z_Xbar}), one still sees a slight pinching/X-shape in the $\feh$ distribution at mean $\feh$ values of ${\sim}-0.25$ dex.

The bottom row of Figure \ref{fig:Z_Xbar} shows the symmetrised mean $\mgfe$ distribution for $\apogee$ and $\atoa$ stars in the $|\mathrm{X}_{\mathrm{bar}}|$-$|\mathrm{Z}|$ plane in slices of $|\mathrm{Y}_{\mathrm{bar}}|$. The $\mgfe$ maps mirror the $\feh$ maps. The mean $\mgfe$ generally decreases towards the plane in all $|\mathrm{Y}_{\mathrm{bar}}|$ slices. For $|\mathrm{Y}_{\mathrm{bar}}|<1$ kpc (bottom left panel of Figure \ref{fig:Z_Xbar}) and $|\mathrm{X}_{\mathrm{bar}}| < 1$ kpc, the rate of decrease of the mean $\mgfe$ is slower and the gradient inverts at small $|\mathrm{Z}|$ such that the inner bulge is slightly more $\mgfe$-rich than its immediate surroundings. A clear X-shape is seen in the mean $\mgfe$ distribution
at roughly $\mgfe\approx0.175$ dex in the $|\mathrm{Y}_{\mathrm{bar}}|<1$ kpc slice. In the region of the X-shape, the $\mgfe$-poor stars dominate at larger $|\mathrm{Z}|$ than they do at larger $|\mathrm{X}_{\mathrm{bar}}|$ or larger $|\mathrm{Y}_{\mathrm{bar}}|$. For larger $|\mathrm{Y}_{\mathrm{bar}}|$, the mean $\mgfe$ distribution becomes increasingly flat.

The $\feh$ and $\mgfe$ distributions are more strongly pinched than the density distribution in the $|\mathrm{Y}_{\mathrm{bar}}|<1$ kpc slice (left panels of Figure \ref{fig:Z_Xbar}). At larger $|\mathrm{Y}_{\mathrm{bar}}|$, the density contours and the $\feh$ and $\mgfe$ contours are in better agreement.

Figure \ref{fig:Ybar_Xbar} shows the symmetrised mean $\feh$ maps in the $|\mathrm{X}_{\mathrm{bar}}|$-$|\mathrm{Y}_{\mathrm{bar}}|$ plane for different slices in $|\mathrm{Z}|$. The stars are restricted to the bar region which we approximate as an ellipse with a semi-major axis and axis ratio of 5 kpc and $0.4$ respectively; see Figure \ref{fig:Z_Xbar_nosym}. For $|\mathrm{Z}|<0.3$ kpc, the center of the bar is more $\feh$-poor than the bar ends. As the distance from the plane increases, this reverses at ${\sim} 0.75$ kpc. At greater heights, we again see that the center of the bar is more $\feh$-poor than the bar ends.

In near infrared star counts, the Galactic bar has a half length of ${\sim}5$ kpc \citep{Wegg_2015}. The b/p bulge extends out to ${\sim}2$ kpc from the GC. The bar region that extends outside the b/p bulge is known as the long bar. \citet{Wegg_2015} showed that the long bar is composed of two bar components, the thin bar with a scale height of $180$ pc, extending to ${\sim}4.6$ kpc, and the super thin bar with a scale height of $45$ pc, reaching ${\sim}5$ kpc. We do not have the resolution to detect an $\feh$ or a $\mgfe$ signature of the super thin bar, however the top panel of Figure \ref{fig:Ybar_Xbar} extents to roughly $1.7$ thin bar scale heights above the plane. From this, we can approximately say that the combined long bar is super solar in $\feh$ (also seen in the top left panel of Figure \ref{fig:Z_Xbar}). The lower left panel of Figure \ref{fig:Z_Xbar} also shows that the region occupied by the long bar is nearly solar in $\mgfe$. This is in contrast to the inner region of the b/p bulge which has a mean sub-solar $\feh$ value and is more $\mgfe$-rich.

\subsection{Abundance Gradients}\label{gradients}
\begin{figure*}
\centering
\includegraphics{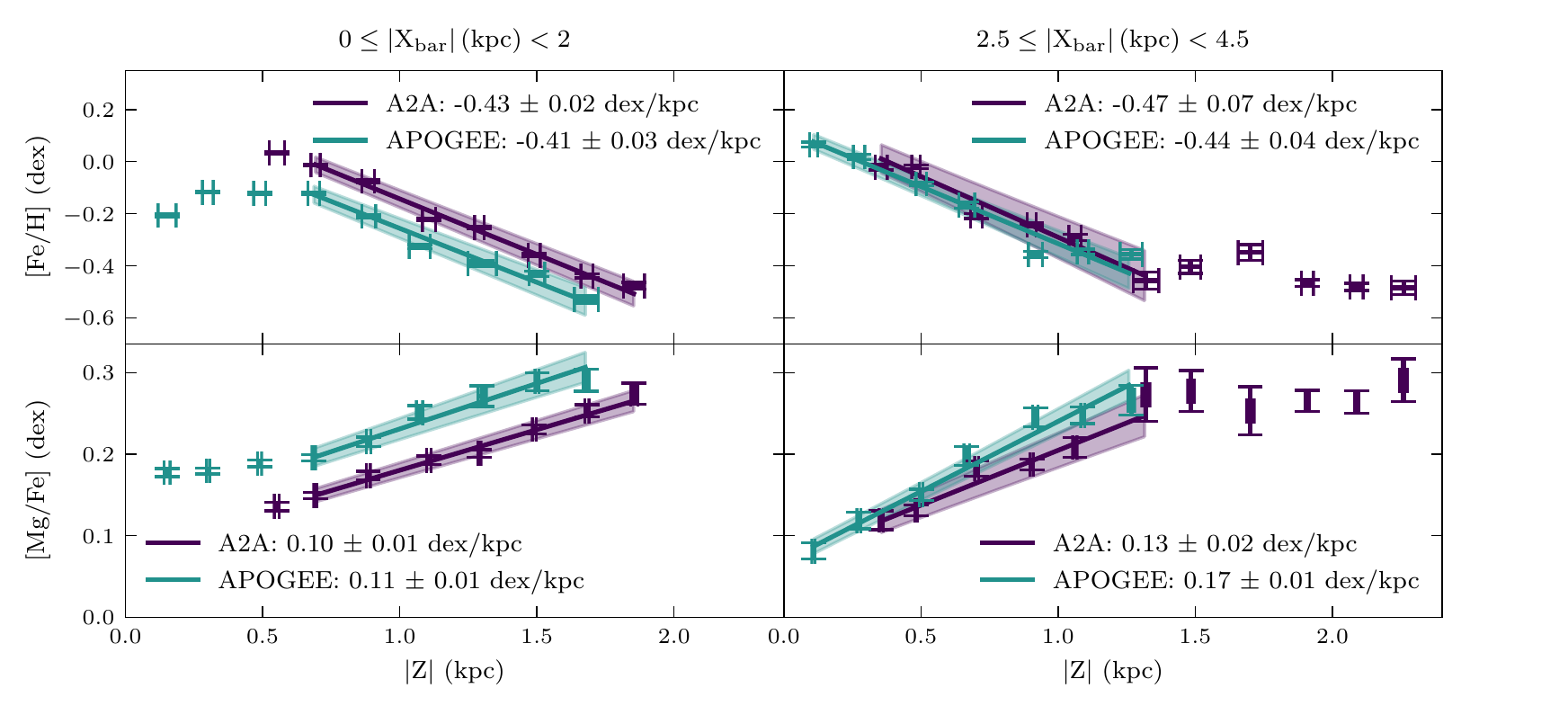}
\caption{SSF-corrected mean $\feh$ (top row) and $\mgfe$ (bottom row) vertical abundance profiles for $\atoa$ and $\apogee$ stars in the inner b/p bulge ($|\mathrm{X_{\mathrm{bar}}}|<2$ kpc, $|\mathrm{Y}_{\mathrm{bar}}|<1 \, \mathrm{kpc}$;  left) and long bar/outer bulge regions ($2.5$ kpc $|\leq\mathrm{X_{\mathrm{bar}}}|<4.5$kpc, $|\mathrm{Y}_{\mathrm{bar}}|< 1 \, \mathrm{kpc}$; right). The $\atoa$ and $\apogee$ gradients of the regions shown by the teal and purple lines are given in the legend of each diagram; error ranges of the linear fits are shown by the shaded regions. For all plots, we require the stars to have $\feh>-1$ dex.}
\label{fig:grad}
\end{figure*}

\begin{figure*}
\centering
\includegraphics{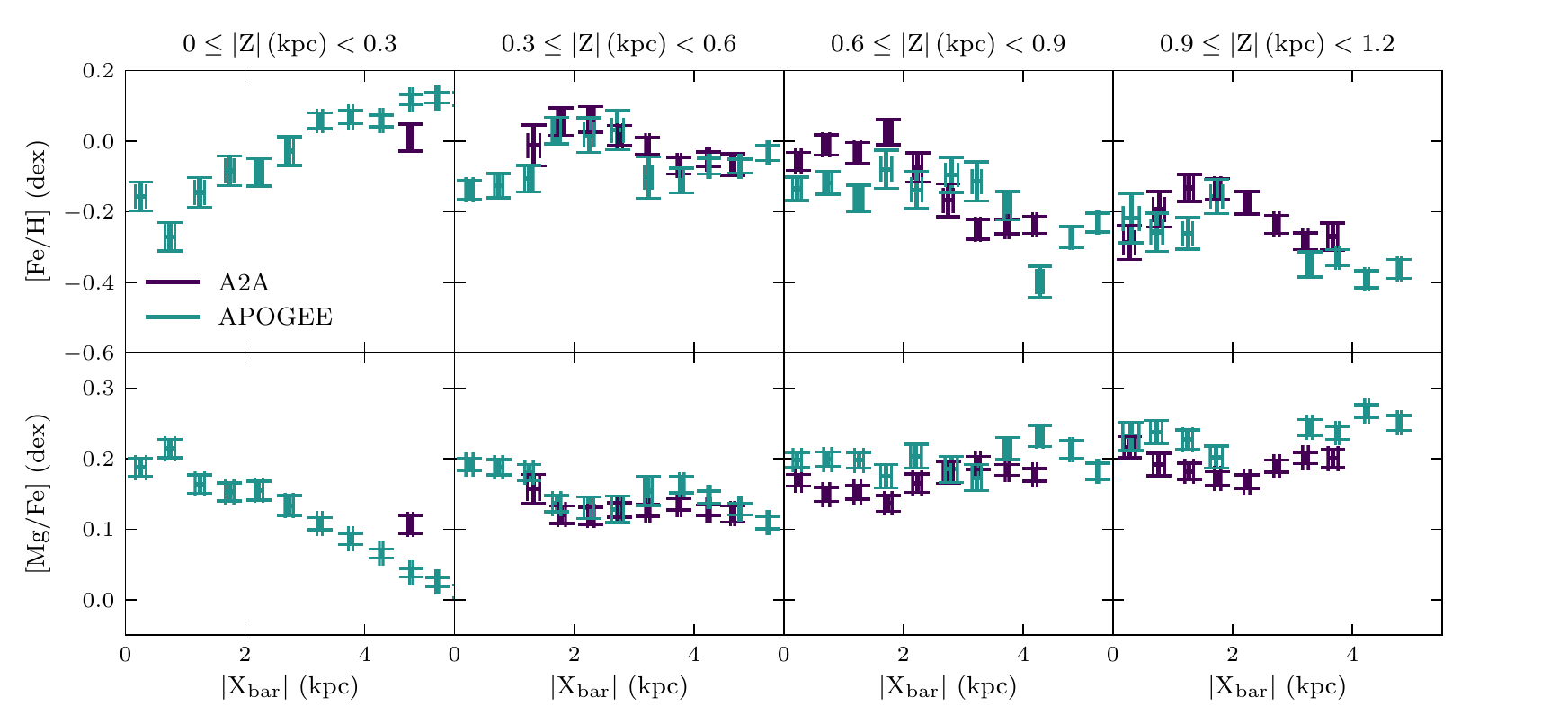}
\caption{SSF-corrected mean $\feh$ (top row) and $\mgfe$ (bottom row) horizontal abundance profiles of $\atoa$ and $\apogee$ stars along the bar and at different heights above the Galactic plane. For all plots, we require the stars to have $\feh>-1$ dex and $|\mathrm{Y}_{\mathrm{bar}}|<1 \, \mathrm{kpc}$.}
\label{fig:hor_grad}
\end{figure*}

Having shown how $\feh$ and $\mgfe$ vary over the bulge, we now quantify the vertical and horizontal abundance gradients in the various bulge regions. In Figure \ref{fig:grad}, we present the mean $\feh$ and $\mgfe$ profiles for $\atoa$ and $\apogee$ stars in the inner bulge (left column) and long bar/outer bulge region (right column) as a function of $|\mathrm{Z}|$. We take the inner bulge to be the region within $|\mathrm{X}_{\mathrm{bar}}|<2 \, \mathrm{kpc}$ and $|\mathrm{Y}_{\mathrm{bar}}|<1 \, \mathrm{kpc}$ and the long bar/outer bulge region to be the region within $2.5 \leq|\mathrm{X}_{\mathrm{bar}}| \, (\mathrm{kpc})< 4.5 \, \mathrm{kpc}$ and $|\mathrm{Y}_{\mathrm{bar}}|< 1 \, \mathrm{kpc}$.  

The $\apogee$ and $\atoa$ $\feh$ and $\mgfe$ gradients agree within the errors in both regions of the bulge. However, the $\feh$ and $\mgfe$ profiles in the inner bulge are offset by roughly $0.1$ dex and $0.05$ dex respectively. These offsets are at least partially due to missing $\feh$-poor, $\mgfe$-rich stars in $\atoa$ as discussed in Section \ref{inconsis}.

The inner bulge has a different vertical $\feh$ profile than the long bar/outer bulge region. For $0.7\lesssim |\mathrm{Z}| \,(\mathrm{kpc}) \lesssim 2$, the inner bulge $\feh$ gradient is ${\sim}-0.41$ dex/kpc; at lower heights, between $0.3 \lesssim |\mathrm{Z}| \,(\mathrm{kpc}) \lesssim 0.7$ it flattens at $\feh \approx -0.12$ dex. This flattening of the $\feh$ gradient is only clear in the $\apogee$ data (the $\atoa$ coverage is too sparse in this area), but was previously seen also by \citet{Rich_2012, Rich_2007} and \citet{Ness_2016}. Below $~0.3$ kpc the mean $\feh$ slightly decreases. This inversion of the mean $\feh$ gradient was also seen in Figure \ref{fig:Z_Xbar}.

The $\feh$ gradient of the outer bulge/long bar region is roughly flat between $1.25 \lesssim |\mathrm{Z}| \,(\mathrm{kpc})\lesssim 2.25$ at a value of ${\sim}-0.44$ dex. For $|\mathrm{Z}|\lesssim1.25$ kpc, the gradient is ${\sim}-0.44$ dex/kpc and has no inner flattening. 

Using stars with $|l|<11\degree$, \citet{Rojas_Arriagada_2020} found the bulge vertical metallicity gradient to be $-0.09$ dex/kpc for $|\mathrm{Z}|<0.7$ kpc and $-0.44$ dex/kpc for $0.7 < |\mathrm{Z}| \,(\mathrm{kpc})< 1.2$. Beyond $|\mathrm{Z}|>1.2$ kpc, \citet{Rojas_Arriagada_2020} found a noisy but flat profile. Assuming the bar is at an angle of $25\degree$ with respect to the Sun, the $11\degree$ limit in Galactic longitude restricts their sample to $\lesssim 2.7$ kpc along the bar, or roughly the region we refer to as the inner bulge. Thus, their vertical metallicity gradient is consistent with our inner bulge vertical metallicity gradient. However, our increased Galactic longitude range allows us to see that the flattening at small $|\mathrm{Z}|$ only occurs in the inner bulge and not in the long bar/outer bulge region.

The $\mgfe$ profiles mirror the $\feh$ profiles. For the inner bulge, the $\mgfe$ profile is roughly flat for $|\mathrm{Z}|\lesssim0.7$ kpc at $\mgfe \approx 0.19$ dex. For $|\mathrm{Z}|\gtrsim0.7$ kpc, the $\mgfe$ gradient is ${\sim}0.11$ dex/kpc. In the long bar/outer bulge region, the $\mgfe$ gradient is ${\sim}0.17$ dex/kpc in $\apogee$ and ${\sim}0.13$ dex/kpc in $\atoa$ between $0 \lesssim |\mathrm{Z}| \,(\mathrm{kpc})\lesssim 1.25$. For $|\mathrm{Z}|\gtrsim1.25$ kpc the $\mgfe$ profile flattens at ${\sim}0.27$ dex.

The top row of Figure \ref{fig:hor_grad} shows the horizontal mean $\feh$ profile of $\atoa$ and $\apogee$ stars along the Galactic bar at different heights above the plane. For $|\mathrm{Z}|\lesssim0.3$ kpc, the radial $\feh$ gradient is steep and positive. However, the stars at $|\mathrm{X}_{\mathrm{bar}}|\gtrsim 2-3$ kpc from both surveys strongly decrease in $\feh$ with increasing height from the plane. For stars at $|\mathrm{X}_{\mathrm{bar}}|\lesssim 2-3$ kpc this effect is less pronounced. For $\atoa$, this decrease in mean $\feh$ within $|\mathrm{X}_{\mathrm{bar}}|\lesssim 2-3$ kpc is stronger at smaller $|\mathrm{X}_{\mathrm{bar}}|$. This reflects the transitions between the relatively $\feh$-poor central bulge, the $\feh$-rich long bar near the Galactic plane, the region of enhanced $\feh$ in the b/p bulge, and the $\feh$-poor region above the long bar in Figure \ref{fig:Z_Xbar}.

The $\mgfe$ horizontal profiles of both surveys are shown in the bottom row of Figure \ref{fig:hor_grad}. For $|\mathrm{Z}|\lesssim0.3$ kpc, the mean radial $\mgfe$ gradient is steep and negative. However as the distance from the plane increases the stars at large $|\mathrm{X}_{\mathrm{bar}}|$ increase in $\mgfe$, causing the profile to flatten. At large $|\mathrm{Z}|$, a minimum in $\mgfe$ is seen around $|\mathrm{X}_{\mathrm{bar}}|\approx 2$ kpc. This profile is due to the lobes of the b/p bulge. 

As was the case for the vertical gradients, there are clear offsets between the $\apogee$ and $\atoa$ horizontal profiles in $\feh$ and $\mgfe$, especially for $|\mathrm{X}_{\mathrm{bar}}|<2$ kpc. This could at least in part be due to the limited $\teff$ range of the reference set (see Section \ref{inconsis}).

\subsection{Shape of Bulge Abundance Distribution Functions}\label{MDFshape}
\begin{figure*}
\centering
\includegraphics{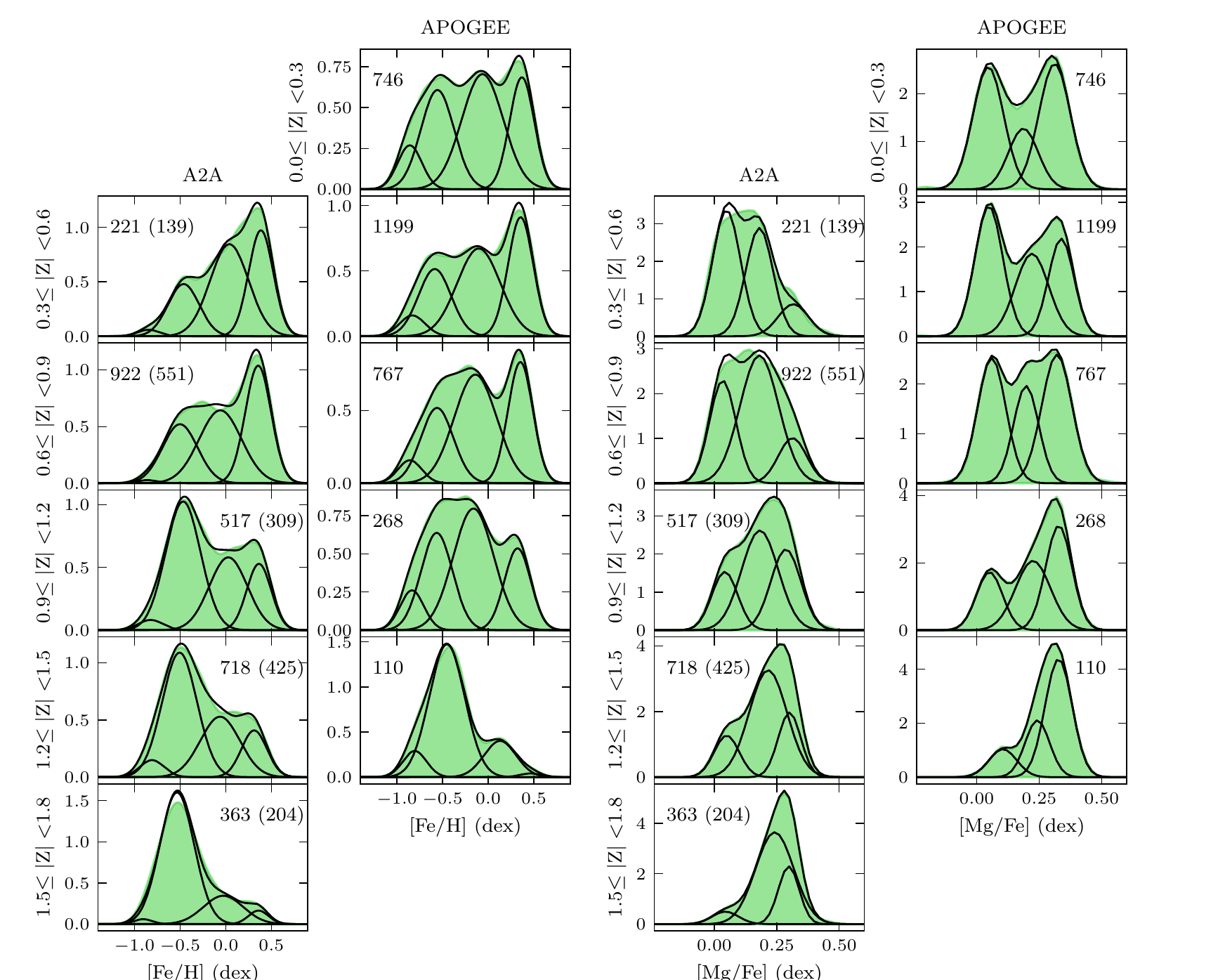} 
\caption{Generalized MDFs (left two columns) and Mg-DFs (right two columns) of $\atoa$ (RC) and $\apogee$ stars at different absolute heights above the plane in the inner bulge, shown as filled distributions in green. The stars are required to have $|\mathrm{X}_{\mathrm{bar}}| < 2 \, \mathrm{kpc}$, $|\mathrm{Y}_{\mathrm{bar}}| < 1 \, \mathrm{kpc}$, and $[\mathrm{Fe}/\mathrm{H}] > -1 \, \mathrm{dex}$. Gaussian mixture decompositions at each height are also shown in black (individual Gaussians and sums). The number of distinct stars composing each distribution is given in each plot. The number in brackets in the $\atoa$ plots gives the number (total weight) of $\atoa$ RC stars.}
\label{fig:inner_mdf}
\end{figure*}

\begin{figure*}
\centering
\includegraphics{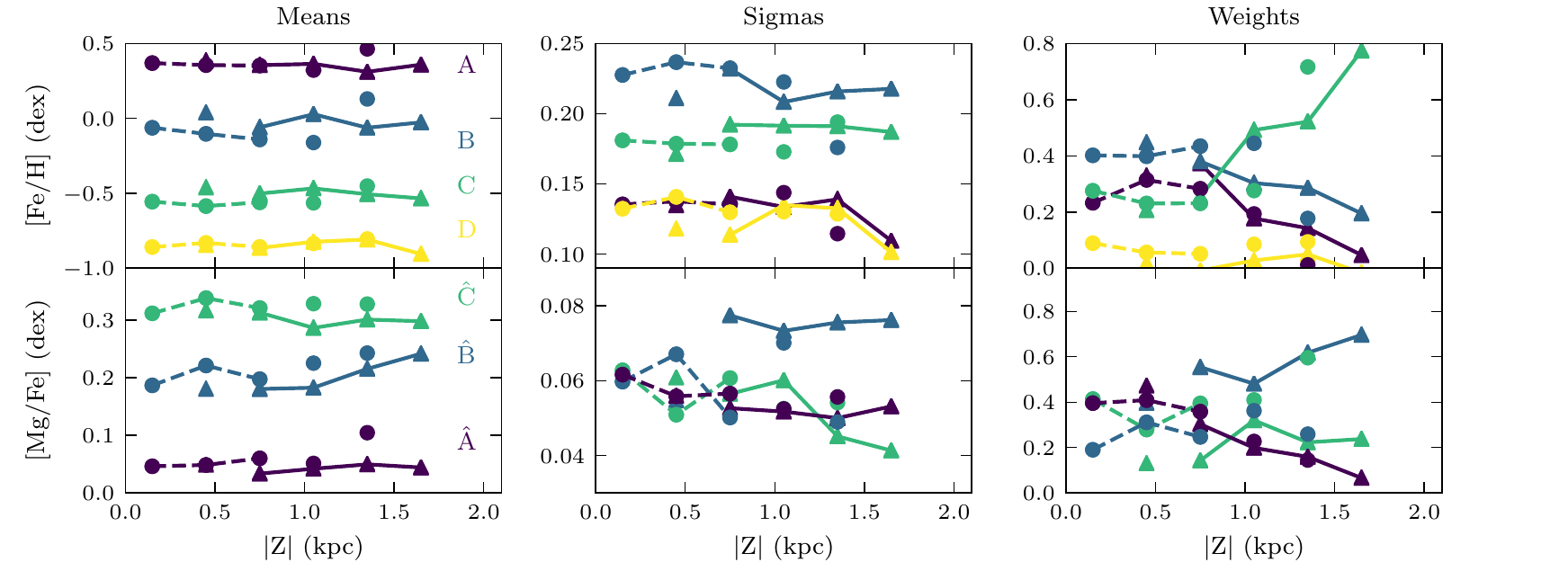} 
\caption{The variation of the Gaussian parameters with height from the Galactic plane ($|\mathrm{Z}|$) for the inner bulge. Top row: parameters from the MDF decompositions. Bottom row: parameters from the Mg-DF decompositions. The lines connect points with at least 300 distinct stars Triangular markers and solid lines: $\atoa$ decompositions. Circular markers and dashed lines: $\apogee$ decompositions.}
\label{fig:inner_mdfparams}
\end{figure*}

\begin{figure*}
\centering
\includegraphics{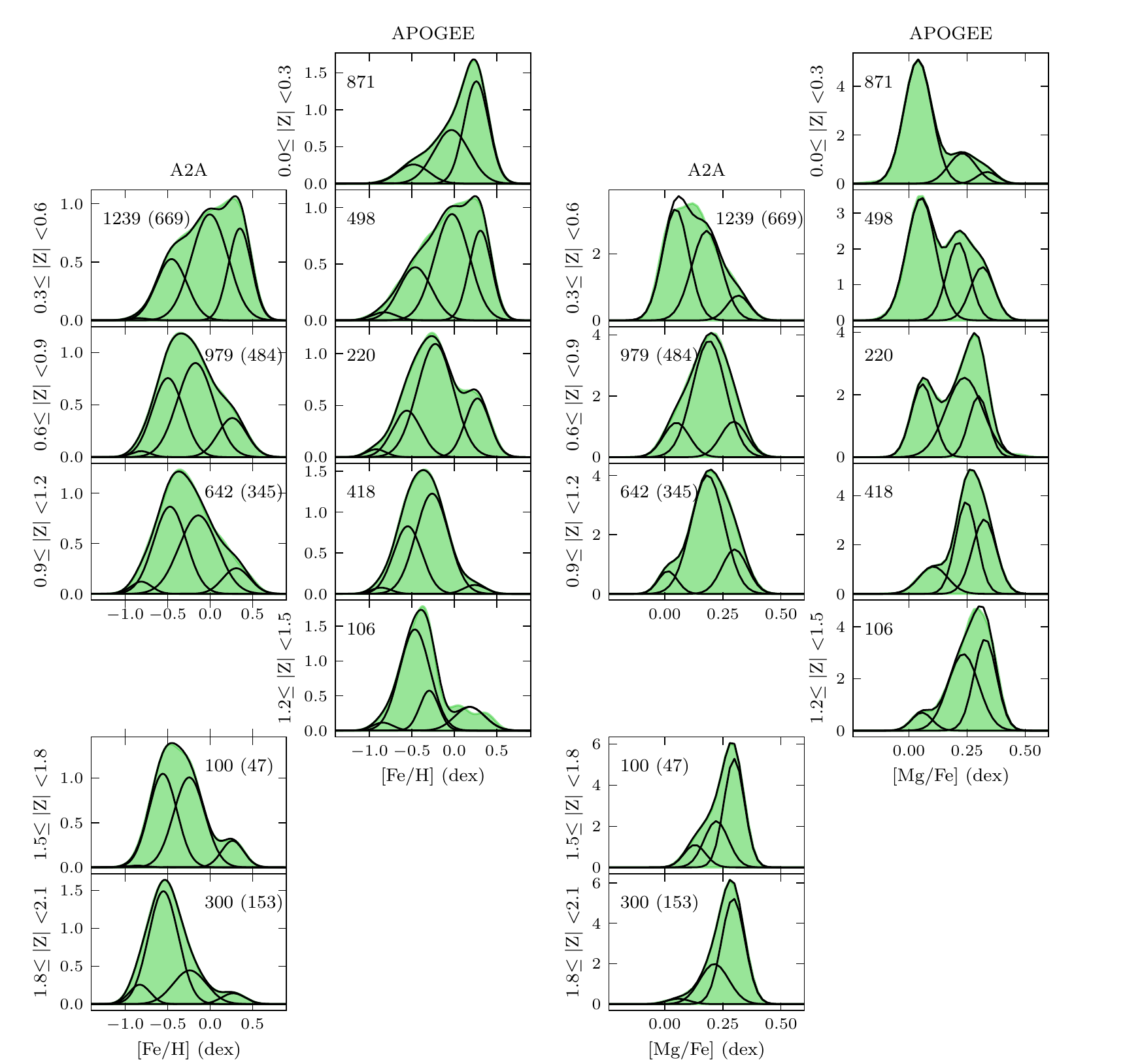} 
\caption{Same as Figure \ref{fig:inner_mdf} except for the long bar/outer bulge region.The stars are required to have $2.5 \, \mathrm{kpc} \leq|\mathrm{X}_{\mathrm{bar}}| < 4.5 \, \mathrm{kpc}$,  $|\mathrm{Y}_{\mathrm{bar}}| < 1 \, \mathrm{kpc}$, and $\feh>-1$ dex.}
\label{fig:outer_mdf}
\end{figure*}

\begin{figure*}
\centering
\includegraphics{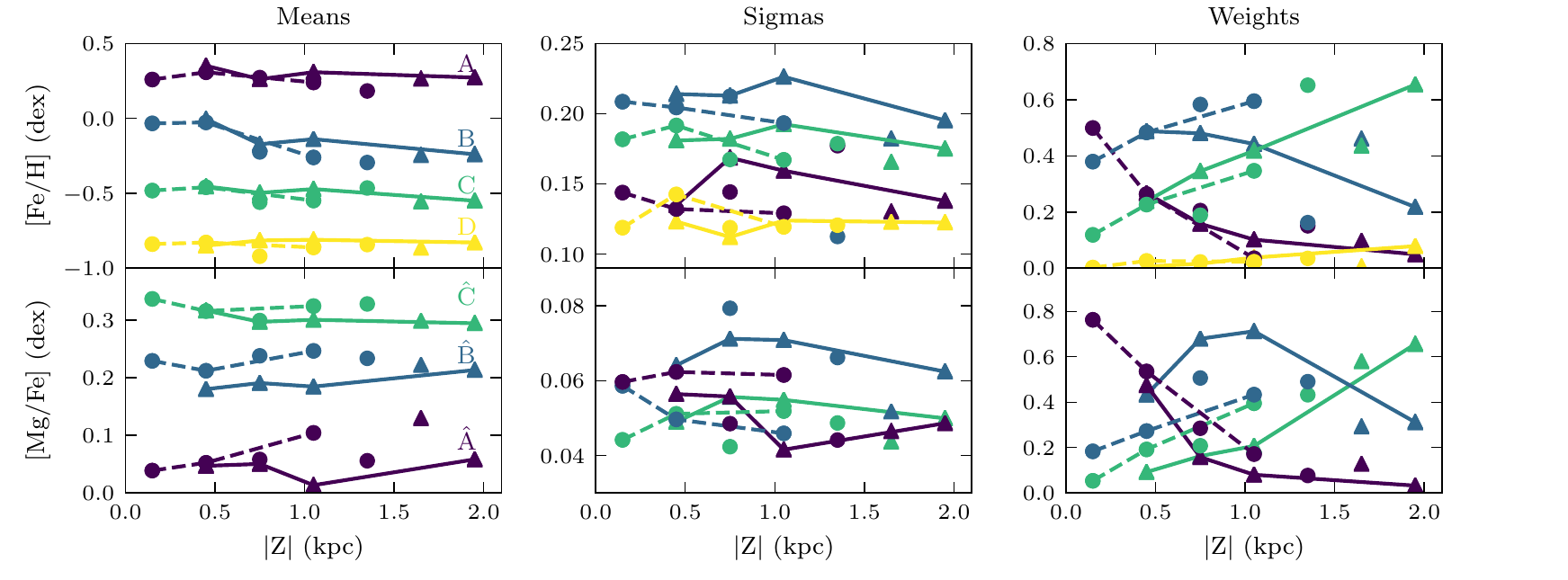}
\caption{Same as Figure \ref{fig:inner_mdfparams} except for the long bar/outer bulge region.}
\label{fig:outer_mdfparams}
\end{figure*}

So far we have only examined how the mean $\feh$ and $\mgfe$ vary with position in the bulge. In this section, we illustrate how the MDFs and Mg-DFs change with position in the bulge.

In Figure \ref{fig:inner_mdf} we plot the generalised MDFs and Mg-DFs of $\atoa$ (RC) and $\apogee$ stars in the inner bulge for $0 < |\mathrm{Z}| \,(\mathrm{kpc})< 1.8$ in bins of width $0.3$ kpc. The bins were chosen such that they each contain at least 100 distinct stars. The Gaussian smoothing of each stars is $0.1$ dex in the MDFs and $0.033$ dex in the Mg-DFs. We see that, while there are some deviations, both surveys show similar trends in their MDFs with $|\mathrm{Z}|$. Far from the plane, the MDFs of both surveys are dominated by a strong peak at ${\sim}-0.4$- ${\sim}-0.5$ dex. As the distance from the plane decreases, a solar and super solar peak in $\feh$ grow and become prominent. The surveys show a stronger difference in their Mg-DFs. In $\apogee$, far from the plane the Mg-DF is dominated by a single peak at ${\sim}0.3$ dex. As the distance from the plane decreases, a second peak at ${\sim}0.05$ dex increases in strength, such that near the Galactic plane, the two peaks are nearly equal in strength. In $\atoa$, the Mg-DF far from the plane is also dominated by a single peak at ${\sim}0.3$ dex. However, as the distance from the plane decreases, the strength of the high $\mgfe$ peak decreases and the stars below ${\sim}0.25$  dex increase in strength. The peak at ${\sim}0.3$ dex, seen in the $\apogee$ Mg-DF, is not prominent in the $\atoa$ Mg-DFs near the plane. 

Using the affine-invariant MCMC sampler, \emph{emcee} \citep{Foreman_Mackey_2013}, we fit a four component Gaussian mixture model (GMM) to each generalized MDF and a three component GMM to each generalized Mg-DF. The Gaussians and their sums are plotted on top of the generalized MDFs and Mg-DFs in Figure \ref{fig:inner_mdf}. To see the variation of the MDF and the Mg-DF Gaussian parameters clearly, we plot the Gaussian means, sigmas, and weights against $|\mathrm{Z}|$ in Figure \ref{fig:inner_mdfparams}. To minimise the effects of noise, we only connect points in Figure \ref{fig:inner_mdfparams} with at least 300 distinct stars. 

The top left plot of Figure \ref{fig:inner_mdfparams} shows the variation of the MDF Gaussian means with $|\mathrm{Z}|$. Both the $\atoa$ and $\apogee$ MDFs are well fit by a super solar $\feh$ Gaussian (A), an intermediate $\feh$ Gaussian (B), an $\feh$-poor Gaussian (C), and a very $\feh$-poor Gaussian (D). The overall variation of the $\feh$ means with latitude is not substantial. The top middle plot shows the MDF sigma variations with $|\mathrm{Z}|$. Gaussian B generally has the largest sigma closely followed by C, and then A and D which are nearly equal in sigma. The top right plot shows the MDF weight variations with $|\mathrm{Z}|$. The weights of all Gaussians are roughly constant below $|\mathrm{Z}|\approx0.7$ kpc, with B having marginally the largest weight. For $|\mathrm{Z}|\gtrsim0.7$ kpc, the most significant metal poor Gaussian C increases, while the other two decrease such that C becomes the most dominant at large $|\mathrm{Z}|$. Gaussian D is the weakest component at all heights as it never reaches over $10\%$ in weight.

The variation of the Gaussian parameters from the three Gaussians fit to the Mg-DFs in the inner bulge is shown in the bottom row of Figure \ref{fig:inner_mdfparams}. The bottom left plot shows the variation of the Gaussian means with $|\mathrm{Z}|$. The Mg-DFs of both surveys are well fit by a $\mgfe$-normal Gaussian (\^{A}), an intermediate $\mgfe$ Gaussian (\^{B}), and a $\mgfe$-rich Gaussian (\^{C}). The bottom middle plot shows the sigma variation of the Gaussians with $|\mathrm{Z}|$. The $\atoa$ Gaussian \^{B} generally has the highest sigma by ${\sim}0.025$ dex. The rest of the Gaussians have nearly equal sigma values. The bottom left plot shows the variations of the weights with $|\mathrm{Z}|$. The Gaussian weights are constant below ${\sim}0.7$ kpc and nearly equal in weight. Above ${\sim}0.7$ kpc, the weight of Gaussian \^{A} decreases with increasing $|\mathrm{Z}|$ such that at ${\sim}1.7$ kpc its weight is nearly zero. Above ${\sim}0.7$ kpc, the behaviors of Gaussians \^{B} and \^{C} strongly differ between the $\atoa$ and $\apogee$ surveys. As $|\mathrm{Z}|$ increases, the $\apogee$ weights of Gaussians \^{C} and \^{B} increase and remain roughly constant respectively, while the $\atoa$ weights of Gaussians \^{C} and \^{B} remain constant and increase respectively.

In Figure \ref{fig:outer_mdf} we perform a similar procedure as in Figure \ref{fig:inner_mdf} but on the long bar/outer bulge region. Using both surveys, we obtain generalised MDFs and Mg-DFs with smoothings of ${\sim} 0.1$ dex and ${\sim}0.033$ dex respectively and their Gaussian decompositions in $|\mathrm{Z}|$ bins of width $0.3$ kpc between $0<|\mathrm{Z}| \, (\mathrm{kpc})<2.1$. As was the case with the inner bulge, the more $\feh$-poor $\mgfe$-rich stars dominate far from the plane while the more $\feh$-rich $\mgfe$-poor stars dominate close to the plane.

The variations of the MDF Gaussian parameters with $|\mathrm{Z}|$ are shown in the top row of Figure \ref{fig:outer_mdfparams}. We only connect points with at least 300 distinct stars to minimise noise.  Similarly to the inner bulge, the top left plot shows that the long bar/outer bulge region is well fit by a super solar $\feh$ Gaussian (A), an intermediate $\feh$ Gaussian (B), an $\feh$-poor Gaussian (C), and a very $\feh$-poor Gaussian (D). The sigma variations are shown in the top middle plot. Gaussian B has the largest sigma value, sequentially followed by Gaussians C,  A and D. The variations of the MDF Gaussian weights are shown in the top right plot. At $|\mathrm{Z}|\gtrsim 1$ kpc, the weights of the Gaussians are similar to those of the inner bulge, with C dominating over B and A. At low $|\mathrm{Z}|$ the weight of the most metal rich Gaussian A is higher than weight of C, the most significant metal poor Gaussian. The transition in weight occurs at lower $|\mathrm{Z}|$ than in the inner bulge. Furthermore, for $|\mathrm{Z}|\lesssim0.7$ kpc, the weight profiles of A and C are not constant, but continue to increase and decrease towards the Galactic plane respectively. This is consistent with the profiles in Figure \ref{fig:grad}. The weight of Gaussian D is weak at all heights, never rising above $10\%$.

In the bottom row of Figure \ref{fig:outer_mdfparams} we plot the variation of the Mg-DFs Gaussian parameters. The left most plot shows the variation of the Gaussian means. The Mg-DFs of both surveys are well fit by a $\mgfe$-normal Gaussian (\^{A}), an intermediate $\mgfe$ Gaussian (\^{B}), and a $\mgfe$-rich Gaussian (\^{C}). We see a significant offset between the Gaussian \^{B} means from both surveys. Furthermore, at $|\mathrm{Z}|\approx 1$ kpc the Gaussians \^{A} means have a large offset. The bottom right plot shows the Gaussian weight variations with $|\mathrm{Z}|$. For both surveys, the $\mgfe$-rich Gaussian \^{C} is strong at high $|\mathrm{Z}|$ and decreases in weight with decreasing $|\mathrm{Z}|$,  while the $\mgfe$-normal Gaussian \^{A} is very weak at high $|\mathrm{Z}|$ and increases in weight with decreasing $|\mathrm{Z}|$. Because of these trends, close to the plane, Gaussian \^{A} dominates and Gaussian \^{C} is near zero in weight. At most heights, the Gaussian \^{B} weight behavior from both surveys deviates with the weight of Gaussian \^{B} from $\atoa$ being much larger than the weight of Gaussian \^{B} from $\apogee$. Accordingly, the weight of Gaussian \^{C} from $\atoa$ is lower than the weight of the correspondingly $\mgfe$-rich Gaussian \^{C} from  $\apogee$ at most heights. 

By comparing the $\apogee$ weight behavior of the MDF and Mg-DF Gaussians in both regions of the bulge, it is clear that Gaussians A, B, and C from the MDF decomposition roughly correspond to Gaussians \^{A}, \^{B}, and \^{C} from the Mg-DF decomposition. However, in the outer bulge, Gaussian \^{A} is higher in weight than Gaussian A, while Gaussian B is higher in weight than Gaussian \^{B}. Therefore, there is mixing between the Gaussians with some stars that compose Gaussians A and B in $\apogee$ being part of Gaussians \^{B} and \^{A} respectively. For $\atoa$, Gaussian A corresponds to Gaussian \^{A}. However, the behaviors of Gaussians B and C from $\atoa$ from the MDF decomposition differ strongly from the behaviors of Gaussians \^{B} and \^{C} from the Mg-DF decomposition. This deviation in Gaussian weight behavior is much stronger in the inner bulge than in the outer bulge. 

Overall, it appears that the $\atoa$ stars do not reach as high in $\mgfe$ as the $\apogee$ stars do. Because of this, the $\apogee$ and $\atoa$ Mg-DFs deviate in shape at the $\mgfe$-rich end. We suspect that this may be due to the limited $\teff$ range of the reference set used to train $\thecannon$ model for the $\atoa$ catalog; see Section \ref{inconsis} for a more detailed discussion on this.

Multiple other surveys \citep{Rojas_Arriagada_2017, Zoccali_2017, Hill_2011, Schultheis_2017, Uttenthaler_2012} have reported the bulge MDF to be bi-modal. The $\feh$-rich Gaussians in many of these surveys have a mean value of ${\sim}0.3$ dex, similar to the mean of our most $\feh$-rich Gaussian in both the inner and outer bulge. Recently, \citet{Rojas_Arriagada_2020} analysed the bulge MDF using a larger sample from $\apogee$ DR16 data and reported that the bulge is best represented by the superposition of three Gaussian components with nearly constant means at $0.32$ dex, $-0.17$ dex, and $-0.66$ dex. While we find that the bulge is best fit by four Gaussians, the fourth most $\feh$-poor Gaussian is not very significant and we mainly include it to improve the fit. In the inner bulge, the mean values of Gaussian A are similar to that of the most $\feh$-rich Gaussian of \citet{Rojas_Arriagada_2020}. However, in the inner bulge the means of Gaussians B and C are generally more $\feh$-rich than their two most $\feh$-poor Gaussians. This is likely due to the inclusion of the fourth $\feh$-poor Gaussian. Alternatively, the difference may arise from differing SSF-correcting methods or differing distance criteria. Thus, we do not believe that our results significantly differ from those of \citet{Rojas_Arriagada_2020}.

\subsection{[Mg/Fe]-[Fe/H] Distribution}\label{mgfedist}
\begin{figure*}
\centering
\includegraphics{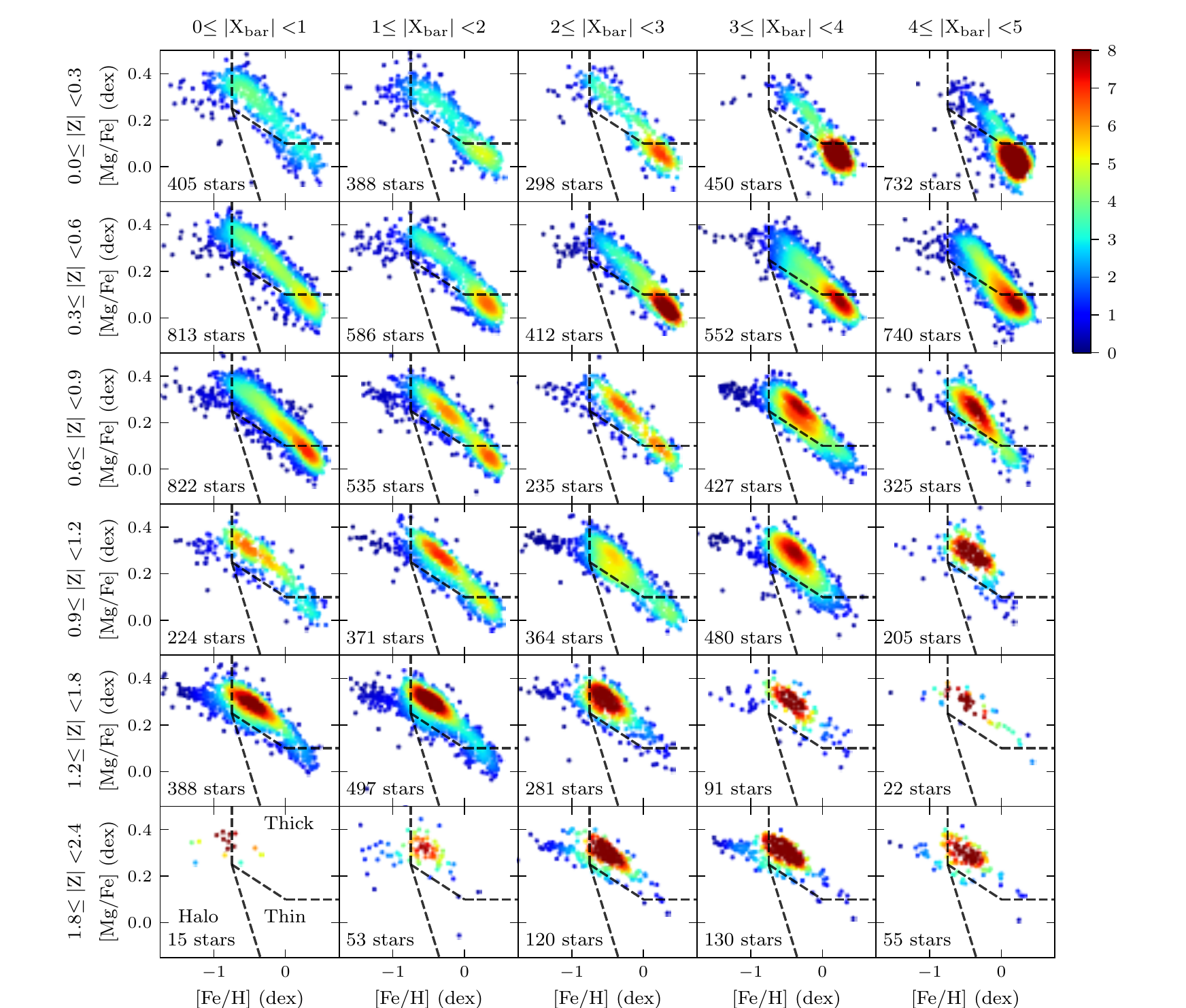}
\caption{The $\mgfe$-$\feh$ distribution of $\atoa$ and $\apogee$ stars in intervals of height from the plane ($|\mathrm{Z}|$, kpc) and distance along the bar ($|\mathrm{X}_{\mathrm{bar}}|$, kpc). The stars are required to have $|\mathrm{Y}_{\mathrm{bar}}|<1$ kpc. The point colour gives the Gaussian kernel density estimate. The dashed lines separate the different regions in the parameter space generally populated by the halo, thin disk and thick disk (defined in the lower-leftmost plot; see text). The number of stars composing each plot ($\atoa$ RC stars + $\apogee$ stars) is given in the lower left corner of each plot.}
\label{fig:alpha_plot}
\end{figure*}

\begin{figure*}
\centering
\includegraphics{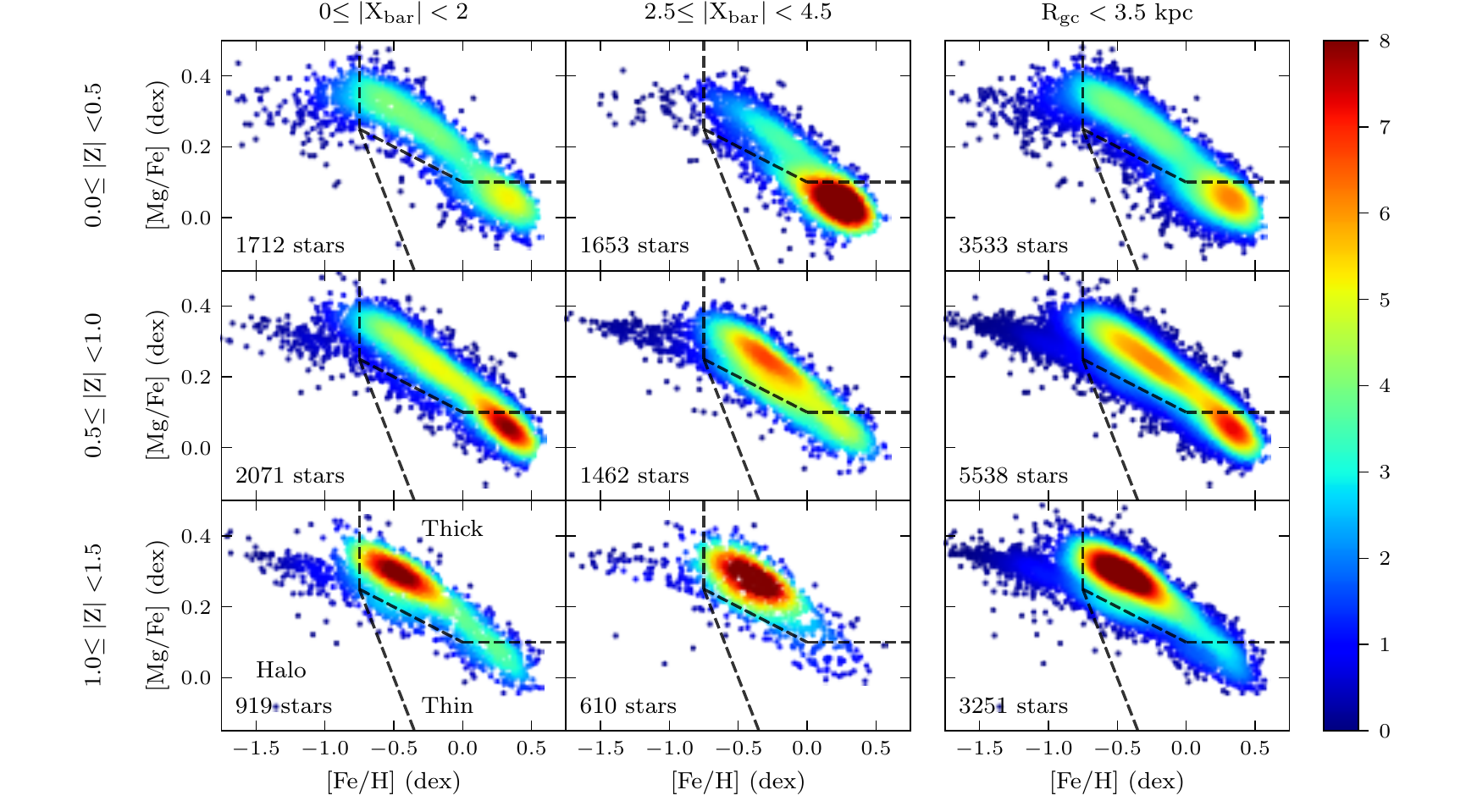}
\caption{The $\mgfe$-$\feh$ distribution for $\atoa$ and $\apogee$ stars in intervals of vertical height (in kpc), for the inner bulge (left column), long bar/outer bulge region (middle column), and within a galactocentric radius of $3.5$ kpc (right column). The stars in the first two columns are required to have $|\mathrm{Y}_{\mathrm{bar}}|<1$ kpc. For plot specifics see the caption of Figure \ref{fig:alpha_plot}.}
\label{fig:alpha_plot_full}
\end{figure*}

In this section we present how the $\mgfe$-$\feh$ distribution changes along the Galactic bar using a similar method to \citet{Hayden_2015} and \citet{Queiroz_2020a}. 

In Figure \ref{fig:alpha_plot}, using stars from both the $\atoa$ and $\apogee$, we plot the $\mgfe$-$\feh$ distribution along the bar in bins of $|\mathrm{Z}|$ and $|\mathrm{X}_{\mathrm{bar}}|$, requiring $|\mathrm{Y}_{\mathrm{bar}}|<1$ kpc. We colour the points by the Gaussian kernel-density estimation using band-widths that obey Scott's rule \citep{Scott_1992}.  

Figure \ref{fig:alpha_plot} shows that the bulge/bar $\mgfe$-$\feh$ distribution has two main maxima, an ``$\feh$ rich-$\mgfe$ poor'' maximum  and an ``$\feh$ poor-$\mgfe$ rich'' maximum. These maxima vary in strength with position along the bar. The ``$\feh$ poor-$\mgfe$ rich'' stars dominate away from the plane while the ``$\feh$ rich-$\mgfe$ poor'' stars dominate close to the plane. This trend is weaker at smaller $|\mathrm{X}_{\mathrm{bar}}|$.

\citet{Hayden_2015} concluded using data from $\apogee$ DR12 \citep{Alam_2015} that the $\alphafe$-$\feh$ distribution of the outer bulge appeared to be a single sequence with ``high $\alphafe$-low $\feh$'' stars dominating far from the plane and ``low $\alphafe$-high $\feh$'' stars dominating close to the plane. Conversely, \citet{Rojas_2019} concluded using data from $\apogee$ DR14 that the inner bulge $\mgfe$-$\feh$ distribution was composed of two sequences, a ``high $\mgfe$-low $\feh$'' sequence and a ``low $\mgfe$-high $\feh$'' sequence, that merge above $\feh=0.15$ dex. The division of the distribution into two sequences was based on there being relatively fewer stars around $(\feh, \mgfe)\!=\!(0.1 \, \mathrm{dex} , 0.15 \, \mathrm{dex})$ which distorted the contours. Most recently, \citet{Queiroz_2020a} concluded using data from $\apogee$ DR16, that the inner bulge $\alphafe$-$\feh$ distribution consists of two sequences that, unlike what was found by \citet{Rojas_2019}, do not merge. Both sequences extend towards each other in $\feh$, but the transition between them is steep and contains very few stars. From Figure \ref{fig:alpha_plot} it is not clear if the two maxima compose two separate sequences (as is the case in the solar neighborhood) or if they are simply the two maxima of a single sequence. \citet{Queiroz_2020b} found, from examining the inner bulge $\alphafe$-$\feh$ and $\mgfe$-$\feh$ distributions using $\apogee$ DR16 data that, while a discontinuity is visible in the $\mgfe$-$\feh$ distribution, it is much stronger and steeper in the $\alphafe$-$\feh$ distribution. Because we use $\mgfe$, it may be harder for us to differentiate whether our data show two merging sequences or a single sequence with two maxima. While we cannot claim separate sequences, our data is consistent with the suggested bi-modality.

Figure \ref{fig:alpha_plot_full} is similar to Figure \ref{fig:alpha_plot} however the distance bins have been expanded. The first two columns give the $\mgfe$-$\feh$ distributions as a function of $|\mathrm{Z}|$ in the inner bulge and long bar/outer bulge region. The third column gives the $\mgfe$-$\feh$ distributions of stars with galactocentric radii ($\mathrm{R}_{\mathrm{gc}}$) less than $3.5$ kpc. Presenting the distribution in these bins has three advantages: it increases number of stars leading to better statistics, allows us to easily compare the inner bulge and long bar/outer bulge $\mgfe$-$\feh$ distributions, and the bins are comparable to those used in \citet{Rojas_2019} and \citet{Queiroz_2020a}. In the first two columns in Figure \ref{fig:alpha_plot_full}, it is only the first panel containing the inner bulge stars closest to the plane that shows a bi-modal distribution with a gap around $(\feh, \mgfe)=(0.1 \, \mathrm{dex}, 0.15 \, \mathrm{dex})$. No other panel shows a clear bi-modal distribution. Close to the plane, the $\mgfe$-$\feh$ distribution of the long bar/outer bulge region differs significantly from that of the inner bulge with most stars residing in the ``$\feh$ rich-$\mgfe$ poor'' maximum. 

In fact, we see that while the ``$\feh$ rich-$\mgfe$ poor'' stars tend to dominate close to the plane and the ``$\feh$ poor-$\mgfe$ rich'' stars tend to dominate far from the plane, these trends are weaker in the inner bulge than in the long bar/outer bulge region. Instead, in the inner bulge, the ``$\feh$ poor-$\mgfe$ rich'' stars extend lower into the plane and the ``$\feh$ rich-$\mgfe$ poor'' stars extend higher out of the plane than they do in the long bar/outer bulge. This is similar to what we saw in Figure \ref{fig:alpha_plot}, although now we see it with greater statistics. Many authors take the bulge to be within  $\mathrm{R}_{\mathrm{gc}}<3.5$ kpc. When we apply this distance cut in the third column of Figure \ref{fig:alpha_plot_full} we again see that the ``$\feh$ rich-$\mgfe$ poor'' stars dominate close to the plane while the ``$\feh$ poor-$\mgfe$ rich'' stars dominate far from the plane. Additionally, there is a clear bi-modality for stars with $|\mathrm{Z}|<0.5$ kpc and $0.5 < |\mathrm{Z}| \, (\mathrm{kpc})< 1$. We confirm the clear bi-modality in the $\feh$-$\mgfe$ distribution in the bulge seen by other authors. However, by dividing the bulge into inner bulge and long bar/outer bulge we see that the bi-modal distribution only occurs in the inner bulge. From Figure \ref{fig:alpha_plot_full} we still can not claim that the two maxima we observe originate from different sequences. 

A large fraction of the stars in the inner galaxy are likely contributed by the thin and thick disks. Additionally, a significant fraction of the halo's stellar mass is thought to reside in the inner galaxy. In Figures \ref{fig:alpha_plot} and \ref{fig:alpha_plot_full} we draw lines that differentiate regions of the $\mgfe$-$\feh$ space generally populated by the halo, thin disk, and thick disk. These lines are drawn by eye and are similar to those in \citet{Adibekyan_2011, Mackereth_2019_2, silva2020coformation}. From examining the stars relative to the lines, we see that stars with abundances similar to the thin disk generally dominate close to the plane while stars with abundances similar to the thick disk generally dominate far from the plane. For small $|\mathrm{X}_{\mathrm{bar}}|$, the ``$\mgfe$ poor-$\feh$ rich'' stars associated with the thin disk dominate at larger $|\mathrm{Z}|$ than they do at larger $|\mathrm{X}_{\mathrm{bar}}|$. This could be a result of the thin disk stars being more efficiently mapped to larger $|\mathrm{Z}|$ during the buckling episodes that built the b/p bulge. We also see that the stars in the very inner bulge ($|\mathrm{X}_{\mathrm{bar}}|<1$ kpc and $|\mathrm{Z}|<0.3$ kpc) are more $\feh$-poor and $\mgfe$-rich than stars in the neighboring bins. As these stars reside mainly in the thick disk region, these stars could be thick disk stars. The thick disk is more centrally concentrated than the thin disk and therefore could dominate over the thin disk in the very center. 

Lastly, how do the Gaussian fits in Section \ref{MDFshape} correspond to the two maxima we see in the $\feh$-$\mgfe$ plane? It is clear from their weight behavior with position, mean $\feh$ values, and their correlations to the Mg-DF Gaussians that the MDF Gaussians A and C mainly populate the ``$\feh$ rich-$\mgfe$ poor'' and ``$\feh$ poor-$\mgfe$ rich'' maxima respectively. Gaussian B from the MDF decomposition appears to be a separate component with strong contamination by the wings of the others.

\subsection{Bulge Chemokinematics}\label{kinematics}
\begin{figure*}
\centering
\includegraphics{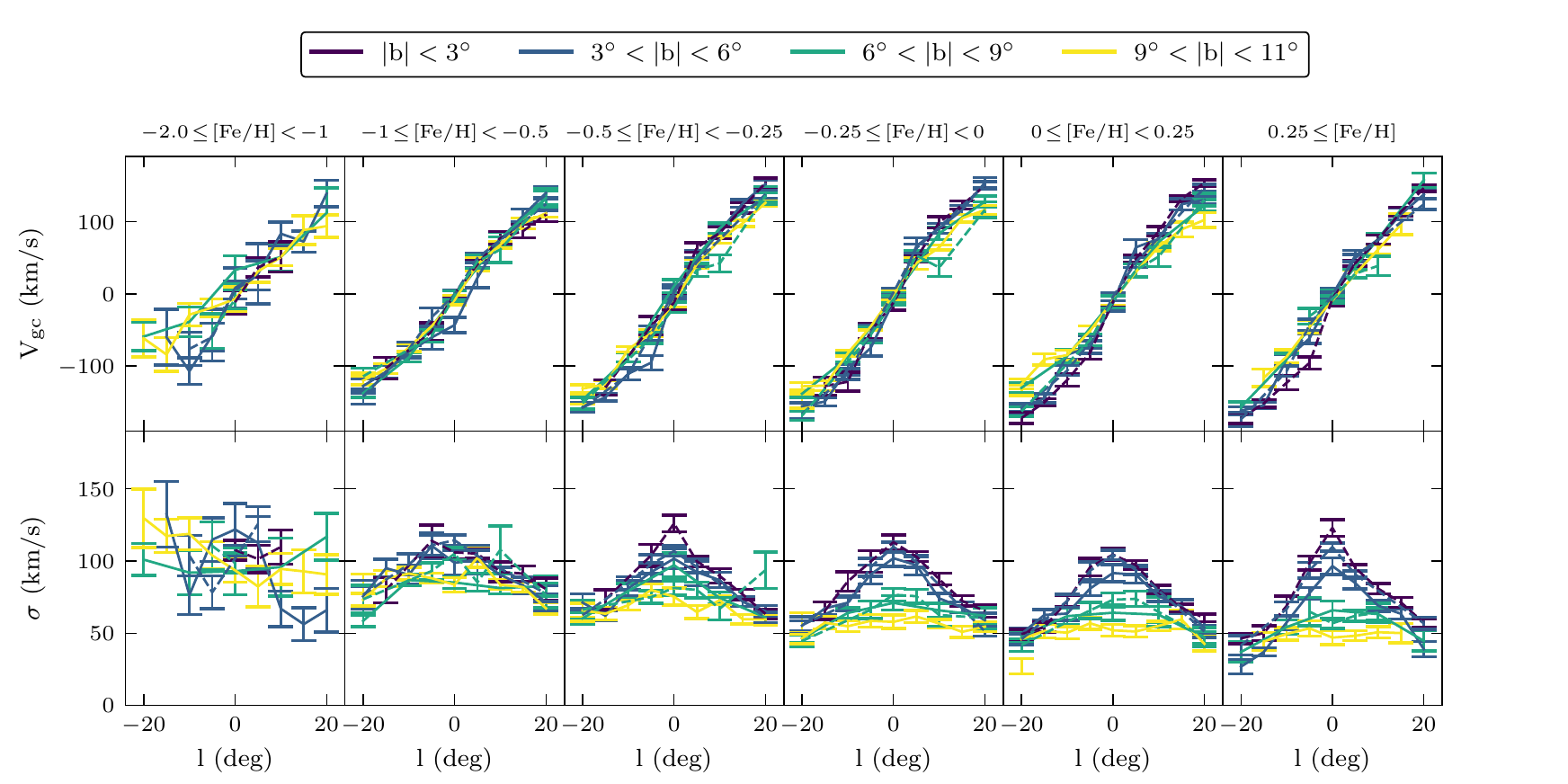}
\caption{Mean radial velocity (top) and velocity dispersion (bottom) versus longitude and latitude in different $\feh$ bins. The line colour indicates the latitude range. Dashed lines connect $\apogee$ fields, solid lines connect $\atoa$ fields. All stars are required to have $\mathrm{R}_{\mathrm{gc}}< 4.5 \, \mathrm{kpc}$.}
\label{fig:Vgc}
\end{figure*}

In this section we look at how the radial velocity and dispersion vary with $\feh$ and field position.

In Figure \ref{fig:Vgc}, we plot the mean velocity (top row) and velocity dispersion (bottom row) of stars against longitude and latitude in different $\feh$ intervals. In order to increase the number of $\apogee$ stars at negative longitudes, we use the HQ $\apogee$ bulge MSp for this analysis, no longer requiring the $\apogee$ stars to have good SSF estimates. Thus, we include $\apogee$ stars from fields represented by both the blue ellipses and red crosses in Figure \ref{fig:fieldloc} out to longitudes of $\pm 22\degree$. Because we are including $\apogee$ stars with no SSF estimates, we do not correct the $\apogee$ catalog to the HQ $\twomass$ catalog. However, we still correct the $\atoa$ catalog to the HQ $\twomass$ catalog it was selected from. Lastly, we restrict the stars to the bulge region by requiring them to have $\mathrm{R}_{\mathrm{gc}}\leq 4.5 \, \mathrm{kpc}$. The error bars in Figure \ref{fig:Vgc} were determined using bootstrapping. 

We see from Figure \ref{fig:Vgc} that for stars with $\feh>-1$ dex, the mean velocity profiles do not significantly vary with $\feh$ and that stars in the bulge rotate cylindrically regardless of $\feh$. There is a weak trend with latitude where stars located closer to the Galactic plane tend to rotate slightly faster than stars located farther from the Galactic plane. Other studies have observed cylindrical rotation in the bulge \citep{Howard_2009, Kunder_2012, Zoccali_2014} and it is a known property of b/p bulges from N-body simulations \citep[e.g.][]{Athanassoula_2002, Saha_2013}. Below $-0.5$ dex, the stars rotate more slowly with decreasing $\feh$. The stars with $\feh<-1$ dex rotate significantly slower than the more $\feh$-rich stars.

The velocity dispersion profiles, shown in bottom row of Figure \ref{fig:Vgc}, behave differently with $\feh$. Far from the Galactic plane, the shapes of the velocity dispersion profiles are flat for $\feh>-1$ dex but the amplitudes increase with decreasing $\feh$ for $-1<\feh \, (\mathrm{dex})<0$. Near the Galactic plane and for $\feh>-0.5$ dex, the shapes of the velocity dispersion profiles are high and peaked around $|\mathrm{l}|=0\degree$, but the change in amplitude is less with decreasing $\feh$. For stars with $\feh>0$ dex, the dispersion does not change significantly with $\feh$. Thus contrary to the velocity profiles, the velocity dispersion profiles of stars with $\feh<0$ dex, do show a clear $\feh$ dependence.

The dispersion profile of stars with $-1<\feh \, (\mathrm{dex})<-0.5$ differs from those with $\feh>-0.5$ dex as the separation between latitudes in the inner longitudes is much weaker and the profiles of stars closer to the plane are flatter. However, we see that, as is the case for the more $\feh$-rich stars, the increase in dispersion is stronger for the stars at latitudes farther from the plane ($|\mathrm{b}|>7.5\degree$). Furthermore, while the dispersion profiles of the stars closer to the plane are flatter than their $\feh$-rich counterparts, the distributions of these stars are still peaked around the central longitudes. Therefore, we conclude that the dispersion profiles of stars with $-1<\feh \, (\mathrm{dex})<-0.5$ are higher dispersion, slightly flatter versions of the profiles of the stars with $\feh>-0.5$ dex.

We find that the stars with $-2<\feh \, (\mathrm{dex})<-1$ have different dispersion profiles than the more $\feh$-rich stars. Apart from the $\atoa$ stars between $3\degree<|\mathrm{b}|<6\degree$, the dispersion profiles of the low latitude stars are not peaked towards the central longitudes. Furthermore, the dispersion of the stars at large absolute longitudes varies significantly between the different latitudes. The range of $\feh$, the lower rotation, and higher dispersion is consistent with these stars being part of the halo or very $\feh$-poor old thick disk.

We examined the velocity and velocity dispersion profiles of $\apogee$ stars in the HQSSF $\apogee$ bulge MSp corrected to the HQ $\twomass$ catalog. We find that while the data are noisier due to the fewer number of stars and fields, we see similar trends between $\feh$, kinematics, and positions as in the case when we do not correct for the SSF. Therefore, we believe that the longitude and latitude bins we have chosen are small enough that the effects of the SSF are minimal. As a further check of the effect of the SSF, we compared out results to those of \citet{Rojas_Arriagada_2020}, who found using $\apogee$ DR16 data that the dispersion profiles of the more metal poor stars with $-1.2<\feh \,(\mathrm{dex})\,<-0.5$ in the bulge ($\mathrm{R}_{\mathrm{gc}}\leq 3.5 \, \mathrm{kpc}$) are flat even for the stars at latitudes close to the plane. We find that we can reproduce the trends they find, especially those at low metallicities, if we use their metallicity, longitude, and latitude bins. However, using their bins, we find that the dispersion of the very $\feh$-rich stars that are close to the plane and near $|\mathrm{l}|=0\degree$ is lower than what they report. We suspect that this is due to discrepancies on how we correct for the $\apogee$ SSF. However, this difference is small and we still observe similar trends as \citet{Rojas_Arriagada_2020} at the high $\feh$ end and therefore do not believe the lack of correction for the SSF will significantly affect the main trends we see or the interpretation of our results. Furthermore, the strong agreement we see between the $\apogee$ and $\atoa$ profiles also supports that the effect of the $\apogee$ SSF is minor.

\section{Discussion}\label{Discussion}
Several authors have shown evidence that the presence of a classical bulge in the Milky Way is at most minimal \citep{Shen_2010, DMatteo_2014, Portail_2017, clarke2019milky, Queiroz_2020b}. Instead, the Milky Way's bulge is likely to be of mainly disk origin. However, whether the bulge formed from an evolving thin disk, a combination of distinct thin and thick disks, or a disk-continuum is currently debated. N-body models of b/p bulges built from single thin disks with inside-out population gradients have reproduced several characteristics observed in the Milky Way's b/p bulge such as cylindrical rotation \citep{Shen_2010} or the vertical metallicity gradient \citep{Martinez_2011}. However, these models also miss many of the finer chemo-kinematic properties seen in the bulge \citep{Di_Matteo_2015, Fragkoudi_2017, DMatteo_2019}.

To put further constraints on the formation of the Galactic bulge, we compare the predictions from a number of models exploring different evolutionary scenarios to our principal results.

\vspace{\baselineskip} 
\noindent\textbf{Bulge Iso-[Fe/H] Contours:}
The bulge's iso-$\feh$ and iso-$\mgfe$ contours are X-shaped and exhibit stronger "pinching" than the bulge's density distribution; see Figure \ref{fig:Z_Xbar}. 

\citet[][DB17]{Debattista_2017} presented an evolving single disk model with continuous star formation that evolved to form a bar which subsequently buckled, forming a b/p bulge. The iso-$\feh$ contours of the resulting b/p bulge are X-shaped and more pinched than its density distribution. \citetalias{Debattista_2017} argued that the stronger pinching of the $\feh$ distribution in their model was the result of kinematic fractionation, which they defined as the separation of stellar populations by the bar due to their different initial radial kinematics. However, \citet[][DM19]{DMatteo_2019} showed that separation due to differences in vertical dispersion and scale-height is similarly important. They created two N-body b/p bulge models using three disks, which either differed in their radial dispersions or in their vertical dispersions and scale-heights, but always had the same radial scale-lengths. Both models produced b/p bulges with iso-$\feh$ contours that were X-shaped and more pinched than their density distributions (see their Figure 5). Lastly, \citet[][F18]{Fragkoudi_2018} built an N-body b/p bulge model initialised from three co-spatial disks each with different kinematics, scale heights, scale lengths, and abundance distributions. The iso-$\feh$ contours of their final model were also X-shaped and more strongly pinched than the density distribution.  

Therefore we cannot use our result at current resolution to differentiate between the models, however the stronger pinching of the iso-$\feh$ contours with respect to the density distribution further verifies that the bulge has a mainly disk origin.

\vspace{\baselineskip}  
\noindent\textbf{Bulge Radial Gradient:}
In the Galactic plane, the Milky Way's bulge has clear radial gradients in $\feh$ and $\mgfe$, such that close to the plane, the inner bulge is more $\feh$-poor and $\mgfe$-rich than the long bar/outer bulge region; see Figures \ref{fig:l_b_femg}-\ref{fig:Ybar_Xbar} and Figure \ref{fig:hor_grad}.

N-body models initialized as a single disk with a strong negative radial metallicity gradient, such as to reproduce the observed vertical gradient in the bulge after bar evolution, inevitably also show a steep radially inward rise in $\feh$ \citep{Di_Matteo_2015, Fragkoudi_2017}.  Unlike these models, the longitudinal abundance gradient of the multi-disk model of \citetalias{Fragkoudi_2018} closely matches the gradients we see in Figure \ref{fig:l_b_femg} (see their Figures 9 and 13; already matched to $\apogee$ DR13 data). This model can reproduce the bulge's final radial $\feh$ gradient because the $\feh$-poor and $\alphafe$-rich initial disks dominate in the central regions, due to their scale lengths being shorter than that of the $\feh$-rich and $\alphafe$-poor disk. On the other hand, the three-disk models of \citetalias{DMatteo_2019}, where all disks were initialised with the same radial scale-lengths, showed similar or larger final $\feh$ in their centres than in their in-plane bar regions (see Figure 5 of \citetalias{DMatteo_2019}). In the star-forming, evolving disk model of \citetalias{Debattista_2017}, the highest metallicities occur in the centre of the final b/p bulge (see their Figures 11, 24, 26). While this model was built from a single disk, before the bar formed the older populations have lower metallicities as well as larger radial and vertical velocity dispersions, and scale heights, than the younger populations which end up with the largest central concentration.

Thus it appears that the critical ingredient in these models for explaining the radial bulge/bar $\feh$ gradient is the shorter radial scale-length of the initial $\feh$-poor disk. However, the \citetalias{Fragkoudi_2018} model assumed a thin disk radial scale-length of $4.8$ kpc, significantly larger than in current observations \citep{Bland_Hawthorn_2016}, and more than twice the thick disk scale-length. This suggests other factors are at play as well.

\citet{Bovy_2019} and \citet{Hasselquist_2020} show age maps for bulge and long bar stars from $\apogee$ DR16. These maps indicate a significant fraction of younger stars in the long bar and surrounding disk for $\mathrm{R}_{\mathrm{gc}}>3.5$ kpc and close to the Galactic plane, which are not prominent at smaller radii and at heights above a few $100$ pc. 
This suggests that after the formation of the bulge, the Galactic bar either experienced substantial additional gas inflow and star formation, or captured a significant population of disk stars, by slowing its pattern rotation and increasing in size. Capturing disk stars could have happened both intermittently, if indeed the bar regularly changes its pattern speed on dynamical time-scales, due to  interaction with spiral arms \citep{Hilmi_2020}, or secularly by angular momentum transfer to the halo \citep{Chiba_2021}. Both mechanisms would be consistent with the dominance of the super-thin bar \citep{Wegg_2015} at these radii. Using the stellar ages from \citet{Mackereth_2019}, and restricting the APOGEE sample to ages $>6$ Gyr or $>8$ Gyr, we still find a horizontal gradient near the disk plane in the remaining bulge sample, but considerably milder. This could be built by a less extreme version of the \citetalias{Fragkoudi_2018} model, with thin and thick disk scale-lengths closer to standard values.

\vspace{\baselineskip} 
\noindent\textbf{Outer Bulge Vertical Gradient:}
The vertical $\feh$ and $\mgfe$ gradients in the long bar/outer bulge region are ${\sim}-0.44$ dex/kpc and ${\sim} 0.13{\sim}0.17$ dex/kpc respectively (see Figures \ref{fig:Z_Xbar} and \ref{fig:grad}). In this region, there is no inner flattening of the vertical gradient near the plane as is the case in the inner bulge.

For their three-disk models \citetalias{DMatteo_2019} showed that while both produce b/p bulges, the region beyond the lobes of the b/p bulge in the model formed from the disks with different radial dispersion has no vertical metallicity gradient (see Figure 6 in \citetalias{DMatteo_2019}). However, in the model built from disks with different vertical scale heights, a vertical gradient remains in the outer bar. The \citetalias{DMatteo_2019} models are not realistic and are instead designed to represent extreme cases. In reality, the in-plane and vertical dispersions of a disk should be correlated such that a disk with higher (lower) radial dispersion should also have higher (lower) vertical dispersion and therefore a larger (smaller) scale height. The \citetalias{Fragkoudi_2018} model is more realistic in that the disks with higher dispersions, which are designed to be more $\feh$-poor, have larger scale heights. Due to the interplay with the disks' scale lengths, the \citetalias{Fragkoudi_2018} model shows a strong vertical gradient in the disk regions.

The single-disk model of \citet{Fragkoudi_2017} with an initial radial $\feh$ gradient did not lead to such a vertical gradient in the outer bar. The evolving disk model, \citetalias{Debattista_2017}, does show a vertical gradient in the disk. In this model the older populations, which tend to be more metal poor, have larger scale heights than the younger populations, which tend to be metal rich.

This comparison supports a multi-disk formation scenario where the disks that form the bulge must differ in vertical velocity dispersion, and therefore scale height, in order to produce a bulge where the region outside the b/p bulge lobes has strong vertical $\feh$ and $\mgfe$ gradients. 

\vspace{\baselineskip} 
\noindent\textbf{The [Fe/H]-[Mg/Fe] Distribution:}
The $\feh$-$\mgfe$ distribution along the bar is roughly linear and composed of two maxima: an ``$\feh$ rich-$\mgfe$ poor'' maximum and an ``$\feh$ poor-$\mgfe$ rich'' maximum. These two maxima vary in strength with position in the bulge such that the $\feh$-rich maximum dominates near the plane while the $\feh$-poor maximum dominates far from the plane; see Figures  \ref{fig:alpha_plot} and \ref{fig:alpha_plot_full}. At small $|\mathrm{X}_{\mathrm{bar}}|$, this trend is also present, although weaker than at larger $|\mathrm{X}_{\mathrm{bar}}|$. There is no clear evidence for two distinct sequences in Figures \ref{fig:alpha_plot} and \ref{fig:alpha_plot_full} \citep[but see also][]{Queiroz_2020b}.

The stronger bi-modality in the $\feh$-$\alphafe$ distribution is often interpreted as two distinct star formation episodes well separated by a period of quenched star formation \citep{Chiappini_1997, Haywood_2018}. The $\apogee$ DR16 $\feh$-$\mgfe$ distribution of the bulge ($\mathrm{R}_\mathrm{gc}<3$ kpc and $|\mathrm{Z}|<0.5$  kpc) was recently modelled by \citet{Lian_2020} using a chemical evolution model with an early period of high star formation which formed the high $\mgfe$ stars, followed by a quick star formation quenching episode and a long lived period of low star formation which formed the low $\mgfe$ stars. This model reproduces the $\feh$-$\mgfe$ distribution of the inner bulge (see Figure \ref{fig:alpha_plot_full}) and the bi-modal Mg-DFs of the inner bulge seen in $\apogee$ (Figure \ref{fig:inner_mdf}) \citep[see also][]{Matteucci_2019}.

\vspace{\baselineskip} 
\noindent\textbf{Kinematics with Metallicity along the [Fe/H]-[Mg/Fe] Distribution:}
The stars composing the two maxima in the $\feh$-$\mgfe$ distribution display different kinematics. Specifically, the stars in the $\feh$-rich maximum (mainly $\feh>0$ dex) are kinematically colder than the stars in the $\feh$-poor maximum (mainly $-1<\feh \, (\mathrm{dex})<-0.25$). Furthermore, the stars composing the $\feh$-rich maximum have little kinematic-$\feh$ dependence while the stars composing the $\feh$-poor maximum rotate slightly slower and increase in dispersion with decreasing $\feh$; see Figure \ref{fig:Vgc}.

\citet{Di_Matteo_2015} showed using an N-body model that if the origin of a b/p bulge is a single disk with a strong negative radial $\feh$ gradient, then the $\feh$-poor stars should rotate faster than, and have warmer, although similarly shaped, velocity dispersion profiles to the $\feh$-rich stars as in this scenario the $\feh$-poor stars originate from larger disk radii. The lack of kinematic dependence on $\feh$ of the stars in the $\feh$-rich maximum and the decrease in mean velocity with decreasing $\feh$ in the $\feh$-poor maximum indicates that the origin of the Galactic bulge cannot be a single disk with a strong negative radial $\feh$ gradient. Instead, kinematic and spatial differences of the maxima support a scenario where the stars composing the maxima originate from at least two separate disks, with the stars composing the $\feh$-rich maximum originating from a colder disk with little to no radial $\feh$ gradient and the stars composing the $\feh$-poor maximum originating from a hotter disk.

\citetalias{Debattista_2017} found that in order for the stars with $\feh<-0.5$ dex in their model to have the high, flat, and largely latitude independent dispersion profiles observed in \citet{Ness_2013_IV}, they needed to add a slowly rotating, low mass ($5\%$ central Milky Way mass), high dispersion component to their model. They associated this component with the stellar halo (see their Figure 21). From our data, the slower rotation and increased dispersion of the stars with $-1 < \feh \, (\mathrm{dex}) < -0.5$ as compared to the stars with  $-0.5 < \feh \, (\mathrm{dex}) < -0.25$ may be the result of contamination by the high dispersion component dominating below $-1$ dex. As the stars in the $\feh$-poor tail are slowly rotating, kinematically hot, and $\mgfe$-rich, we tentatively associate them with the stellar halo. A contamination by halo stars is further supported by \citet{Lucey_2021} who, from examining the chemo-kinematics of metal poor bulge stars, found that the fraction of halo interlopers in the bulge increases with decreasing metallicity between $-3<\feh \, (\mathrm{dex})<0.5$.

\vspace{\baselineskip} 
In all, our data paints a consistent picture for the origin of the b/p bulge: at least two initial disks with differing dispersions, scale heights, and scale lengths underwent spatial and kinematic fractionation resulting in the b/p bulge of the Milky Way that we observe today. Due to their kinematics as well as their $\feh$-$\mgfe$ distributions, we associate these disks with the thin and thick disks. We also associate the very $\feh$-poor, $\mgfe$-rich, kinematically hot tail of the bulge stars with the stellar halo.

\section{Summary and Conclusions}\label{Conclusion}
In this work we used the data driven method, $\thecannon$, to put stars from the $\argos$ survey onto the parameter and abundance scales of the $\apogee$ survey. After doing so, we could directly combine the two surveys and gain a deeper and more reliable coverage of the Galactic bulge. 

In the first half of the paper, we describe how we applied $\thecannon$ to the $\argos$ stars to obtain the $\atoa$ survey. To show that we have successfully placed the $\atoa$ stars  on the $\apogee$ label scales, we performed three validation tests: the pick-one-out test (Figure \ref{fig:pickoneout}), spectrum reconstruction (Figure \ref{fig:model_data}), and stellar $\teff$-$\logg$ distribution (Figure \ref{fig:hrd}). These tests showed that it is possible to perform $\thecannon$ label transfer using a moderate number ($204$) of reference set stars and still obtain labels with good precisions. After performing the validation tests, we accounted for the SSF of each survey such that after it is corrected for, we statistically obtain the HQ $\twomass$ catalogs they were selected from. After this, we compared the SSF-corrected MDFs and Mg-DFs of the $\apogee$ and $\atoa$ surveys and found that the different spatial regions probed by each survey cause a clear spatial bias, however when the distribution functions are compared in fixed distance bins, the MDFs and Mg-DFs agree except at the high $\mgfe$ and low $\feh$ ends where $\apogee$ observes more stars. This may be due to trends between $\teff$ and the abundances in $\apogee$ and the fact that the reference set only covers a limited $\teff$ range on the $\apogee$ scale. 

In the second half of the paper, we used stars from both the $\atoa$ and $\apogee$ surveys to investigate the abundance structure of the bulge. The results we found include:
\begin{enumerate}[label=(\roman*),widest=99,itemindent=*,leftmargin=0pt]
  \item The $\feh$ and $\mgfe$ maps built using $\apogee$ and $\atoa$ data show strong X-shapes that are more pinched than the density distribution given by the currently best dynamical model of the Milky Way's bulge/bar from \citet{Portail_dyn_2017}. The stronger pinching in the $\feh$ and $\mgfe$ maps than in the density map supports a mainly disk origin for the Galactic bulge.
  
\vspace{\baselineskip} 
  \item The inner bulge and outer bulge/long bar region have different chemical properties. While the inner bulge $\feh$ and $\mgfe$ profiles are nearly flat within $0.7$ kpc of the Galactic plane and then steepen to ${\sim} -0.41$ dex/kpc and ${\sim} 0.11$ dex/kpc respectively, the vertical $\feh$ and $\mgfe$ profiles of the outer bulge/long bar region are steep near the plane at ${\sim} -0.44$ dex/kpc and ${\sim} 0.13{\sim}0.17$ dex/kpc respectively, with a flat distribution for $|\mathrm{Z}|>1.25$ kpc. Close to the plane the inner bulge is sub-solar in $\feh$ and $\mgfe$-rich while the flat bar is nearly solar in $\feh$ and $\mgfe$-normal.
  
\vspace{\baselineskip} 
  \item The $\feh$-$\mgfe$ distributions in the bulge and long bar have two main maxima, an ``$\feh$ rich-$\mgfe$ poor'' maximum and an ``$\feh$ poor-$\mgfe$ rich'' maximum. The ``$\feh$ rich-$\mgfe$ poor'' maximum dominates close to the plane, has lower dispersion, and shows no significant mean radial velocity dependence on $\feh$. The `$\feh$ poor-$\mgfe$ rich'' maximum dominates far from the plane, has higher dispersion, and its $\feh$-poorer stars rotate slightly slower on average than its $\feh$-richer stars.
  
\vspace{\baselineskip} 
  \item The most $\feh$-poor stars ($\feh<-1$ dex) rotate slowly and have high flat dispersion profiles. We associate these stars with the stellar halo. This is also supported by the distribution of these $\feh$-poor stars in the $\feh$-$\mgfe$ plane in Figure \ref{fig:alpha_plot}.
 
\end{enumerate}

The positive horizontal $\feh$ gradient in the bulge close to the Galactic plane and the negative vertical $\feh$ gradient in the long bar region favour models in which the bulge/bar formed from initial thin and thick disks with different vertical and radial scale-lengths \citep{Fragkoudi_2018}. This multi-disk origin is further supported by the higher pinching of the abundance distributions in the bulge as compared to the density distributions and the differing kinematics of the low and high $\feh$ stars. However, the large thin-disk scale-lengths required by these models, together with younger estimated mean ages in the outer bar \citep{Bovy_2019, Hasselquist_2020} suggest that the Galactic bar may have captured younger, more metal-rich stars well after its formation.

\begin{acknowledgements}
J. Bland-Hawthorn is funded by an ARC Laureate Fellowship that partially supports the $\galah$ survey science team.\\

Funding for the Sloan Digital Sky Survey IV has been provided by the Alfred P. Sloan Foundation, the U.S. Department of Energy Office of Science, and the Participating Institutions. SDSS-IV acknowledges
support and resources from the Center for High-Performance Computing at
the University of Utah. The SDSS web site is www.sdss.org.

SDSS-IV is managed by the Astrophysical Research Consortium for the 
Participating Institutions of the SDSS Collaboration including the 
Brazilian Participation Group, the Carnegie Institution for Science, 
Carnegie Mellon University, the Chilean Participation Group, the French Participation Group, Harvard-Smithsonian Center for Astrophysics, 
Instituto de Astrof\'isica de Canarias, The Johns Hopkins University, 
Kavli Institute for the Physics and Mathematics of the Universe (IPMU) / 
University of Tokyo, the Korean Participation Group, Lawrence Berkeley National Laboratory, 
Leibniz Institut f\"ur Astrophysik Potsdam (AIP),  
Max-Planck-Institut f\"ur Astronomie (MPIA Heidelberg), 
Max-Planck-Institut f\"ur Astrophysik (MPA Garching), 
Max-Planck-Institut f\"ur Extraterrestrische Physik (MPE), 
National Astronomical Observatories of China, New Mexico State University, 
New York University, University of Notre Dame, 
Observat\'ario Nacional / MCTI, The Ohio State University, 
Pennsylvania State University, Shanghai Astronomical Observatory, 
United Kingdom Participation Group,
Universidad Nacional Aut\'onoma de M\'exico, University of Arizona, 
University of Colorado Boulder, University of Oxford, University of Portsmouth, 
University of Utah, University of Virginia, University of Washington, University of Wisconsin, 
Vanderbilt University, and Yale University.
\end{acknowledgements}

\section*{ORCID iDs}
\noindent S.M. Wylie \orcid{0000-0001-9116-6767} \href{https://orcid.org/0000-0001-9116-6767}{https://orcid.org/0000-0001-9116-6767}\\
\noindent O.E. Gerhard \orcid{0000-0003-3333-0033} \href{https://orcid.org/0000-0003-3333-0033}{https://orcid.org/0000-0003-3333-0033}\\
\noindent M.K. Ness \orcid{0000-0001-5082-6693}
\href{https://orcid.org/0000-0001-5082-6693}{https://orcid.org/0000-0001-5082-6693}\\
\noindent J.P. Clarke \orcid{0000-0002-2243-178X} \href{https://orcid.org/0000-0002-2243-178X}{https://orcid.org/0000-0002-2243-178X}\\

\bibliography{mknbib}

\begin{appendix}
\section{Effect of log(g) Limits on the A2A Red Clump Sample Completeness}\label{aprccomp}
\begin{figure}
\centering
\includegraphics[width=\columnwidth]{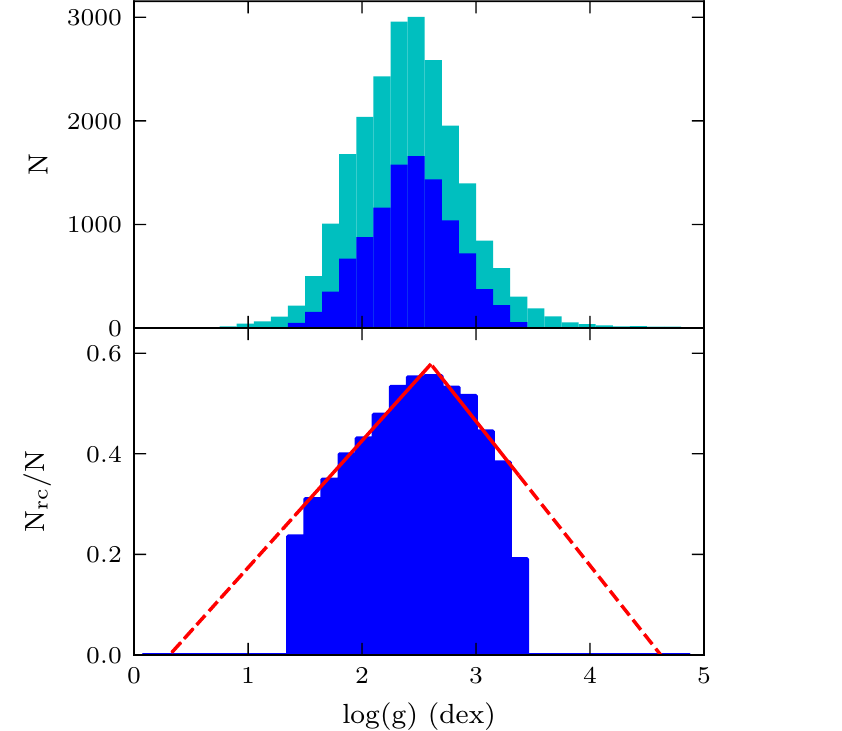} 
\caption{Top: The number of stars as a function of $\argos$ $\logg$. The cyan histogram gives the number of stars after the $\teff$, $\mgfe$, $\feh$, and colour limits (\ref{first:a}, \ref{third:a}, \ref{third:b}, and \ref{second}) have been applied. The blue histogram gives the number of $\atoa$ RC stars per $\argos$ $\logg$ bin. Bottom: Fraction of $\atoa$ RC stars as a function of $\argos$ gravity (ratio of blue to cyan histogram in the top plot). The red lines are linear fits to the distribution.} 
\label{fig:sf_gravity}
\end{figure}

We can approximate the effect of the $\logg$ limits (\ref{third:b}) on the completeness of the $\atoa$ RC stars. In the top plot of Figure \ref{fig:sf_gravity}, the cyan histogram shows the number of $\argos$ stars as a function of $\argos$ $\logg$. The colour cut in Equation \ref{second} and reference set limits \ref{first:a}, \ref{third:a}, \ref{third:b} have been applied. The gravity reference set limit \ref{third:b} has not been applied. The blue histogram in the top plot of Figure \ref{fig:sf_gravity} gives the number of $\atoa$ RC stars at each $\argos$ $\logg$. Notice that because the blue histogram is built from $\atoa$ stars, the $\logg$ limits (\ref{third:b}) have been applied and the blue histogram does not extend as far in $\logg$ as the cyan histogram. In the bottom plot of Figure \ref{fig:sf_gravity}, we take the ratio of the two histograms to get the fraction of $\atoa$ RC stars at each $\argos$ $\logg$. At $\logg = 2.6$ dex, $55.5\%$ of the stars in the cyan histogram are $\atoa$ RC stars. The fraction decreases linearly outward in both directions to $31\%$ at $1.6$ dex and $38.0\%$ at $3.2$ dex. We ignore the final bins because the $\logg$ limits fall within these bins causing the number of RC stars to be artificially lower. Assuming the linear trend on both sides continues past the limits, we fit a line to each side and extrapolate beyond the cuts to get the fraction of RC stars at each $\argos$ $\logg$. We then multiply the number of stars given by the cyan histogram in the outer bins by the fractions we get from extrapolating to get the approximate number of RC stars removed by the $\logg$ limits. Using this method, we find that approximately $200$ RC stars are removed by the $\logg$ limits. Under the assumption that, to the first order, the other limits (\ref{first:a}, \ref{third:a}, and \ref{third:b}) affect the RC and non-RC stars equally, then $92\%$ of the RC stars originally observed by $\argos$ are accounted for in the $\atoa$ catalog.

\section{Abundance Trends in APOGEE and the Effect on A2A}\label{abundtrends}
\begin{figure}
\centering
\includegraphics{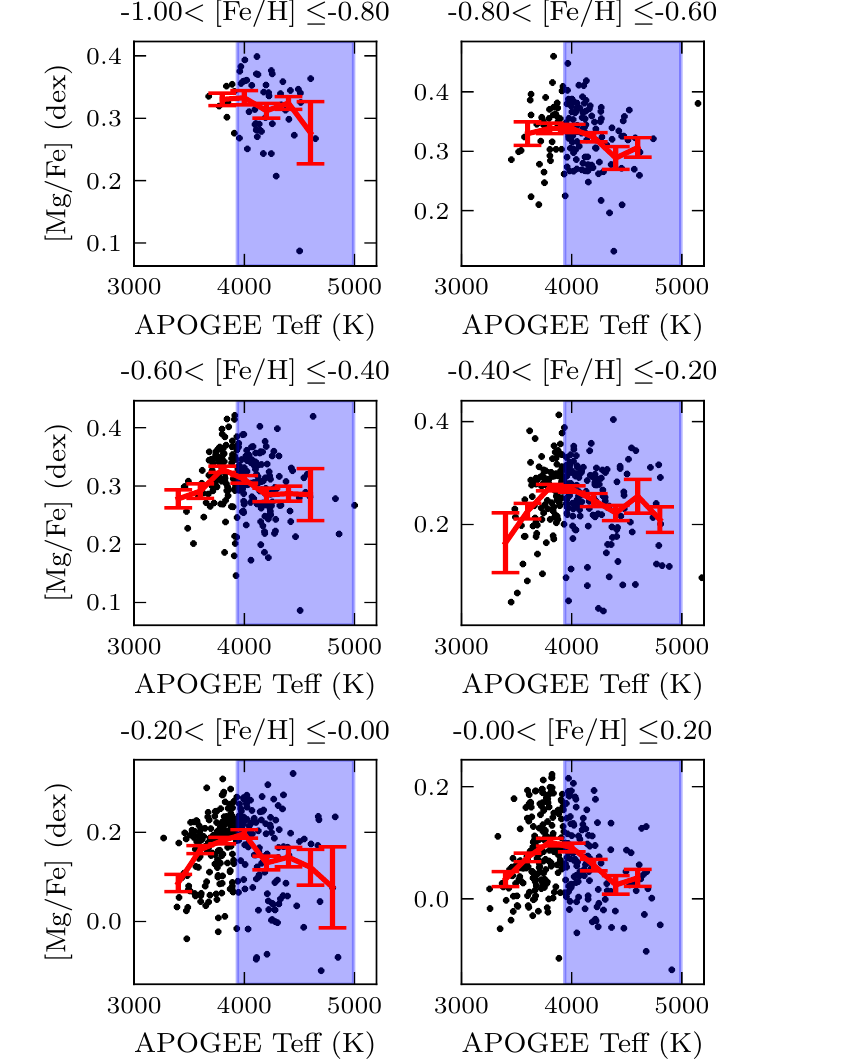}
\caption{The ASPCAP $\teff$-$\mgfe$ distribution of $\apogee$ stars in different $\feh$ bins with $6<\mathrm{Ds} \, (\mathrm{kpc})<8$, $|\mathrm{Z}|<1$ kpc, and $100<\mathrm{SNR}<200$. The red line gives the running mean of the distribution while the blue shaded area gives the $\teff$ range spanned by the reference set on the $\apogee$ scale.}
\label{fig:tmgtrend}
\end{figure}

\begin{figure}
\centering
\includegraphics[width=\columnwidth]{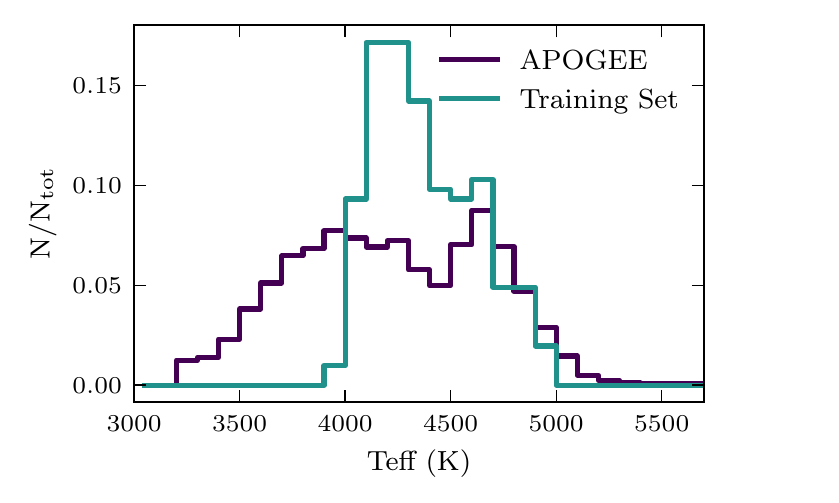} 
\caption{$\teff$ distributions of the HQSSF $\apogee$ bulge MSp and of the reference set used to put the $\atoa$ survey on the $\apogee$ parameter scale.}
\label{fig:teffcompare}
\end{figure}

It has been found in $\apogee$ that the ASPCAP $\teff$ is correlated with some of the ASPCAP abundances due to the physical stellar models \citep{jonsson_2018, Jofre_2019}. We show the correlation between $\teff$ and $\mgfe$ of stars in bins of $\feh$ in Figure \ref{fig:tmgtrend}. For all plots, the stars are restricted to narrow ranges in distance, SNR, and height from the plane. The reference set we use to train $\thecannon$ model to build the $\atoa$ catalog does not span the entire $\teff$ range covered by the HQSSF $\apogee$ bulge MSp; see Figure \ref{fig:teffcompare}.

\end{appendix}
\end{document}